\documentclass[11pt,final]{article}
\usepackage{amsfonts}
\usepackage{amsmath}
\usepackage{amssymb}
\usepackage{amsthm}
\usepackage{cite}
\usepackage[margin=1.0in]{geometry}
\usepackage{graphicx}
\usepackage{color}

\usepackage[labelfont=bf, labelsep=period, font=small]{caption}
\usepackage{subcaption}

\usepackage{bm}
\usepackage{enumitem}

\definecolor{cite_blue}{rgb}{0,0,0.97}
\definecolor{link_red}{rgb}{0.7,0,0}

\def\bE{{\mathbb{E}}}

\def\eps{\varepsilon}

\def\spp{\,\,}

\def\bsigma{\bm{\sigma}}

\title{Stochastic and Coarse-Grained Two-Dimensional Modeling of Directional Particle Movement}

\date{\today}

\author{
William Ott\thanks{\texttt{ott@math.uh.edu}}~,~~Ilya Timofeyev\thanks{\texttt{ilya@math.uh.edu}}~,~~Thomas Weber\thanks{\texttt{spathiwa@math.uh.edu}}
\\
{\small \itshape University of Houston}
}

\begin{document}
\maketitle
\begin{abstract}
We study the evolution of interacting groups of pedestrians in two-dimensional geometries.
We introduce a microscopic stochastic model that includes floor fields modeling the global flow of individual groups as well as local interaction rules.
From this microscopic model we derive an analytically-tractable system of conservation laws that governs the evolution of the macroscopic pedestrian densities.
Numerical simulations show good agreement between the system of conservation laws and the microscopic model, though the latter is slightly more diffusive.
We conclude by deriving second-order corrections to the system of conservation laws.
\end{abstract}


\section{Introduction}
\label{sec:intro}

Agent-based modeling has developed substantially in recent years.
Pedestrian dynamics and evacuation theory have partially motivated this active research area.
Microscopic agent-based models can incorporate complex, realistic rules for pedestrian interactions.
Such models consequently enhance our understanding of many aspects of evacuation theory - optimal location of exits and obstacles, exit times, et cetera (see for example~\cite{blad01, bksz01, kirch02, kirch03, lwrz02, guhu081, Varas2007631, Zheng20111042, KNT05,KYN12}). 
On the other hand, microscopic models can be computationally costly, and complex microscopic rules typically render such models analytically intractable.
This situation arises in many areas of modern nonlinear science.

Recent efforts have therefore focused on establishing better connections between realistic microscopic dynamics and coarse models. 
For example, recent papers on car traffic~\cite{soka06,duso07,also08}, 
pedestrian dynamics~\cite{burger1,burger2},
bacterial movement~\cite{lega13, lewe14}, and
convection modeling~\cite{bsbg13} address this challenge.
Coarse models offer improved computational efficiency relative to their agent-based microscopic counterparts.
Further, rich analytical machinery (e.g. conservation law theory) may be brought to bear on systems of partial differential equations (PDEs).

In this paper, we establish a detailed connection between a realistic, two-dimensional, agent-based, microscopic pedestrian flow model and corresponding coarse PDE descriptions of the dynamics.
We build on~\cite{ckpt14}, wherein Timofeyev et al. introduce a prototypical microscopic pedestrian flow model in dimension one, and then derive corresponding coarse PDE descriptions.

There exist a vast number of models for pedestrian dynamics (see~\cite{bedo11, bpt12, helbing2001, cfl09, css00, skkkrs09, schchni10, belgib15, burger1, burger2} for recent reviews in this and related areas). 
The majority of microscopic models
include both long-distance (global) interactions, and local rules for adjusting behavior due to the presence of other agents.
The following examples represent the main modeling tendencies in this field: 
Social-force local interactions~\cite{HM95}; 
avoidance schemes wherein models explicitly track the position and velocity of each agent, and agents tend to change their velocities to avoid possible collisions~\cite{degond13,avoidmodel1,avoidmodel2};
lattice models wherein agent motion is governed by 
a floor-field function, used to specify the probability of transition between lattice sites~\cite{bksz01,kirch02, kirch03}; 
and lattice models where agents avoid each other by local side-stepping rules~\cite{burger1, burger2}.

Unlike the vehicular traffic setting, pedestrian dynamics received little modeling attention until recent decades.
There exist no canonical coarse PDE models that have earned universal acceptance within the pedestrian dynamics community.
Nevertheless, interest in evacuation theory and crowd dynamics have led to the development of many coarse models in recent years.
To give a few examples:
Various fluid-like models and kinetic PDEs have been proposed~\cite{ardm11, helbing92, krss09, Hughes2002, hughes03, MMA624, coscan08, beldog08, bpt12}. 
In~\cite{burger1, burger2}, the authors derive 2D PDE models for evolving pedestrian densities from lattice models for interacting groups of pedestrians.
This work demonstrates the emergence of lane formation: An agent tends to follow agents from her own group, and avoids agents from other groups via local side-stepping rules.
In~\cite{degond13}, a kinetic approach is used to derive a coarse PDE model for density and average velocity from a microscopic avoidance model.
Our current work draws inspiration directly from~\cite{ckpt14}, wherein the authors derive a 1D PDE model for interacting groups from a microscopic lattice model that features slowdown interactions. 

Here, we study interacting groups of pedestrians in two-dimensional geometries.
We develop both a 2D microscopic stochastic lattice model and corresponding coarse PDE descriptions.
Instead of targeting various evacuation scenarios, we focus on capturing 
interactions between different groups and on how these interactions \lq\lq{}propagate\rq\rq{} into the 
the coarse PDE analog for the temporal evolution of the pedestrian densities.
The stochastic microscopic model is an exclusion process on a lattice, featuring a local interaction mechanism similar to the one-dimensional mechanism developed in~\cite{ckpt14}.
In particular, agents from different groups may simultaneously occupy the same lattice site, but the exclusion principle applies to agents from the same group.
Moreover, an agent will slow down when members of other groups are present: We explicitly formulate slowdown interactions in our modeling.
We combine these local interaction rules with a global modeling component that specifies the overall direction of motion for each group.
Our global modeling is analogous to the static floor-field approach used in~\cite{bksz01, kirch02}.

The model developed in this paper is particularly suitable for simulating scenarios wherein pedestrian groups move in transverse directions, such as flow through complex intersections and crossings.
We assume the exclusion principle holds within each group in order to simplify our derivations, but we can relax this assumption.
For instance, it would be natural to assume both intra-group and inter-group slowdown interactions, with the former far stronger than the latter.
Although we demonstrate our approach for only two groups, our model easily extends to many groups and to more complex local slowdown interactions.

The work of~\cite{burger1, burger2} is close to ours in setting and spirit:
The authors also consider stochastic particle motion on a 2D lattice and derive a coarse-grained PDE model for the evolution of pedestrian densities.
Nevertheless, important differences exist.
The modeling in~\cite{burger1, burger2} specifically emphasizes lane formation:
Only two groups of pedestrians compete, and these groups move in opposite directions.
Their microscopic stochastic lattice model features interaction rules that have been specifically tailored to allow for emergent lane formation.
In particular, agents within each group tend to follow one another, while agents from opposing groups employ a side-stepping mechanism to avoid upcoming traffic.
By contrast, our microscopic setup fundamentally differs from this follow/side-step pairing.
We pair global floor-field motion for each group~\cite{bksz01, kirch02, kirch03, guhu081} with a local slowdown mechanism that modulates agent movement as a function of the local configuration of agents from other groups~\cite{ckpt14}.
Our modeling framework is extensible:
We can handle any number of groups and complex interaction rules.
We can allow our governing stochastic rules to explicitly depend on time.
Consequently, our framework is suitable for the study of decision-making in complex environments.
We finish by noting that in~\cite{burger1, burger2}, the authors rigorously analyze their coarse-grained PDE model, whereas we do not.

We view the following as a primary contribution of this work:
We systematically analyze a new model that combines two interaction mechanisms - the global floor-field mechanism and a local slowdown mechanism.
In particular, we demonstrate that a coarse deterministic model accurately reproduces the behavior of the stochastic microscopic model.
Importantly, the global floor-field mechanism does not \lq\lq{}interfere\rq\rq{} with the local slowdown mechanism, and the assumptions required for the derivation of the coarse PDE model still hold when we combine the two interaction mechanisms.
This demonstrates that we may potentially combine global rules with a variety of local rules to produce more realistic microscopic models and deterministic PDE counterparts that faithfully capture the microscopic dynamics.
In particular, future possibilities include combining floor-field models with look-ahead potentials~\cite{soka06} to model decision-making using a tunable field of vision, and
combining floor-field models with multiple types of local interactions (such as side-stepping~\cite{bksz01,kirch02,kirch03,guhu081}, slowdown~\cite{ckpt14}, and social-force interactions~\cite{HM95}). 

The remainder of the paper is organized as follows. In Section \ref{sec:stoch}, we introduce the microscopic stochastic model for two interacting groups of pedestrians. In Section \ref{sec:meso}, we derive a deterministic mesoscopic model (the mesoscopic model is defined on the same lattice as the microscopic stochastic model).
We derive the coarse macroscopic PDE model and briefly discuss its main properties in Section \ref{sec:pde}.
In Section \ref{sec:num}, we present numerical simulations comparing the behavior of the microscopic stochastic model to that of the deterministic mesoscopic model.
Finally, we derive second-order corrections to the coarse macroscopic PDE model in Section \ref{sec:diff}.

\section{Stochastic Model}
\label{sec:stoch}

We develop a stochastic model that describes the dynamics of multiple groups of agents interacting on a lattice of any dimension.
Importantly, our stochastic model combines two mechanisms:
Group-specific floor-field functions specify the global motion of the various groups, and agents belonging to different groups interact.
Different groups interact with one another via slowdown interactions, similar to those considered in~\cite{ckpt14} on a one-dimensional lattice.

For the sake of clarity and brevity, we consider two groups of agents interacting on a lattice of dimension two. 
This allows us to focus on stochastic model construction, as well as the mesoscopic and macroscopic limits, without burdening the exposition with the lengthy formulas that would result from imposing more complex interaction rules between at least three groups.

Our stochastic model is a continuous-time Markov chain on a lattice.
In particular, agents from each of the two groups move on the $N_1\times N_2$ lattice $\mathcal{L}$.
We represent agents in the first and second groups by variables $\sigma^A_{j,k}(t) \equiv \sigma^A(j,k,t)$ and  $\sigma^B_{j,k}(t) \equiv \sigma^B(j,k,t)$, respectively, 
where $(j,k)$ is the position on the lattice and $t$ denotes time:
\begin{align*}
	\sigma^A_{j,k}(t) &= \left\{\begin{array}{ll} 1,& \text{if at time }t\text{ cell }(j,k)\text{ is occupied by an agent from group A;}\\
		0, & \text{otherwise,}\end{array}\right.\\
	\sigma^B_{j,k}(t) &= \left\{\begin{array}{ll} 1,& \text{if at time }t\text{ cell }(j,k)\text{ is occupied by an agent from group B;}\\
		0, & \text{otherwise.}\end{array}\right.
\end{align*}
Let $\sigma^{A} (t)$ and $\sigma^{B} (t)$ denote the collections of occupation values $\sigma_{j,k}^{A} (t)$ and $\sigma_{j,k}^{B} (t)$, respectively, over lattice sites $(j,k) \in \mathcal{L}$.
To simplify the presentation, we notationally suppress the dependence of $\sigma^A$ and $\sigma^B$ on time in what follows, unless we wish to explicitly emphasize time.

When developing an agent-based model for directional particle movement, the key challenge is the mathematical formulation of agent velocities and agent-agent interaction rules.
Here, we adopt the following general principles.
\begin{enumerate}[leftmargin=*, labelindent=\parindent, itemsep=0ex, label=\textbf{(P\arabic*)}, ref=P\arabic*]
\item
\label{li:floor-fields}
(\textbf{Floor fields})
In isolation, agent transition probabilities are determined by a velocity field that depends on the group identity of the agent.
\item
\label{li:exclusion}
(\textbf{Exclusion principle})
Two agents from the same group cannot simultaneously occupy a single lattice site.
\item
\label{li:slowdown}
(\textbf{Slowdown interactions})
Agent movement depends on the local configuration of agents from the other group.
In particular:
\begin{enumerate}[topsep=0ex, label=\textbf{(\alph*)}]
\item
Agents from different groups can occupy the same lattice site at the same time.
\item
Agent transition probabilities decrease when at least one agent from the other group occupies the same cell, or a neighboring cell in the direction of motion.
The amount of decrease depends on the local configuration.
\end{enumerate}
\item
\label{li:no-diagonal}
(\textbf{Lattice motion})
Agent motion is horizontal or vertical.
Consequently, diagonal motion can only result from two transitions (vertical followed by horizontal or vice-versa).
\end{enumerate}

With these general principles in place, we now precisely formulate our stochastic model.
Since the model is a continuous-time Markov chain, we must describe the admissible transitions and the corresponding transition probabilities.
We do so as follows.
\begin{itemize}
\item
  \textit{Admissible transitions.}
  We assign a floor velocity field (defined on the lattice $\mathcal{L}$) to each group of agents.
  Importantly, different groups may have different floor fields.
  For a given agent located at a given lattice site, the horizontal and vertical projections of the corresponding floor field vector specify the only two possible transitions for this agent (in accord with (\ref{li:no-diagonal})).
  (Only one transition is possible if the floor field vector is itself horizontal or vertical.)
  For our stochastic model, the admissible transitions are those specified by the floor fields, subject to the additional constraints imposed by the exclusion principle (\ref{li:exclusion}).

\item
  \textit{Transition probabilities.}
  For each group of agents, we define an \textit{interaction velocity field} that specifies transition rates for the stochastic model.
  These interaction velocity fields are obtained by modifying the floor fields (depending on local system configuration) to account for slowdown interactions, in accord with (\ref{li:slowdown}).
  For each admissible horizontal or vertical transition, the length of the corresponding projection of the associated interaction velocity vector gives the transition rate.
  Multiplying the transition rates by short time intervals $\Delta t$ produces the transition probabilities.
\end{itemize}


\subsection*{Floor velocity fields.}

Our two-dimensional stochastic model is significantly more flexible than the one-dimensional model in~\cite{ckpt14}, in terms of domain geometry, number of interacting groups, and complexity of motion.
When constructing stochastic models in dimension one, specifying direction of movement is straightforward (\textit{e.g.}~\cite{ckpt14,burger2}).
Such one-dimensional models can produce PDEs for bidirectional wave propagation (\textit{e.g.}~\cite{goatin,ardm11}).
In dimension at least two, however, there exists a rich set of possibilities for agent motion.
A flexible modeling framework should allow direction of motion to vary in space, in order to handle scenarios such as obstacle avoidance and motion in complex geometries.
To achieve such flexibility, we use floor velocity fields (see \textit{e.g.}~\cite{bksz01,kirch02}) to specify direction of motion for each group in isolation (as indicated in~(\ref{li:floor-fields}).
Importantly, different groups may obey different floor velocity fields.

For the sake of clarity, we consider the situation when each group is moving toward 
its own \emph{target exit point}.
We denote these exit points as $(j_0^{A}, k_0^{A})$ and $(j_0^{B}, k_0^{B})$ for groups A and B, respectively.
Interesting dynamics occur when the target points differ, a case we examine in detail.
We assume that target exit points do not move over time, although one can easily extend our model to include moving target exit points.
Let $\phi^A(j,k, j_0^A, k_0^A)$ and $\phi^B(j,k, j_0^B, k_0^B)$ denote the floor fields defined over $(j,k) \in \mathcal{L}$.
(We often omit the explicit dependence of these fields on the target exit points.)
We assume that $\phi^{A} = (\phi_{1}^{A}, \phi_{2}^{A})$ and $\phi^{B} = (\phi_{1}^{B}, \phi_{2}^{B})$ arise from potential functions $\psi^{A}$ and $\psi^{B}$:
\begin{equation}
\label{eq:phi}
\phi(j,k) = \left\{\begin{array}{rl}-\frac{\nabla \psi(j,k)}{||\nabla\psi(j,k)||_{1}},&||\nabla\psi(j,k)||_{1} \neq 0,\\ \bm{0},& \text{otherwise.}\end{array}\right.,
\end{equation}
We focus in this work on the quadratic potential
\begin{equation}
\label{eq:quadpsi}
\psi (j,k) = (j-j_{0})^{2} + (k-k_{0})^{2}.
\end{equation}

Our theoretical framework does not require that the floor fields be gradient fields, nor does it require target exit points.
In fact, our framework places no restrictions on the structure of the floor fields.
We simply use the quadratic potential~\eqref{eq:quadpsi} for simulations.

\subsection*{Interaction velocity fields.}

With the floor fields in place, we define a velocity field for each group that captures interaction effects.
Used to determine the probabilities that an agent moves to neighboring cells, these \textit{interaction velocity fields} account for the presence or absence of agents from the other group in nearby cells (slowdown interactions). 
In particular, the interaction velocity field $V^A(j,k,\sigma^{B}) = (V^A_1(j,k,\sigma^{B}), V^A_2(j,k,\sigma^{B}))$ for agents in 
group $A$ depends on the local configuration of group B agents.
This velocity field is defined as follows:
\begin{align}
	V^A_1(j,k,\sigma^{B}) &= \begin{cases}
			c_0\phi^A_1(j,k), & \text{if } \sigma^B_{j-1,k} = \sigma^B_{j,k} = \sigma^B_{j+1,k} = 0,\\
			\phi^A_1(j,k)[c_1H(-\phi^A_1(j,k)) + c_0H(\phi^A_1(j,k))], & \text{if } \sigma^B_{j-1,k} = 1, \spp \sigma^B_{j,k} = \sigma^B_{j+1,k} = 0,\\
			\phi^A_1(j,k)[c_0H(-\phi^A_1(j,k)) + c_1H(\phi^A_1(j,k))], & \text{if } \sigma^B_{j-1,k} = \sigma^B_{j,k} = 0, \spp \sigma^B_{j+1,k} = 1,\\
			c_1\phi^A_1(j,k), & \text{if } \sigma^B_{j-1,k} = \sigma^B_{j+1,k} = 1, \spp \sigma^B_{j,k} = 0,\\
			\phi^A_1(j,k)[c_2H(-\phi^A_1(j,k)) + c_3H(\phi^A_1(j,k))], & \text{if } \sigma^B_{j-1,k} = 0, \spp \sigma^B_{j,k} = \sigma^B_{j+1,k} = 1,\\
			\phi^A_1(j,k)[c_3H(-\phi^A_1(j,k)) + c_2H(\phi^A_1(j,k))], & \text{if } \sigma^B_{j-1,k} = \sigma^B_{j,k} = 1, \spp \sigma^B_{j+1,k} = 0,\\
			c_2\phi^A_1(j,k), & \text{if } \sigma^B_{j-1,k} = \sigma^B_{j+1,k} = 0, \spp \sigma^B_{j,k} = 1,\\
			c_3\phi^A_1(j,k), & \text{if }   \sigma^B_{j-1,k} = \sigma^B_{j+1,k} = \sigma^B_{j,k} = 1,
\end{cases}\label{vel1}\\
	V^A_2(j,k,\sigma^{B}) &= \begin{cases}
                   c_0\phi^A_2(j,k), & \text{if } \sigma^B_{j,k-1} = \sigma^B_{j,k} = \sigma^B_{j,k+1} = 0,\\
			\phi^A_2(j,k)[c_1H(-\phi^A_2(j,k)) + c_0H(\phi^A_2(j,k))], & \text{if } \sigma^B_{j,k-1} = 1, \spp \sigma^B_{j,k} = \sigma^B_{j,k+1} = 0,\\
			\phi^A_2(j,k)[c_0H(-\phi^A_2(j,k)) + c_1H(\phi^A_2(j,k))], & \text{if } \sigma^B_{j,k-1} = \sigma^B_{j,k} = 0, \spp \sigma^B_{j,k+1} = 1,\\
			c_1\phi^A_2(j,k), & \text{if } \sigma^B_{j,k-1} = \sigma^B_{j,k+1} = 1, \spp \sigma^B_{j,k} = 0,\\
			\phi^A_2(j,k)[c_2H(-\phi^A_2(j,k)) + c_3H(\phi^A_2(j,k))], & \text{if } \sigma^B_{j,k-1} = 0, \spp \sigma^B_{j,k} = \sigma^B_{j,k+1} = 1,\\
			\phi^A_2(j,k)[c_3H(-\phi^A_2(j,k)) + c_2H(\phi^A_2(j,k))], & \text{if } \sigma^B_{j,k-1} = \sigma^B_{j,k} = 1, \spp \sigma^B_{j,k+1} = 0,\\
			c_2\phi^A_2(j,k), & \text{if } \sigma^B_{j,k-1} = \sigma^B_{j,k+1} = 0, \spp \sigma^B_{j,k} = 1,\\
			c_3\phi^A_2(j,k), & \text{if } \sigma^B_{j,k-1} = \sigma^B_{j,k+1} = \sigma^B_{j,k} = 1.
\end{cases}\label{vel2}
\end{align}

Expressions~\eqref{vel1} and \eqref{vel2} are rather involved, because we must explicitly 
consider all possible configurations for the second group, $B$, in the vicinity
of the lattice site $(j,k)$.
However, when only one group of pedestrians is present
(\textit{e.g.} $\sigma_{j,k}^{B} (t) = 0$ for all $(j,k)$ and all $t$), the interaction velocity field assumes the simple form
\begin{align*}
& V^A_1(j,k) = c_0\phi^A_1(j,k) = c_0 \frac{j_0 - j}{|j-j_{0}| + |k-k_{0}|}, \\
& V^A_2(j,k) = c_0\phi^A_2(j,k) = c_0 \frac{k_0 - k}{|j-j_{0}| + |k-k_{0}|},
\end{align*}
where we have used the quadratic potential~\eqref{eq:quadpsi} as a particular example.
The horizontal component of this velocity, $V^A_1(j,k)$, is positive if $j < j_0$ and negative if $j > j_0$. 
This means that the particle moves toward the target coordinate, $j_0$.
The vertical component $V^A_2(j,k)$ behaves analogously.

Returning to the general setup, we define the interaction velocity field for group B, $V^{B}$, by invoking the substitutions $A \to B$ and $B \to A$ throughout~\eqref{vel1} and \eqref{vel2}.

The functions $\phi_{1}^{A}$ and $\phi_{2}^{A}$ in~\eqref{vel1} and \eqref{vel2} give the horizontal and vertical components of the floor field, respectively, and $H(x)$ is the Heaviside function with $H(0) = 0$. 
The Heaviside function is used to determine the direction of movement for the agent in cell $(j,k)$ (\textit{i.e.} the direction in which the velocity given by $\phi_1(j,k)$ or $\phi_2(j,k)$ is positive).
Further, it is used to ensure that a slowdown occurs if and only if an agent from the other group occupies the same cell as the agent in cell $(j,k)$, or the adjacent cell in the desired direction of movement, or both. Our formulation ensures that agents from the other group positioned \emph{behind} the agent in cell $(j,k)$ do not contribute to slowdown.

The velocity scalings $c_{0}$, $c_{1}$, $c_{2}$, and $c_{3}$ in $V^{A}$ quantify slowdown linked to the local configuration around a given agent.
For instance, consider the horizontal motion of an agent from group A in cell $(j,k)$; this is described by the interaction velocity component $V^A_1(j,k,\sigma^{B})$.
If there are no agents from group B in $(j,k)$ and horizontally adjacent
cells $(j-1,k)$ and $(j+1,k)$, then no slowdown occurs in the horizontal direction - the velocity scaling is $c_{0}$. 
However, if an agent from group B also occupies cell $(j,k)$, but no agents from group B appear in horizontally adjacent cells, then a slowdown should occur because the group A agent in cell $(j,k)$ must interact with the group B agent at the same location.
In this case, the velocity scaling is $c_{2}$.
Slowdown also results from the presence of group B agents in the direction of motion.
For instance, suppose that the floor velocity component $\phi_1^A(j,k)$ is negative, meaning the group A agent in cell $(j,k)$ is moving toward $(j-1,k)$.
The presence of a group B agent in cell $(j-1,k)$ should induce a slowdown.
If cell $(j-1,k)$ is occupied by a group B agent but cell $(j,k)$ is not, then the velocity scaling is $c_{1}$.
The velocity scaling is $c_{3}$ if both $(j-1,k)$ and $(j,k)$ are occupied by agents from group B.

We assume that the velocity scalings satisfy the natural relationship
\begin{equation}\label{ccond}
c_3 < c_2 \leqslant c_1 < c_0.
\end{equation}
With respect to $V_{1}^{A} (j, k, \sigma^{B})$, inequalities~\eqref{ccond} reflect the fact that velocity scaling $c_0$ corresponds to local absence of agents from group B, velocity scalings $c_1$ and $c_2$ correspond to interaction with only one agent from group B, and velocity scaling $c_3$ corresponds to interaction with two agents from group B.

\subsection*{Transition probabilities for the stochastic model.}

The interaction velocity fields $V^A(j,k,\sigma^{B})$ and $V^B(j,k,\sigma^{A})$ incorporate~(\ref{li:floor-fields}) and~(\ref{li:slowdown}).
We use them to specify transition probabilities for an agent from group $A$ (or group $B$, respectively) at the current position $(j,k)$.
Agent motion must be horizontal or vertical~(\ref{li:no-diagonal}), and must be consistent with the corresponding interaction velocity field.
That is, the group $A$ agent can move left if $V_{1}^{A} (j, k, \sigma^{B}) < 0$, right if $V_{1}^{A} (j, k, \sigma^{B}) > 0$, down if $V_{2}^{A} (j, k, \sigma^{B}) < 0$, and up if $V_{2}^{A} (j, k, \sigma^{B}) > 0$.
We enforce the exclusion principle~(\ref{li:exclusion}) for agents in the same group.

Assuming this framework, the probability of transition $(j,k) \to (j \pm 1,k)$ for a member of group A during a small time 
interval $\Delta t$ is given by
\begin{align}
\label{PA}
	P^A_{(j,k) \to (j \pm 1,k)} &= \pm\Delta t \phi^A_1(j,k)H(\pm\phi^A_1(j,k))\sigma^A_{j,k}(1-\sigma^A_{j\pm 1, k}) \times \\
\nonumber
		&\quad \left[c_0(1-\sigma^B_{j,k})(1-\sigma^B_{j\pm 1,k}) + c_1(1-\sigma^B_{j,k})\sigma^B_{j\pm 1,k} + c_2\sigma^B_{j,k}(1-\sigma^B_{j\pm 1,k}) + c_3\sigma^B_{j,k}\sigma^B_{j\pm 1,k}\right],
\end{align}
while the probability of transition $(j,k) \to (j,k\pm 1)$ for a member of group A is given by
\begin{align}
\label{PB}
	P^A_{(j,k) \to (j,k \pm 1)} &= \pm\Delta t \phi^A_2(j,k)H(\pm\phi^A_2(j,k))\sigma^A_{j,k}(1-\sigma^A_{j, k \pm 1}) \times \\
\nonumber
		&\quad \left[c_0(1-\sigma^B_{j,k})(1-\sigma^B_{j,k\pm 1}) + c_1(1-\sigma^B_{j,k})\sigma^B_{j,k\pm 1} + c_2\sigma^B_{j,k}(1-\sigma^B_{j,k\pm 1}) + c_3\sigma^B_{j,k}\sigma^B_{j,k\pm 1}\right].
\end{align}
The diagonal entries of the transition probability matrix are defined in a standard manner as
\begin{equation*}
  P^{A}_{(j,k) \to (j,k)} = 1 - P^{A}_{(j,k) \to (j + 1,k)} - P^{A}_{(j,k) \to (j - 1,k)} - P^{A}_{(j,k) \to (j,k + 1)} - P^{A}_{(j,k) \to (j,k - 1)},
\end{equation*}
so that the entries in each row sum to $1$.
In~\eqref{PA} and \eqref{PB}, the Heaviside functions ensure that the probability of transition to a neighboring horizontal (vertical) cell is nonzero only if the horizontal (vertical) projection of the floor field points in the direction of the neighboring cell.
The $\sigma^A_{j,k}(1-\sigma^A_{j\pm 1, k})$ and $\sigma^A_{j,k}(1-\sigma^A_{j, k \pm 1})$ terms express the exclusion principle: The probability of transition is nonzero only if the current cell is occupied and the target cell is not occupied by a member of the same group. 

We assume periodic boundary conditions for simplicity.
It would be easy to specify, and interesting to analyze, more exotic boundary conditions.

Our stochastic model is $\bsigma(t) = (\sigma^{A} (t), \sigma^{B} (t))$, the finite-state, continuous-time Markov chain with transition probabilities given by~\eqref{PA} and \eqref{PB}.
Note that the dimension of the state space is quite high, as is typical for spatially extended stochastic processes.

\subsection*{Remarks.}

As explained before, agent motion must be consistent with the corresponding interaction velocity field - the angle between the random direction of motion and this field must be strictly less than $\pi / 2$.
When the floor fields assume the gradient form~\eqref{eq:phi} associated with potential~\eqref{eq:quadpsi}, each agent must move toward its target exit point (down the gradient).
It would be interesting to allow motion \textit{away} from the target exit point.
Such motion could be useful for obstacle avoidance and evacuation scenarios.

Our exclusion principle~(\ref{li:exclusion}) applies only to agents from the same group. 
This modeling issue has been treated various ways in the pedestrian dynamics literature.
When modeling car traffic, the exclusion principle can be applied to all agents (even from different groups) because all agents move in the same direction.
However, vehicle traffic and pedestrian traffic differ fundamentally.
Two groups of pedestrians can move toward each other and should be allowed to pass through one another.
Such behavior occurs at complex intersections and crossings, for example.

Existing work has focused on two pass-through mechanisms - (i) the side-stepping (or avoidance) mechanism~\cite{bksz01,kirch02,kirch03,guhu081}, and (ii) the slowdown mechanism~\cite{ckpt14}.
Both mechanisms are probably relevant for pedestrian flow, especially at intermediate and high densities.
Look-ahead mechanisms would add realism as well (\textit{e.g.}~\cite{soka06,degond13,hst14}).
In the present work, we have opted to focus on the slowdown mechanism.
Note that setting $c_1=c_2=c_3=0$ in our stochastic model would invoke the complete exclusion principle with respect to agents from both groups, thereby essentially disallowing the groups to pass through one another.

\subsection*{Some extensions.}

Before examining mesoscopic and macroscopic descriptions, we discuss generalizations.
First, our modeling framework naturally extends to lattices of any dimension and to at least three groups of interacting agents.
Second, the interaction mechanisms between different groups can be more complex.
For instance, agents can be allowed to switch between groups, either probabilistically or deterministically.
This mechanism is relevant when modeling obstacle avoidance, or evacuation scenarios with multiple exits. 
Third, our model can be extended to include \lq\lq{}chemical reaction\rq\rq{} mechanisms, as is done in the reaction diffusion master equation framework (see \textit{e.g.}~\cite{rdma1,rdma2,rdma3} for recent work).
This is relevant for problems wherein transport of reacting particles is of interest. 
We intend to explore these generalizations both theoretically and computationally in future work.

\section{Mesoscopic Deterministic Model}
\label{sec:meso}

The stochastic model described in Section~\ref{sec:stoch} can be viewed as a continuous-time Markov chain on a state space of extremely high dimension.
Developing analytical understanding of such complex, spatially extended models is a daunting task.
The inclusion of additional complex interaction rules, such as group switching (see \textit{e.g.}~\cite{lega13, lewe14}) and look-ahead interactions~\cite{soka06}, amplifies the level of analytical difficulty. 
Moreover, numerical simulations require considerable computational resources, especially for many 
interacting groups.
These analytical and computational challenges commonly arise for spatially extended stochastic systems.

Alternatively, one can derive coarse equations that accurately represent the bulk statistical 
properties of spatially extended stochastic models. 
These PDEs are amenable to analysis and can illuminate the mechanisms that drive the dynamics.
Further, they offer numerical advantages of interest when addressing practical problems, such as improved efficiency and scalability.
We adopt this philosophy in the current paper and derive coarse dynamical equations for the evolution of the agent group densities.

In order to derive dynamical equations for the densities, we proceed in a manner similar to the 
approach outlined in~\cite{soka06,ckpt14,hst14}.
In particular, the process $\bsigma_t = ( \sigma^A_t, \sigma^B_t )$ constitutes a continuous-time Markov chain (note that we move time to the subscript position for this section).
Consequently, we consider the generator $L$ of the stochastic process $\bm{\sigma}_{t}$ given by
\begin{equation*}
L \Psi = \lim_{\Delta t \to 0} \frac{\mathbb{E} [\Psi(\bsigma_{\Delta t}) | \bsigma_{0}] - \Psi(\bsigma_0)}{\Delta t}.
\end{equation*}
Here $\bsigma_0$ is the initial configuration, $\bsigma_{\Delta t}$ is the configuration at time $\Delta t$, $\Psi$ is any test function, and the expectation is taken over all possible transitions from $\bsigma_0$ to $\bsigma_{\Delta t}$. 
We consider simple cases when test functions are defined as the value of the process at a particular location.
In particular, we consider $\Psi(\bsigma)= \sigma^A(j,k)$ and $\Psi(\bsigma)= \sigma^B(j,k)$.
In the case that $\Psi(\bsigma)= \sigma^A(j,k)$, we can write the action of the generator as
\begin{equation}
\begin{aligned}
\label{e:generator-on-process}
	L \sigma^A (j,k) = \; & \frac{P^A_{(j-1,k) \to (j,k)} - P^A_{(j,k) \to (j+1,k)} + P^A_{(j+1,k) \to (j,k)} - P^A_{(j,k) \to (j-1,k)}}{\Delta t} +\\
		& \frac{P^A_{(j,k-1) \to (j,k)} - P^A_{(j,k) \to (j,k+1)} + P^A_{(j,k+1) \to (j,k)} - P^A_{(j,k) \to (j,k-1)}}{\Delta t}.
\end{aligned}
\end{equation}
Roughly speaking, the formula above describes the time derivative for the evolution of the expected value of the process $\sigma^A (j,k)$ at the target cell $(j,k)$.
Note that the right side of~\eqref{e:generator-on-process} involves only four transitions 
$(j,k) \to (j\pm1,k)$ and $(j,k) \to (j,k\pm1)$
\lq\lq{}from\rq\rq{} the target cell and four transitions 
$(j\pm1,k) \to (j,k)$ and $(j,k\pm1) \to (j,k)$ \lq\lq{}into\rq\rq{} the target cell.
These are the only transitions which affect the value of the process $\sigma^{A}$ in cell $(j,k)$, and thus affect the expected value 
$\mathbb{E} [\sigma^A(j,k)]$. 
Note further that since only motion compatible with the floor field is allowed, at most four of the terms on the right side of~\eqref{e:generator-on-process} are nonzero at each cell $(j,k)$, depending on the signs of the two components $\phi_{1}^{A} (j,k)$ and $\phi_{2}^{A} (j,k)$ of the floor field.
For example, if $\phi_{1}^{A} (j,k) > 0$, then the outflow transition $(j,k) \to (j+1,k)$ is possible, but the outflow transition $(j,k) \to (j-1,k)$ is not.
Vertical transitions behave analogously.
Terms on the right side of~\eqref{e:generator-on-process} that are incompatible with the direction of motion specified by the floor field automatically disappear because of the use of the Heaviside function in~\eqref{PA} and~\eqref{PB}.

The fundamental property of the generator, namely
\[
	\frac{d}{dt}\mathbb{E} [\Psi] = \mathbb{E} [L \Psi],
\]
yields a differential equation for the time evolution of the agent density $\rho^A_{j,k}(t) \equiv \bE [\sigma^A(j,k,t)]$ when applied to the test function $\Psi(\bsigma)= \sigma^A(j,k)$:
\begin{align}
\label{rhoA1}
	\frac{d}{dt}\rho^A_{j,k} &= \mathbb{E}\Big[\phi^A_1(j-1,k)H(\phi^A_1(j-1,k))\sigma^A_{j-1,k}(1-\sigma^A_{j,k}) \\
\nonumber
		&\qquad\quad \times\left[c_0(1-\sigma^B_{j-1,k})(1-\sigma^B_{j,k}) + c_1(1-\sigma^B_{j-1,k})\sigma^B_{j,k} + c_2\sigma^B_{j-1,k}(1-\sigma^B_{j,k}) + c_3\sigma^B_{j-1,k}\sigma^B_{j,k}\right]\\
\nonumber
		&\qquad {}- \phi^A_1(j,k)H(\phi^A_1(j,k))\sigma^A_{j,k}(1-\sigma^A_{j+1,k})\\
\nonumber
		&\qquad\quad \times\left[c_0(1-\sigma^B_{j,k})(1-\sigma^B_{j+1,k}) + c_1(1-\sigma^B_{j,k})\sigma^B_{j+1,k} + c_2\sigma^B_{j,k}(1-\sigma^B_{j+1,k}) + c_3\sigma^B_{j,k}\sigma^B_{j+1,k}\right]\\
\nonumber
		&\qquad {}- \phi^A_1(j+1,k)H(-\phi^A_1(j+1,k))\sigma^A_{j+1,k}(1-\sigma^A_{j,k})\\
\nonumber
		&\qquad\quad \times\left[c_0(1-\sigma^B_{j+1,k})(1-\sigma^B_{j,k}) + c_1(1-\sigma^B_{j+1,k})\sigma^B_{j,k} + c_2\sigma^B_{j+1,k}(1-\sigma^B_{j,k}) + c_3\sigma^B_{j+1,k}\sigma^B_{j,k}\right]\\
\nonumber
		&\qquad {}+ \phi^A_1(j,k)H(-\phi^A_1(j,k))\sigma^A_{j,k}(1-\sigma^A_{j-1,k})\\
\nonumber
		&\qquad\quad \times\left[c_0(1-\sigma^B_{j,k})(1-\sigma^B_{j-1,k}) + c_1(1-\sigma^B_{j,k})\sigma^B_{j-1,k} + c_2\sigma^B_{j,k}(1-\sigma^B_{j-1,k}) + c_3\sigma^B_{j,k}\sigma^B_{j-1,k}\right]\\
\nonumber
		&\qquad {}+ \phi^A_2(j,k-1)H(\phi^A_2(j,k-1))\sigma^A_{j,k-1}(1-\sigma^A_{j,k})\\
\nonumber
		&\qquad\quad \times\left[c_0(1-\sigma^B_{j,k-1})(1-\sigma^B_{j,k}) + c_1(1-\sigma^B_{j,k-1})\sigma^B_{j,k} + c_2\sigma^B_{j,k-1}(1-\sigma^B_{j,k}) + c_3\sigma^B_{j,k-1}\sigma^B_{j,k}\right]\\
\nonumber
		&\qquad {}- \phi^A_2(j,k)H(\phi^A_2(j,k))\sigma^A_{j,k}(1-\sigma^A_{j,k+1})\\
\nonumber
		&\qquad\quad \times\left[c_0(1-\sigma^B_{j,k})(1-\sigma^B_{j,k+1}) + c_1(1-\sigma^B_{j,k})\sigma^B_{j,k+1} + c_2\sigma^B_{j,k}(1-\sigma^B_{j,k+1}) + c_3\sigma^B_{j,k}\sigma^B_{j,k+1}\right]\\
\nonumber
		&\qquad {}- \phi^A_2(j,k+1)H(-\phi^A_2(j,k+1))\sigma^A_{j,k+1}(1-\sigma^A_{j,k})\\
\nonumber
		&\qquad\quad \times\left[c_0(1-\sigma^B_{j,k+1})(1-\sigma^B_{j,k}) + c_1(1-\sigma^B_{j,k+1})\sigma^B_{j,k} + c_2\sigma^B_{j,k+1}(1-\sigma^B_{j,k}) + c_3\sigma^B_{j,k+1}\sigma^B_{j,k}\right]\\
\nonumber
		&\qquad {}+ \phi^A_2(j,k)H(-\phi^A_2(j,k))\sigma^A_{j,k}(1-\sigma^A_{j,k-1})\\
\nonumber
		&\qquad\quad \times\left[c_0(1-\sigma^B_{j,k})(1-\sigma^B_{j,k-1}) + c_1(1-\sigma^B_{j,k})\sigma^B_{j,k-1} + c_2\sigma^B_{j,k}(1-\sigma^B_{j,k-1}) + c_3\sigma^B_{j,k}\sigma^B_{j,k-1}\right]\Big].
\end{align}
In differential equation~\eqref{rhoA1}, the majority of terms come from the velocity equations \eqref{vel1} and \eqref{vel2}, as well as the requirement to describe all possible transitions \lq\lq{}from\rq\rq{} and \lq\lq{}into\rq\rq{} the target cell $(j,k)$. 
Terms of the type $\sigma^A_{j,k}(1-\sigma^A_{j+1,k})$ restrict agent movement, allowing an agent to move only from the cell it occupies to a cell unoccupied by a member of the same group. Terms of the type 
$(1-\sigma^B_{j,k})$ and $\sigma^B_{j,k}$
describe absence or presence of an agent from group B in cell $(j,k)$, respectively. 
For instance, the term $(1-\sigma^B_{j,k})\sigma^B_{j,k-1}$ is one when no agent from group B is present 
in cell $(j,k)$, but an agent from group B occupies cell $(j,k-1)$. In this case, there should be a 
slowdown in the vertical direction, which is reflected in the velocity scaling $c_1$ used for this term.

Differential equation \eqref{rhoA1} above is exact, but not closed. In order to derive a closed-form equation for $\rho^A_{j,k}$, we assume that the joint measure on $\bsigma_t$ is 
approximately a product measure (see the discussion in \cite{soka06}), and that all mixed moments can be well-approximated by the \lq\lq{}approximate independence\rq\rq{} closure (see \cite{hst14} for a detailed study of this assumption in a related traffic model). 
In particular, this implies that $\sigma^A(i,j)$, $\sigma^A(k,l)$ are approximately independent if $(i,j)\ne(k,l)$, and 
$\sigma^A(i,j)$, $\sigma^B(k,l)$ are approximately independent for all $(i,j)$, $(k,l)$.
Then, higher-order moments can be represented as products of expectations, and the closed-form equation for $\rho^A_{j,k}$ is given by
\begin{align}
\label{rhoA2}
	\frac{d}{dt}\rho^A_{j,k} &= \phi^A_1(j-1,k)H(\phi^A_1(j-1,k))\rho^A_{j-1,k}(1-\rho^A_{j,k})\\
\nonumber
		&\qquad\quad \times\left[c_0(1-\rho^B_{j-1,k})(1-\rho^B_{j,k}) + c_1(1-\rho^B_{j-1,k})\rho^B_{j,k} + c_2\rho^B_{j-1,k}(1-\rho^B_{j,k}) + c_3\rho^B_{j-1,k}\rho^B_{j,k}\right]\\
\nonumber
		&\qquad {}- \phi^A_1(j,k)H(\phi^A_1(j,k))\rho^A_{j,k}(1-\rho^A_{j+1,k})\\
\nonumber
		&\qquad\quad \times\left[c_0(1-\rho^B_{j,k})(1-\rho^B_{j+1,k}) + c_1(1-\rho^B_{j,k})\rho^B_{j+1,k} + c_2\rho^B_{j,k}(1-\rho^B_{j+1,k}) + c_3\rho^B_{j,k}\rho^B_{j+1,k}\right]\\
\nonumber
		&\qquad {}- \phi^A_1(j+1,k)H(-\phi^A_1(j+1,k))\rho^A_{j+1,k}(1-\rho^A_{j,k})\\
\nonumber
		&\qquad\quad \times\left[c_0(1-\rho^B_{j+1,k})(1-\rho^B_{j,k}) + c_1(1-\rho^B_{j+1,k})\rho^B_{j,k} + c_2\rho^B_{j+1,k}(1-\rho^B_{j,k}) + c_3\rho^B_{j+1,k}\rho^B_{j,k}\right]\\
\nonumber
		&\qquad {}+ \phi^A_1(j,k)H(-\phi^A_1(j,k))\rho^A_{j,k}(1-\rho^A_{j-1,k})\\
\nonumber
		&\qquad\quad \times\left[c_0(1-\rho^B_{j,k})(1-\rho^B_{j-1,k}) + c_1(1-\rho^B_{j,k})\rho^B_{j-1,k} + c_2\rho^B_{j,k}(1-\rho^B_{j-1,k}) + c_3\rho^B_{j,k}\rho^B_{j-1,k}\right]\\
\nonumber
		&\qquad {}+ \phi^A_2(j,k-1)H(\phi^A_2(j,k-1))\rho^A_{j,k-1}(1-\rho^A_{j,k})\\
\nonumber
		&\qquad\quad \times\left[c_0(1-\rho^B_{j,k-1})(1-\rho^B_{j,k}) + c_1(1-\rho^B_{j,k-1})\rho^B_{j,k} + c_2\rho^B_{j,k-1}(1-\rho^B_{j,k}) + c_3\rho^B_{j,k-1}\rho^B_{j,k}\right]\\
\nonumber
		&\qquad {}- \phi^A_2(j,k)H(\phi^A_2(j,k))\rho^A_{j,k}(1-\rho^A_{j,k+1})\\
\nonumber
		&\qquad\quad \times\left[c_0(1-\rho^B_{j,k})(1-\rho^B_{j,k+1}) + c_1(1-\rho^B_{j,k})\rho^B_{j,k+1} + c_2\rho^B_{j,k}(1-\rho^B_{j,k+1}) + c_3\rho^B_{j,k}\rho^B_{j,k+1}\right]\\
\nonumber
		&\qquad {}- \phi^A_2(j,k+1)H(-\phi^A_2(j,k+1))\rho^A_{j,k+1}(1-\rho^A_{j,k})\\
\nonumber
		&\qquad\quad \times\left[c_0(1-\rho^B_{j,k+1})(1-\rho^B_{j,k}) + c_1(1-\rho^B_{j,k+1})\rho^B_{j,k} + c_2\rho^B_{j,k+1}(1-\rho^B_{j,k}) + c_3\rho^B_{j,k+1}\rho^B_{j,k}\right]\\
\nonumber
		&\qquad {}+ \phi^A_2(j,k)H(-\phi^A_2(j,k))\rho^A_{j,k}(1-\rho^A_{j,k-1})\\
\nonumber
		&\qquad\quad \times\left[c_0(1-\rho^B_{j,k})(1-\rho^B_{j,k-1}) + c_1(1-\rho^B_{j,k})\rho^B_{j,k-1} + c_2\rho^B_{j,k}(1-\rho^B_{j,k-1}) + c_3\rho^B_{j,k}\rho^B_{j,k-1}\right].
\end{align}
The differential equation for the density of group B, $\rho^B_{j,k}$, has the same form as~\eqref{rhoA2}, but with group designation exchanged (i.e. $A \leftrightarrow B$).

The equations for $\rho^A_{j,k}$ and $\rho^B_{j,k}$ constitute a coupled system, defined on the same lattice $\mathcal{L}$ as the microscopic model. 
Since the derivation of this mesoscopic system involves the
\lq\lq{}approximate independence\rq\rq{} closure assumption, one needs to verify approximate independence numerically over the relevant range of model parameters in concrete situations.
A detailed numerical investigation of this 
assumption in a related car traffic model with look-ahead interaction rules has been carried out in~\cite{hst14}.
In Section~\ref{sec:num}, we verify approximate independence indirectly for our model by comparing ensemble simulations of the stochastic model from Section~\ref{sec:stoch} with the behavior of mesoscopic model~\eqref{rhoA2}.

\section{Macroscopic PDE Model}
\label{sec:pde}

We now treat sites $(j,k) \in \mathcal{L}$ as square cells with fixed side length $h > 0$. Let $\Omega$ denote the subdomain of $\mathbb{R}^2$ corresponding to the lattice $\mathcal{L}$, where the number of cells depends on $h$. 
We derive a system of conservation law PDEs for the evolution of agent densities by passing to the $h \to 0$ limit (number of cells tends to infinity), and simultaneously rescaling time as $t \to ht$.

We rewrite differential equation~\eqref{rhoA2} for the density $\rho_{j,k}^{A}$ in the following flux form, taking the time rescaling into account:
\begin{equation}
\label{e:flux-form}
	\frac{d\rho_{j,k}^{A}}{dt} = - \frac{F^A_{j,j+1} - F^A_{j-1,j} + G^A_{k,k+1} - G^A_{k-1,k}}{h},
\end{equation}
where the horizontal flux is defined by
\begin{align*}
	F^A_{j,j+1} &= \phi^A_1(j,k)H(\phi^A_1(j,k))\rho^A_{j,k}(1-\rho^A_{j+1,k})\\
		&\qquad\quad \times\left[c_0(1-\rho^B_{j,k})(1-\rho^B_{j+1,k}) + c_1(1-\rho^B_{j,k})\rho^B_{j+1,k} + c_2\rho^B_{j,k}(1-\rho^B_{j+1,k}) + c_3\rho^B_{j,k}\rho^B_{j+1,k}\right]\\
		&\quad {}+ \phi^A_1(j+1,k)H(-\phi^A_1(j+1,k))\rho^A_{j+1,k}(1-\rho^A_{j,k})\\
		&\qquad\quad \times\left[c_0(1-\rho^B_{j+1,k})(1-\rho^B_{j,k}) + c_1(1-\rho^B_{j+1,k})\rho^B_{j,k} + c_2\rho^B_{j+1,k}(1-\rho^B_{j,k}) + c_3\rho^B_{j+1,k}\rho^B_{j,k}\right],
\end{align*}
and the vertical flux is given by
\begin{align*}
	G^A_{k,k+1} &= \phi^A_2(j,k)H(\phi^A_2(j,k))\rho^A_{j,k}(1-\rho^A_{j,k+1})\\
		&\qquad\quad \times\left[c_0(1-\rho^B_{j,k})(1-\rho^B_{j,k+1}) + c_1(1-\rho^B_{j,k})\rho^B_{j,k+1} + c_2\rho^B_{j,k}(1-\rho^B_{j,k+1}) + c_3\rho^B_{j,k}\rho^B_{j,k+1}\right]\\
		&\quad {}+ \phi^A_2(j,k+1)H(-\phi^A_2(j,k+1))\rho^A_{j,k+1}(1-\rho^A_{j,k})\\
		&\qquad\quad \times\left[c_0(1-\rho^B_{j,k+1})(1-\rho^B_{j,k}) + c_1(1-\rho^B_{j,k+1})\rho^B_{j,k} + c_2\rho^B_{j,k+1}(1-\rho^B_{j,k}) + c_3\rho^B_{j,k+1}\rho^B_{j,k}\right].
\end{align*}
Multiplying the flux form~\eqref{e:flux-form} by $\varphi_{j,k} := \varphi(jh, kh)$, where $\varphi \in C_0^1(\bar\Omega)$ is a test function, and using the summation by parts property over $\Omega$ yields
\[
	\sum_{j,k} \varphi_{j,k}\frac{d\rho_{j,k}^{A}}{dt} = \sum_{j,k}\left(F^A_{j,j+1}\frac{\varphi_{j+1,k}-\varphi_{j,k}}{h} + G^A_{k,k+1}\frac{\varphi_{j,k+1}-\varphi_{j,k}}{h}\right).
\]

We define pedestrian densities on $\Omega$ as follows. Reusing the notation $\rho^A$ for convenience, define the function $\rho^A(x,y,t)$ as a continuous piecewise-linear interpolation of $\rho^A_{j,k}(t)$.
Taking the $h \to 0^+$ limit and noting that both $\rho^A$ and $\frac{d\rho^A_{j,k}}{dt}$ are bounded, we obtain a weak formulation of a limiting PDE:
\[
	\iint\limits_\Omega \varphi(x,y)\frac{\partial}{\partial t}\rho^A(x,y,t) \, dx \, dy = \iint\limits_\Omega \left(F^A(\rho^A, \rho^{B}) \frac{\partial}{\partial x}\varphi + G^A(\rho^A, \rho^{B}) \frac{\partial}{\partial y}\varphi\right) dx \, dy,
\]
where $F^A$ and $G^A$ are defined as the corresponding limits of $F^A_{j,j+1}$ and $G^A_{k, k+1}$, i.e.,
\begin{align*}
	F^A(\rho^A, \rho^{B}) &= \phi^A_1\rho^A(1-\rho^A)\left[(c_0-c_1-c_2+c_3)(\rho^B)^2 + (c_1 + c_2 - 2c_0)\rho^B + c_0\right],\\
	G^A(\rho^A, \rho^{B}) &= \phi^A_2\rho^A(1-\rho^A)\left[(c_0-c_1-c_2+c_3)(\rho^B)^2 + (c_1 + c_2 - 2c_0)\rho^B + c_0\right].
\end{align*}
Notice that if we turn off the slowdown interaction mechanism by setting all of the $c_{i}$ equal to $c_{0}$, then the bracketed expressions in the flux equations reduce to $c_{0}$.
We write the limiting system of PDEs in differential form as
\begin{align*}
	\rho^A_t + [\phi^A_1f(\rho^A)g(\rho^B)]_x + [\phi^A_2f(\rho^A)g(\rho^B)]_y &= 0,\\
	\rho^B_t + [\phi^B_1f(\rho^B)g(\rho^A)]_x + [\phi^B_2f(\rho^B)g(\rho^A)]_y &= 0,
\end{align*}
where
\begin{equation}
\label{fg}
	f(u) = u(1-u), \quad g(u) = (c_0-c_1-c_2+c_3)u^2 + (c_1 + c_2 - 2c_0)u + c_0.
\end{equation}
In vector form, the limiting PDE system is therefore
\begin{equation}
\label{PDEs}
	\frac{\partial}{\partial t}\left[\begin{array}{c}\rho^A\\ \rho^B\end{array}\right] + \frac{\partial}{\partial x}\left[\begin{array}{c}\phi^A_1f(\rho^A)g(\rho^B)\\ \phi^B_1f(\rho^B)g(\rho^A)\end{array}\right] + \frac{\partial}{\partial y}\left[\begin{array}{c}\phi^A_2f(\rho^A)g(\rho^B)\\ \phi^B_2f(\rho^B)g(\rho^A)\end{array}\right] = \mathbf{0}.
\end{equation}

The limiting PDE system~\eqref{PDEs} connects naturally with several classical models for traffic flow.
In particular, the flux function $f(u)$ in~\eqref{fg} is the classical Greenshield potential.
In the absence of slowdown ($c_{0} = c_1 = c_2 = c_3$), we recover a two-dimensional model of Lighthill-Whitham-Richards type~\cite{lwr1,lwr2} for two non-interacting groups of agents.
Model~\eqref{PDEs} generalizes to more interacting groups and falls 
into the class of $n$-populations models (as discussed in~\cite{bgc2003}, for instance).

The system of PDEs in~\eqref{PDEs} is a system of conservation laws, but it is only conditionally hyperbolic. Indeed, the hyperbolicity of the system depends not only on the values of the densities, but on the floor velocity fields and the values of the scaling constants $c_0$, $c_1$, $c_2$, and $c_3$ as well. We now analyze these dependencies in detail.

In what follows, we introduce a slowdown parameter $\alpha \geqslant 1$ that quantifies interaction strength between members of different groups.
We then study the practical slowdown regime obtained by scaling the $c_{i}$ as
\begin{equation}
\label{vel}
c_1 = c_2 = \frac{c_0}{\alpha}, \quad c_3 = \frac{c_0}{2 \alpha}.
\end{equation}

\subsection*{Hyperbolicity Conditions}
The conservation law system~\eqref{PDEs} is hyperbolic when the matrix
\[
	A = \gamma_1\left[\begin{array}{cc}\phi^A_1f'(\rho^A)g(\rho^B)&\phi^A_1f(\rho^A)g'(\rho^B)\\ \phi^B_1f(\rho^B)g'(\rho^A)&\phi^B_1f'(\rho^B)g(\rho^A)\end{array}\right] + \gamma_2\left[\begin{array}{cc}\phi^A_2f'(\rho^A)g(\rho^B)&\phi^A_2f(\rho^A)g'(\rho^B)\\ \phi^B_2f(\rho^B)g'(\rho^A)&\phi^B_2f'(\rho^B)g(\rho^A)\end{array}\right]
\]
is diagonalizable with real eigenvalues for all $\gamma_1,\gamma_2 \in \mathbb{R}$. 
We let $\boldsymbol\gamma = (\gamma_1,\gamma_2)$ and rewrite $A$ as
\begin{equation}\label{eigenmat}
	A = \left[\begin{array}{cc}(\boldsymbol\gamma \cdot \phi^A) f'(\rho^A)g(\rho^B)&(\boldsymbol\gamma \cdot \phi^A) f(\rho^A)g'(\rho^B)\\ (\boldsymbol\gamma \cdot \phi^B) f(\rho^B)g'(\rho^A)&(\boldsymbol\gamma \cdot \phi^B) f'(\rho^B)g(\rho^A)\end{array}\right].
\end{equation}
The eigenvalue equation for $A$ implies that system~\eqref{PDEs} is hyperbolic if
\begin{align}
	0 &\leqslant \left[(\boldsymbol\gamma \cdot \phi^A) f'(\rho^A)g(\rho^B)\right]^2 + \left[(\boldsymbol\gamma \cdot \phi^B) f'(\rho^B)g(\rho^A) \right]^2  \nonumber\\
		&\quad {}- 2(\boldsymbol\gamma \cdot \phi^A)(\boldsymbol\gamma \cdot \phi^B)f'(\rho^A)f'(\rho^B)g(\rho^A)g(\rho^B)  \label{eq:hypercon}\\
		&\quad {}+ 4(\boldsymbol\gamma \cdot \phi^A)(\boldsymbol\gamma \cdot \phi^B)f(\rho^A)f(\rho^B)g'(\rho^A)g'(\rho^B)  \nonumber
\end{align}
for all unimodular $\boldsymbol\gamma \in \mathbb{R}^2$, and $A$ is diagonalizable whenever the right side of~\eqref{eq:hypercon} is zero.
($A$ has distinct real eigenvalues whenever the right side of~\eqref{eq:hypercon} is positive.)

It is difficult to analyze inequality~\eqref{eq:hypercon} in general. 
Consequently, we consider several important cases to gain insight into the nature of the conditional hyperbolicity of~\eqref{PDEs}.

\subsubsection*{Case 1: $\phi^A = \phi^B$}
When $\phi^A = \phi^B$, the two groups of agents follow the same floor velocity field. 
This could happen, for instance, in an evacuation scenario with only one exit point. 
When the two floor fields are equal, the hyperbolicity condition~\eqref{eq:hypercon} simplifies to
\begin{equation}
\label{eq:hypercon1}
0 \leqslant (\boldsymbol{\gamma} \cdot \phi)^{2} \left( \left[ f'(\rho^A)g(\rho^B) - f'(\rho^B)g(\rho^A) \right]^2 
+ 4f(\rho^A)f(\rho^B)g'(\rho^A)g'(\rho^B) \right).
\end{equation}
The function $h_1(x,y)$ defined by 
\begin{equation*}
	h_1(x,y) = \left[ f'(x)g(y) - f'(y)g(x) \right]^{2} + 4f(x)f(y)g'(x)g'(y)
\end{equation*}
is positive on $(0,1) \times (0,1)$ for all values $\alpha \geqslant 1$ of the slowdown parameter in~\eqref{vel}.
Consequently, system~\eqref{PDEs} never exhibits a region of non-hyperbolicity when $\phi^{A} = \phi^{B}$.

The case $\phi^A = \phi^B$ is not equivalent to the single-group case.
This is so because while our exclusion principle~(\ref{li:exclusion}) forbids two agents from any single group from co-occupying a single lattice cell, agents from different groups may do so.
Said another way, the total pedestrian density (from both groups), $\rho^A + \rho^B$, can exceed $1$.

The single-group case is equivalent to assuming $\phi^{A} = \phi^{B}$ and $c_{1} = c_{2} = c_{3} = 0$.
The latter assumption results in complete exclusion - two agents cannot co-occupy any lattice cell, regardless of group identity.
The continuum dynamics in the single-group case reduce to the scalar PDE
\begin{equation*}
\rho_t + 
\left[c_0 \phi_1f(\rho)\right]_x + 
\left[c_0 \phi_2f(\rho)\right]_y = 0.
\end{equation*}
It is easy to show that this equation is always hyperbolic.

\subsubsection*{Case 2: $\phi^A = -\phi^B$}
When $\phi^A = -\phi^B$, pedestrians from opposing groups follow floor velocity fields that point in opposite directions at every point on the lattice. 
This configuration would be appropriate, for instance, when two groups of pedestrians traverse a hallway in opposite directions. Indeed, this scenario is analogous to the traffic model studied in~\cite{ckpt14}. The hyperbolicity condition~\eqref{eq:hypercon} for this case simplifies to
\begin{align} 
	0 &\leqslant (\boldsymbol{\gamma} \cdot \phi)^{2} \Big( \left[ f'(\rho^A)g(\rho^B)\right]^2 + \left[f'(\rho^B)g(\rho^A) \right]^2 + 2 f'(\rho^A)f'(\rho^B)g(\rho^A)g(\rho^B)\nonumber\\
		&\qquad \qquad \qquad {}- 4f(\rho^A)f(\rho^B)g'(\rho^A)g'(\rho^B) \Big).
\end{align}
Define the function $h_2(x,y)$ by 
\begin{equation*}
	h_2(x,y) = \left[ f'(x)g(y)\right]^2 + \left[f'(y)g(x) \right]^2  + 2 f'(x)f'(y)g(x)g(y) - 4f(x)f(y)g'(x)g'(y).
\end{equation*}

\begin{figure}[t]
	\begin{subfigure}[t]{0.5\textwidth}
		\centering
		\includegraphics[width=2.7in]{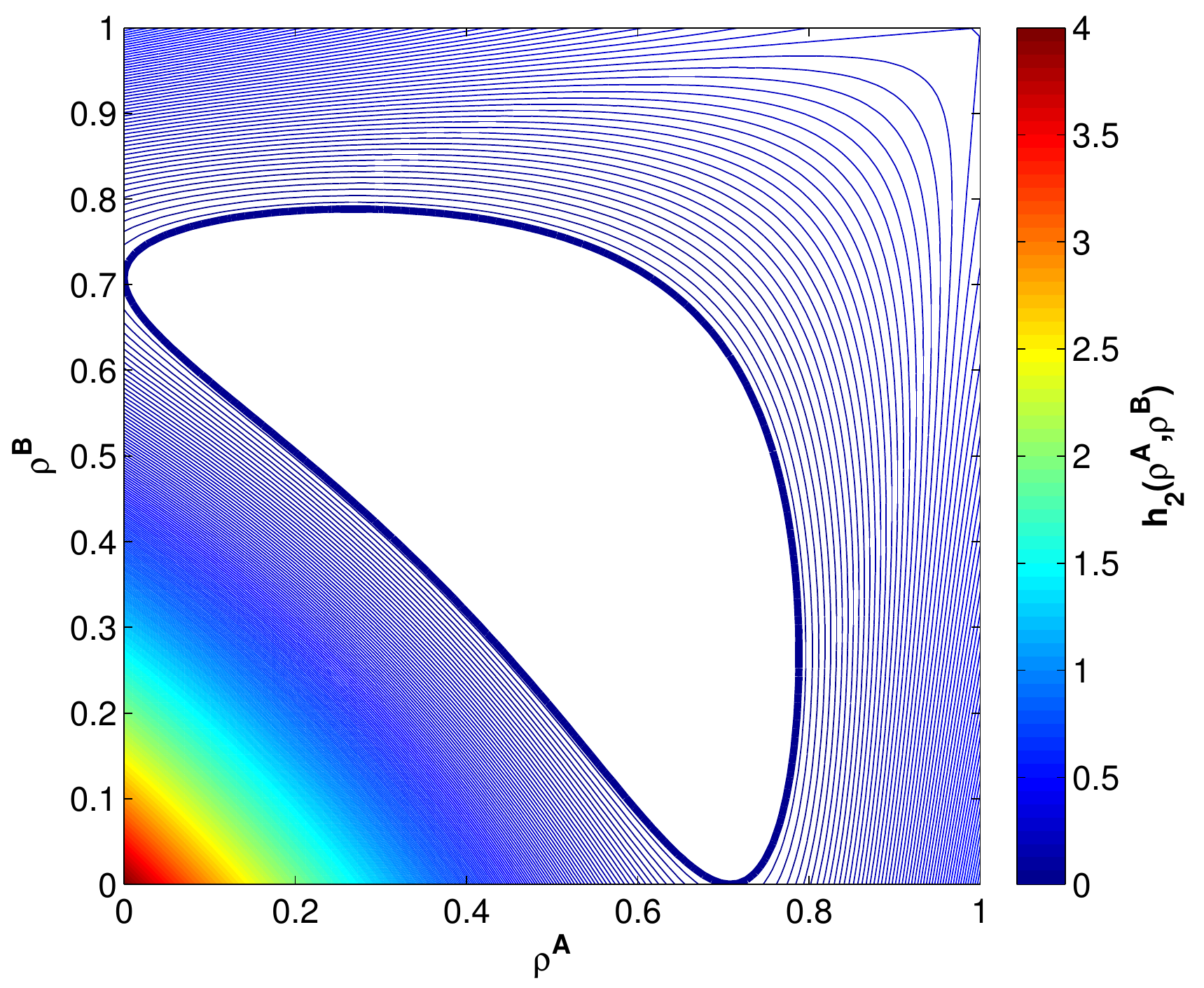}
		\caption{$\alpha = 2$}
	\end{subfigure}
	\begin{subfigure}[t]{0.5\textwidth}
		\centering
		\includegraphics[width=2.7in]{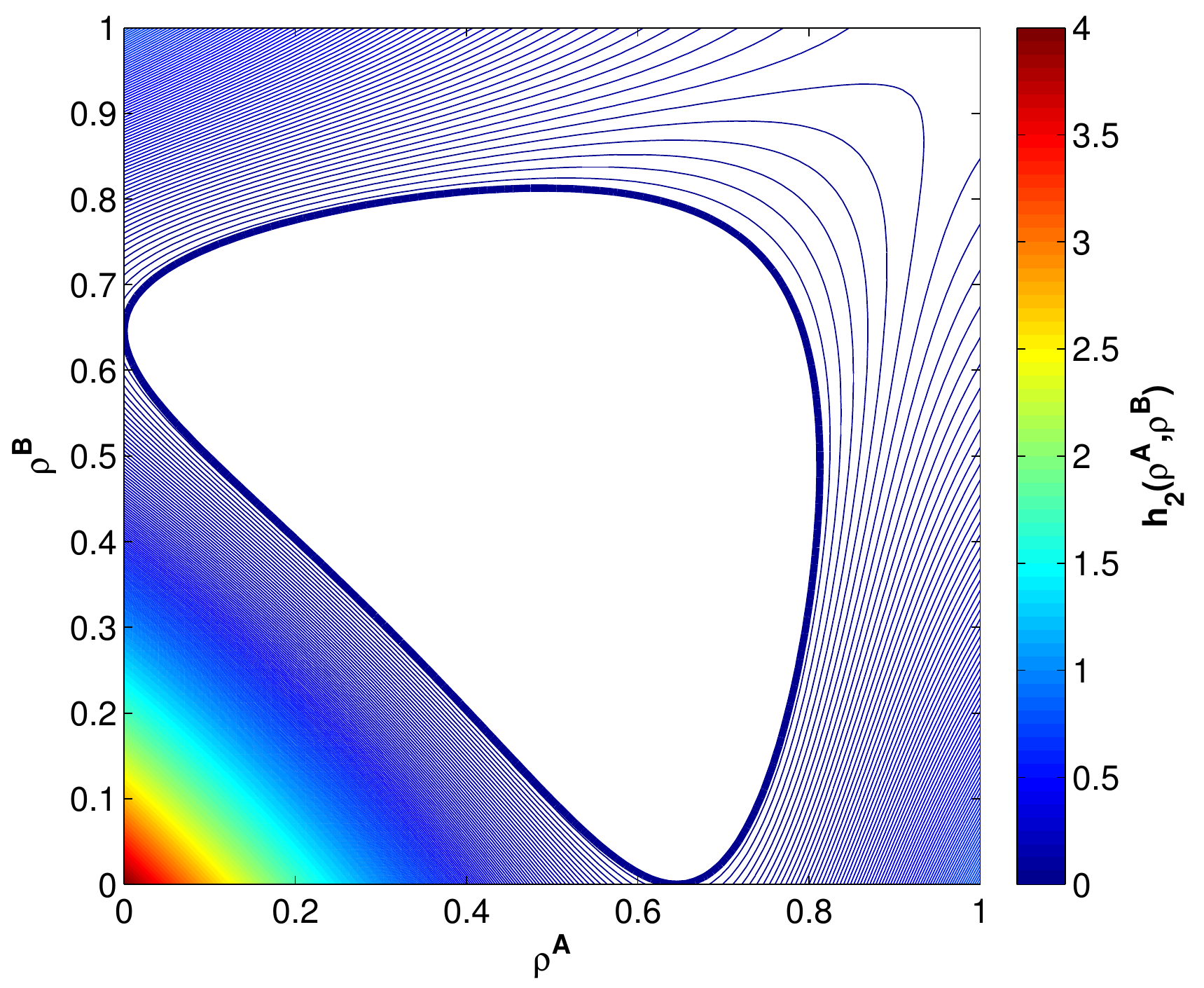}
		\caption{$\alpha = 4$}
	\end{subfigure}\\
	\begin{subfigure}[t]{0.5\textwidth}
		\centering
		\includegraphics[width=2.7in]{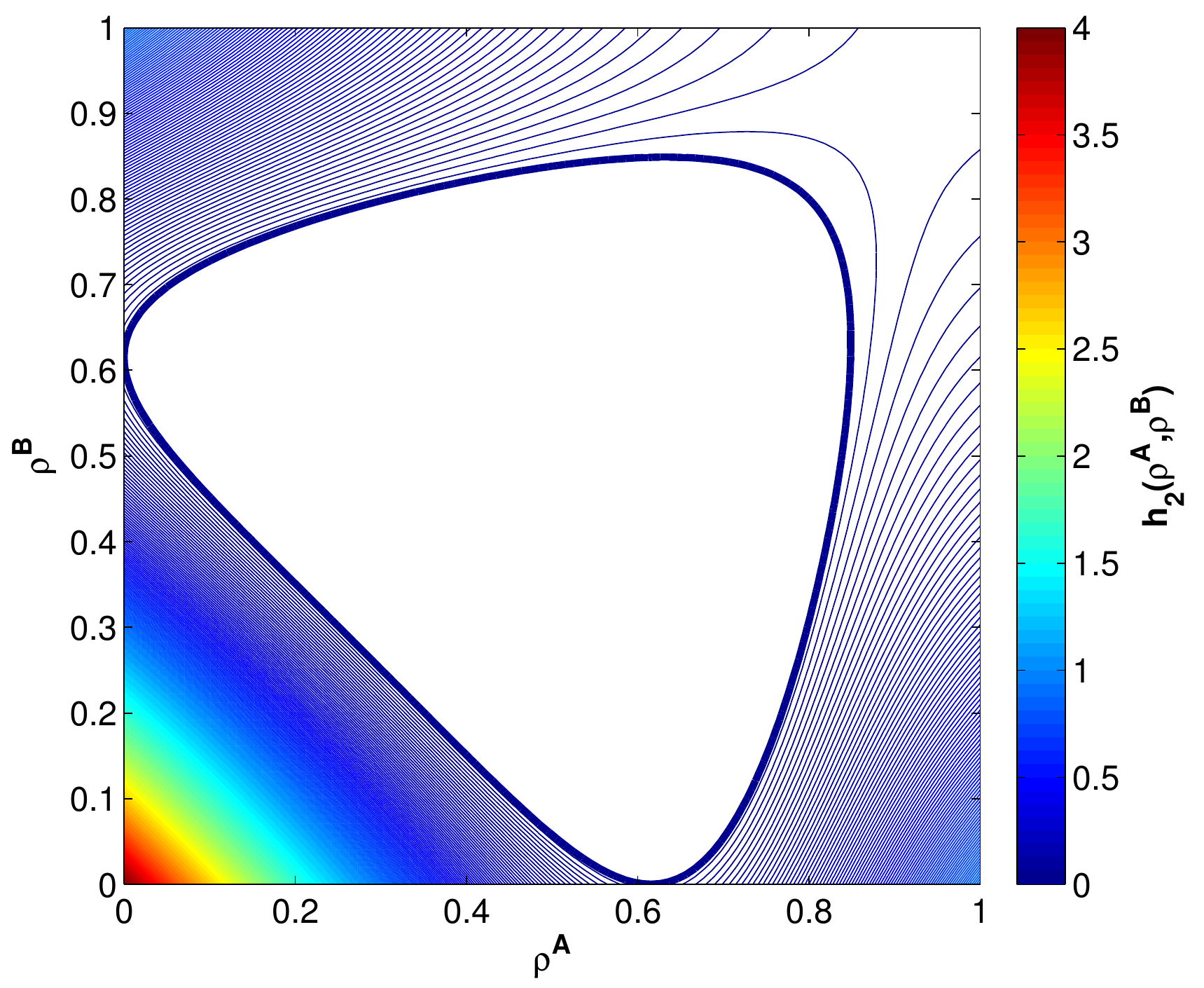}
		\caption{$\alpha = 8$}
	\end{subfigure}
	\begin{subfigure}[t]{0.5\textwidth}
		\centering
		\includegraphics[width=2.7in]{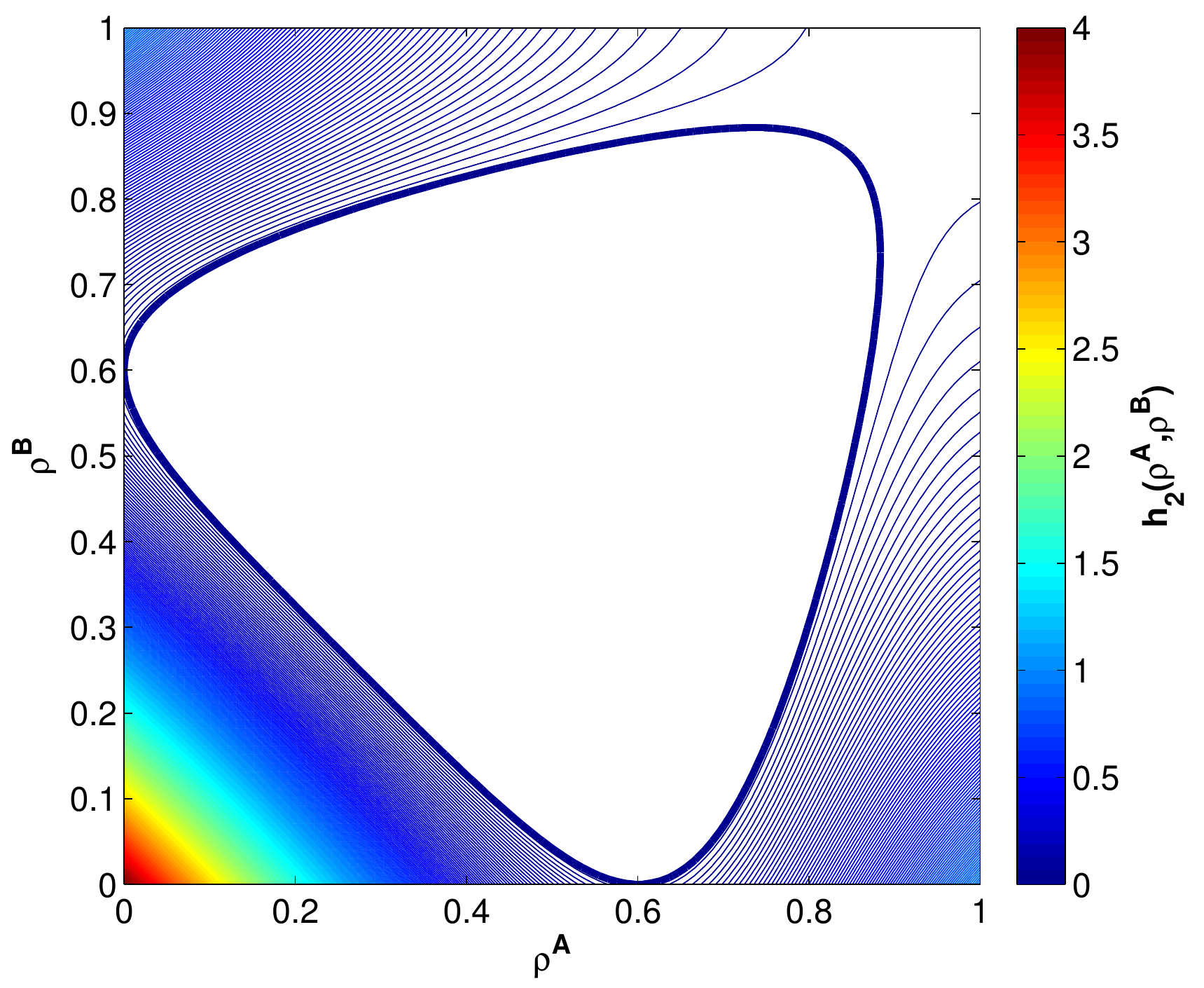}
		\caption{$\alpha = 16$}
	\end{subfigure}\\
	\begin{subfigure}[t]{0.5\textwidth}
		\centering
		\includegraphics[width=2.7in]{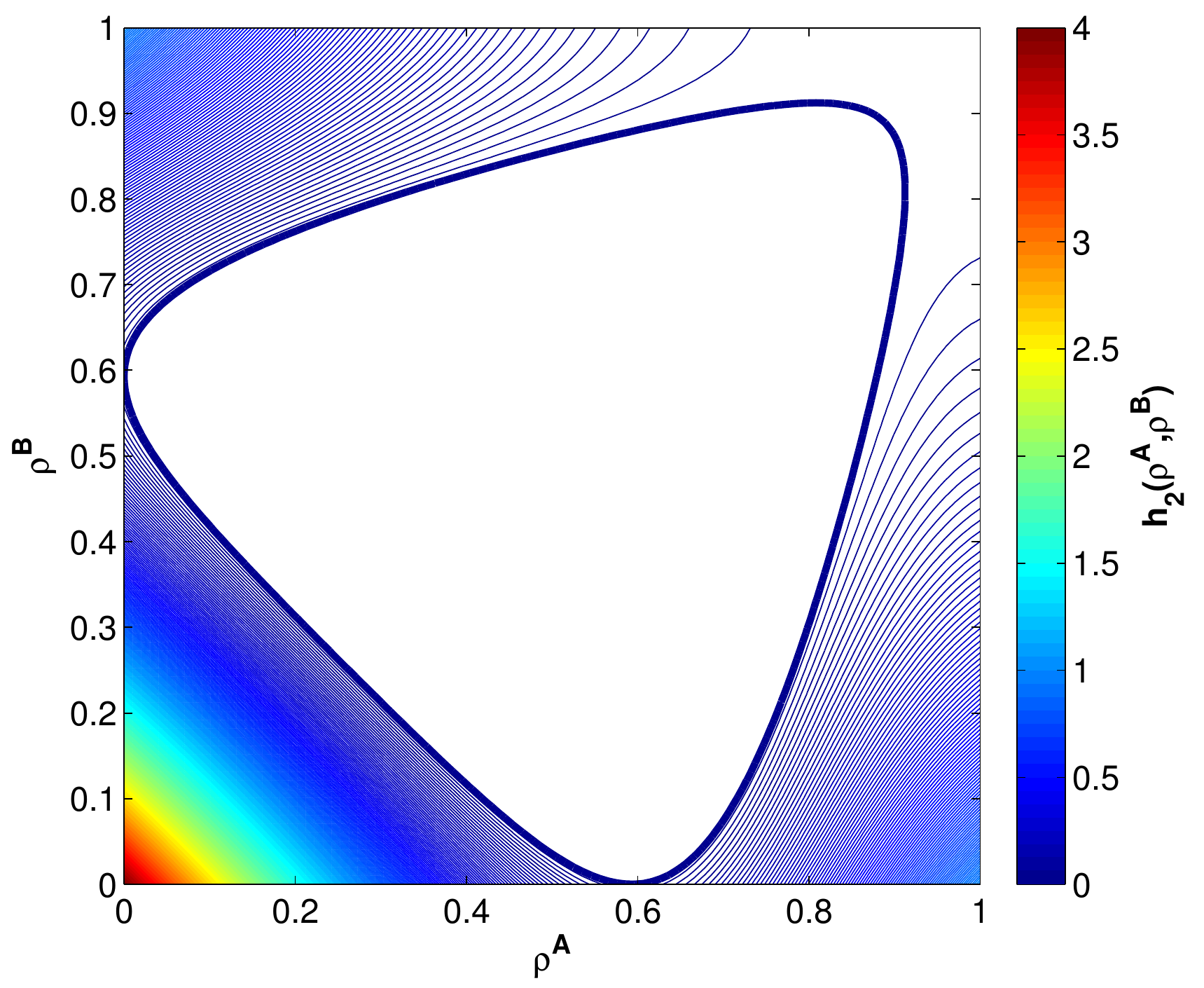}
		\caption{$\alpha = 32$}
	\end{subfigure}
	\begin{subfigure}[t]{0.5\textwidth}
		\centering
		\includegraphics[width=2.7in]{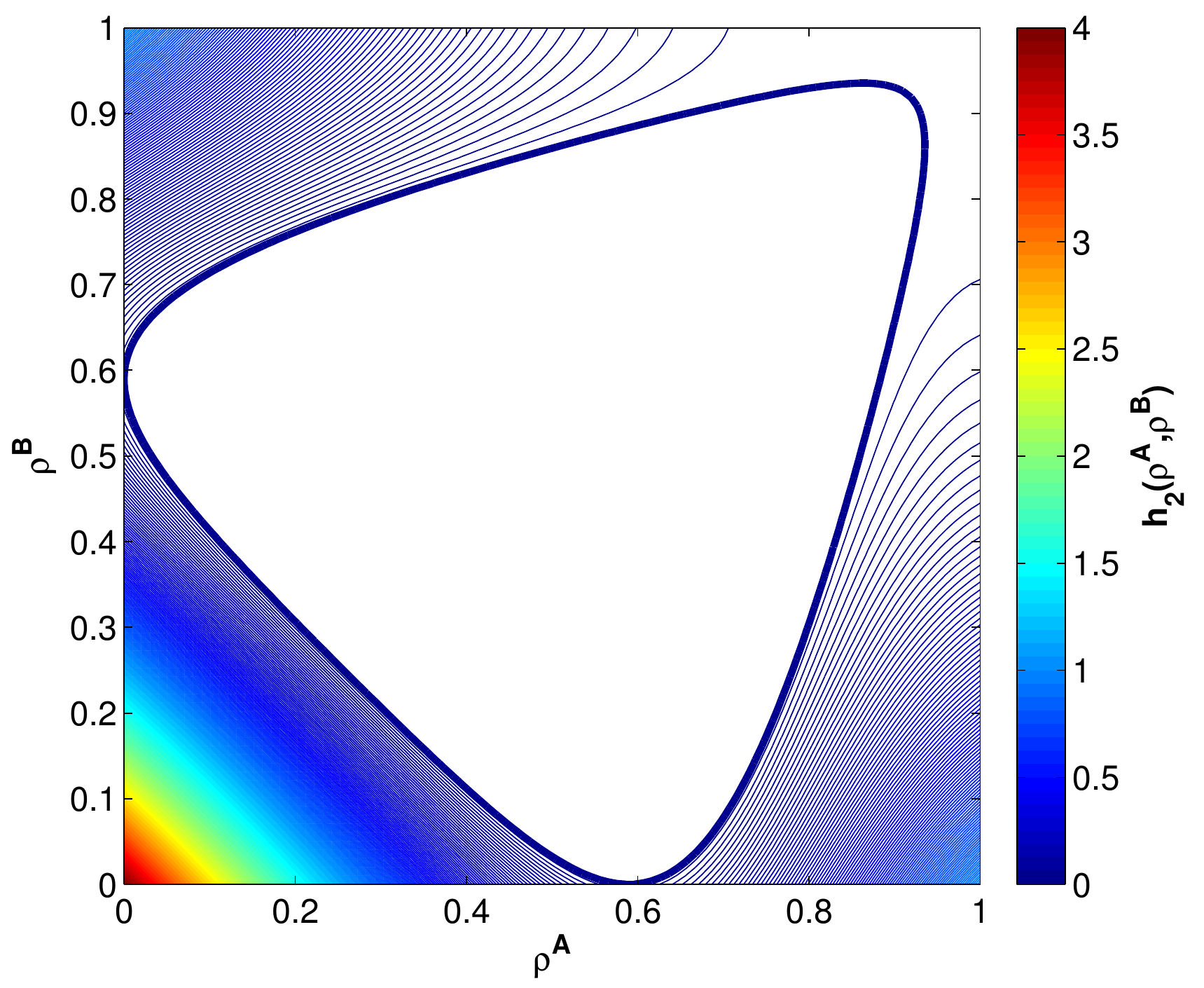}
		\caption{$\alpha = 64$}
	\end{subfigure}
	\caption{Regions of non-hyperbolicity for conservation law system~\eqref{PDEs} emerge when the floor velocity fields satisfy $\phi^{A} = - \phi^{B}$. For various values of the slowdown parameter $\alpha$, we plot the function $h_{2}$ implicated in the hyperbolicity condition for~\eqref{PDEs}. White shading indicates the region of non-hyperbolicity (where $h_{2} (\rho^{A}, \rho^{B}) < 0$). As $\alpha$ increases, so too does the likelihood of entering a non-hyperbolic region.}
	\label{hypmin1}
\end{figure}

We plot the value of $h_2(\rho^A,\rho^B)$ over $0 \leqslant \rho^A \leqslant 1$ and $0 \leqslant \rho^B \leqslant 1$ for various values of $\alpha$ in Figure~\ref{hypmin1}, marking a bold curve where $h_{2} (\rho^A,\rho^B) = 0$ and omitting contour levels below zero. 
Thus in each plot, the central white region represents those values of $\rho^A$ and $\rho^B$ for which $h_2(\rho^A,\rho^B) < 0$.
%
Figure~\ref{hypmin1} shows that when $\phi^A = -\phi^B$, large regions of non-hyperbolicity emerge for system~\eqref{PDEs}.
The size and shape of these regions depend on $\alpha$: As $\alpha$ (and therefore the strength of the slowdown interaction) increases, system~\eqref{PDEs} becomes more likely to enter a non-hyperbolic regime.

\section{Numerical Simulations}
\label{sec:num}

Here we compare the behavior of the stochastic model described in Section~\ref{sec:stoch} to that of the limiting system of conservation laws~\eqref{PDEs}.
We do so by comparing direct numerical simulation of the stochastic model with a finite-difference scheme for~\eqref{PDEs} given by the mesoscopic model in~\eqref{rhoA2}.

\subsection{Simulations with Uniform Initial Density}
\label{sec:numsumu}

In the first set of simulations, we consider uniform initial densities on non-overlapping squares. In particular,  
we consider two groups of agents moving toward each other on a $200 \times 200$ lattice with
initial conditions for the two groups given by
\[
	\sigma^A_{j,k} = \begin{cases} 1, & 81 \leq j \leq 100 \text{ and } 81 \leq k \leq 100,\\
				0, & \text{otherwise}, \end{cases}
\] 
\[
	\sigma^B_{j,k} = \begin{cases} 1, & 101 \leq j \leq 120 \text{ and } 101 \leq k \leq 120,\\
				0, & \text{otherwise}. \end{cases}
\]
We take the velocity potentials to be
\[
	\psi^A(j,k) = (180 - j)^2 + (180 - k)^2 {\rm ~~and~~}
	\psi^B(j,k) = (21 - j)^2 + (21 - k)^2.
\]
Computing the gradients of these potentials and then normalizing in $\ell^{1}$ yields the floor velocity fields for the two groups of agents:
\begin{align*}
	\phi^A(j,k) &= \left( \frac{180 - j}{|180 - j| + |180 - k|},  \frac{180 - k}{|180 - j| + |180 - k|}\right),
\\
	\phi^B(j,k) &= \left( \frac{21 - j}{|21 - j| + |21 - k|},  \frac{21 - k}{|21 - j| + |21 - k|}\right).
\end{align*}
Group $A$ moves toward the point $(180,180)$, while the target point for group $B$ is $(21,21)$.  
Consequently, the two groups pass through each other in order to reach their corresponding target points.
We set the velocity scaling parameter $c_0 = 1$ and consider two cases of differing slowdown interaction strengths: $\alpha = 2,4$ in~\eqref{vel}.

\begin{figure}[t]
\centering
\includegraphics[width=68mm]{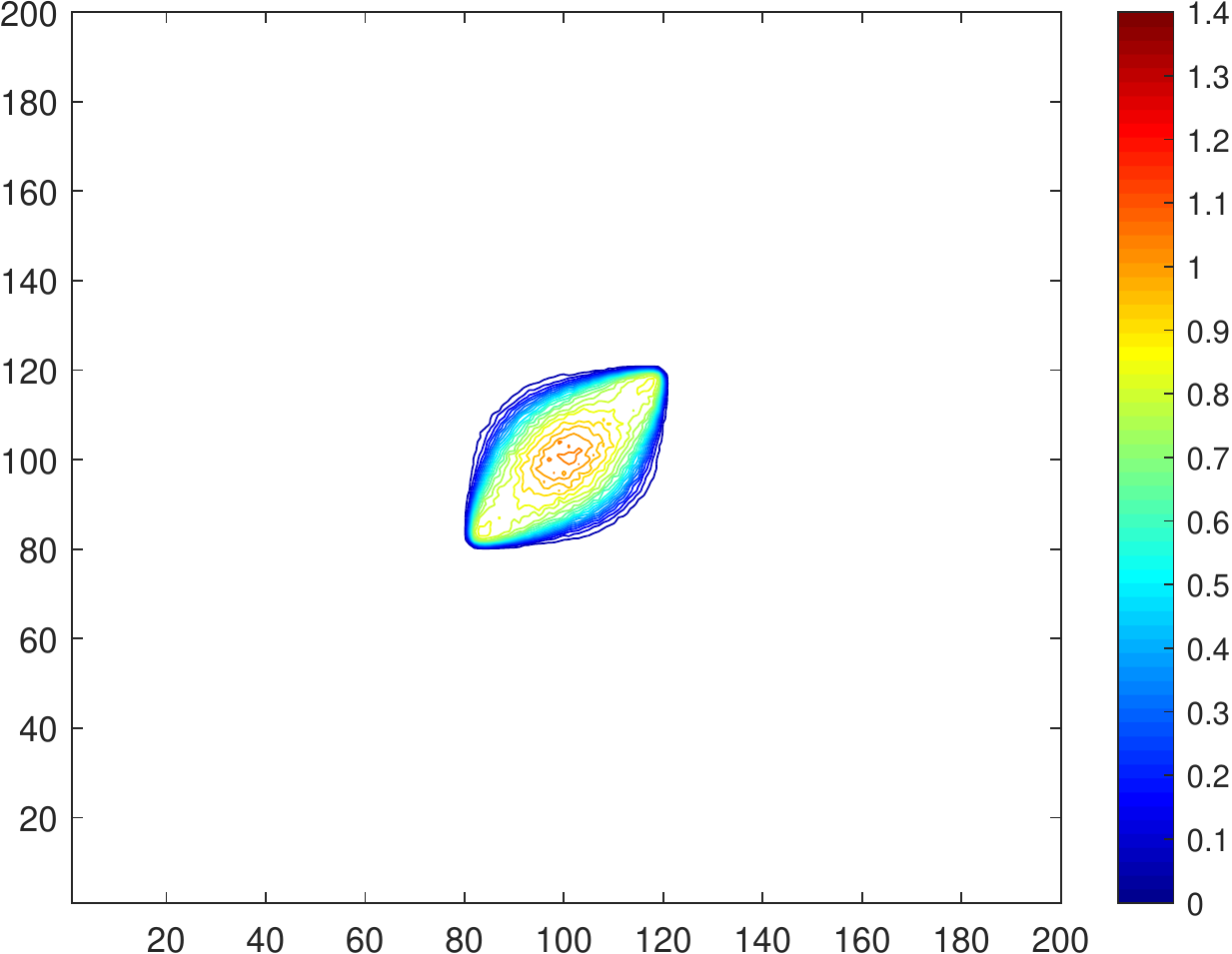}
\hspace{1em}
\includegraphics[width=68mm]{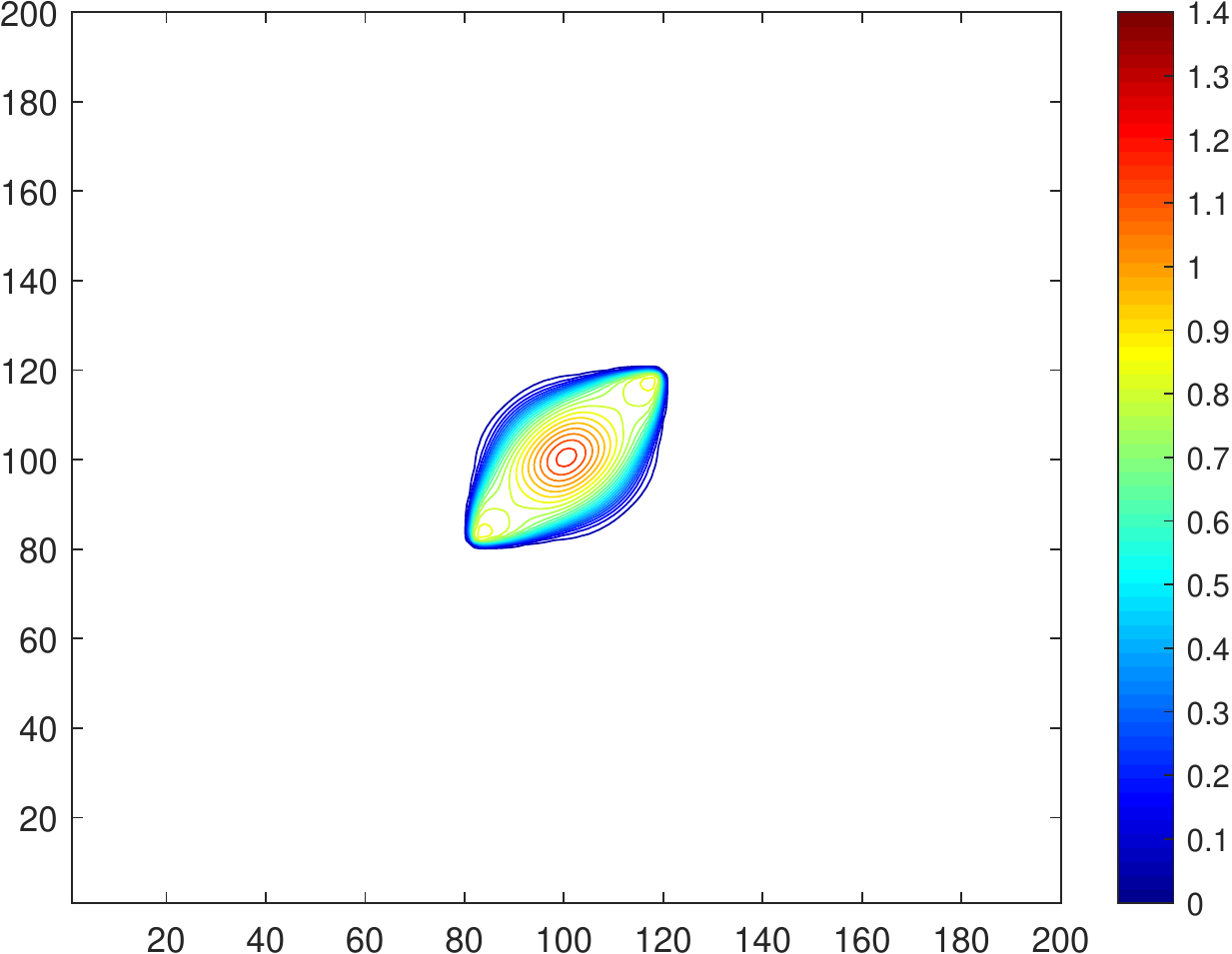}\\
\includegraphics[width=68mm]{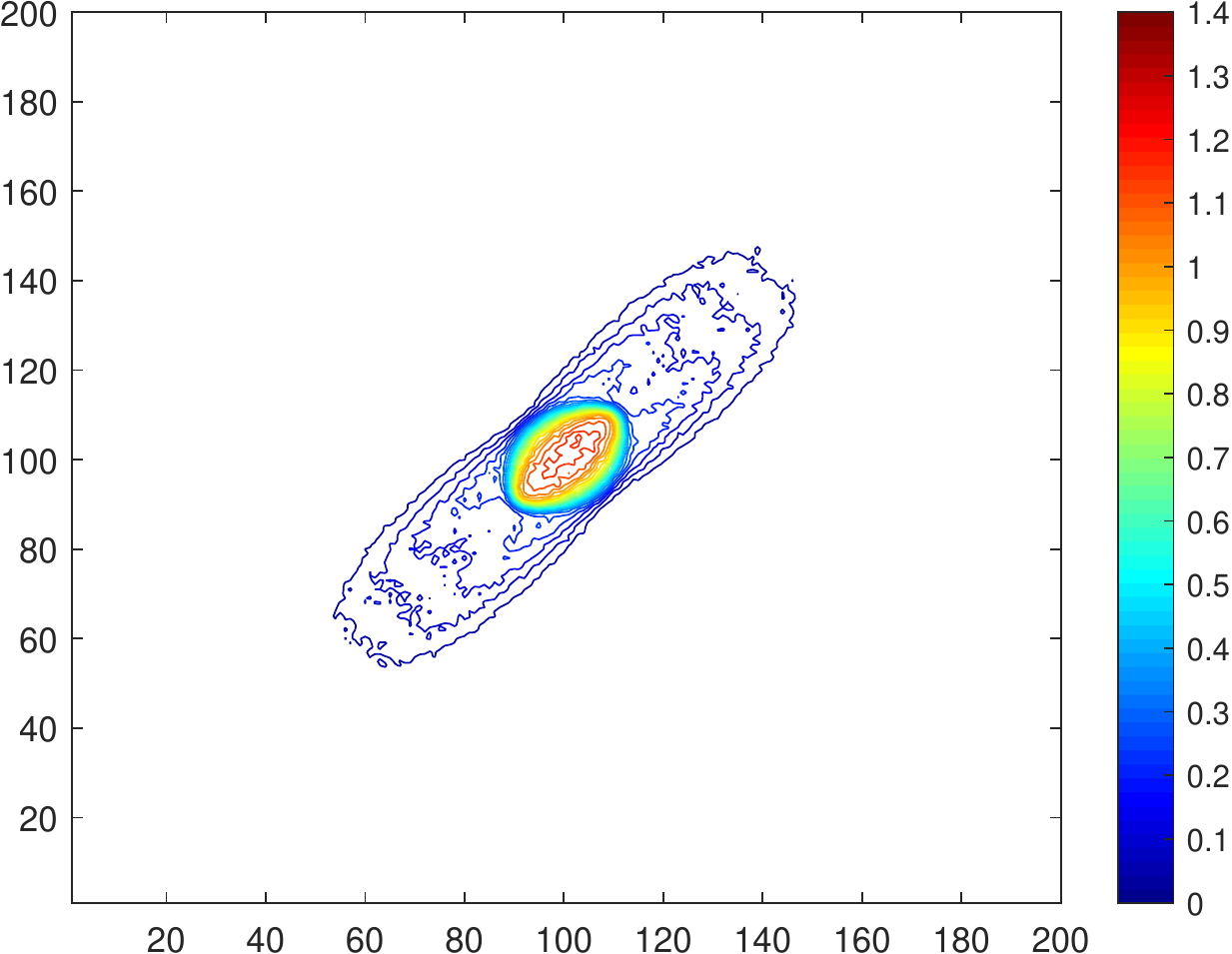}
\hspace{1em}
\includegraphics[width=68mm]{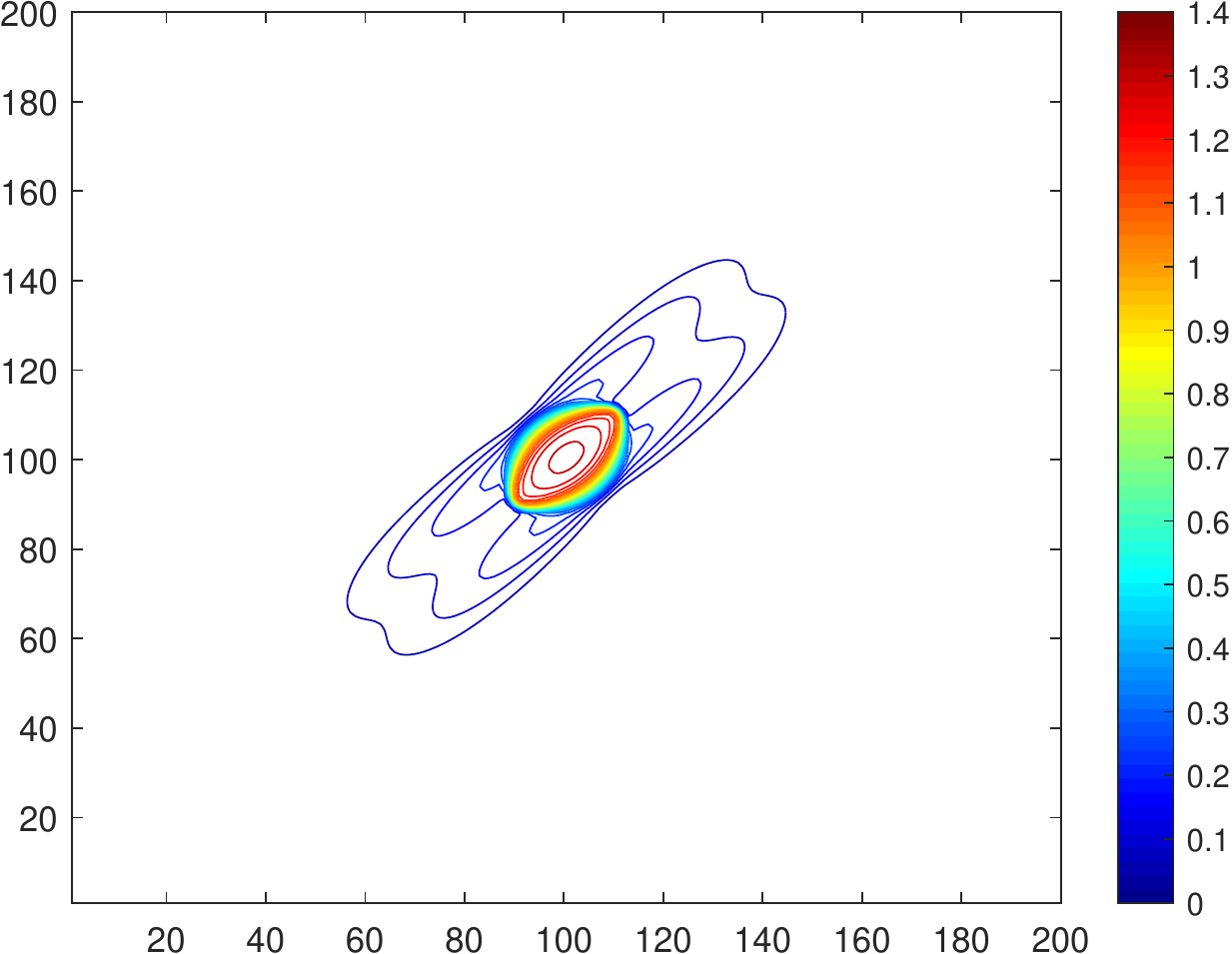}\\
\includegraphics[width=68mm]{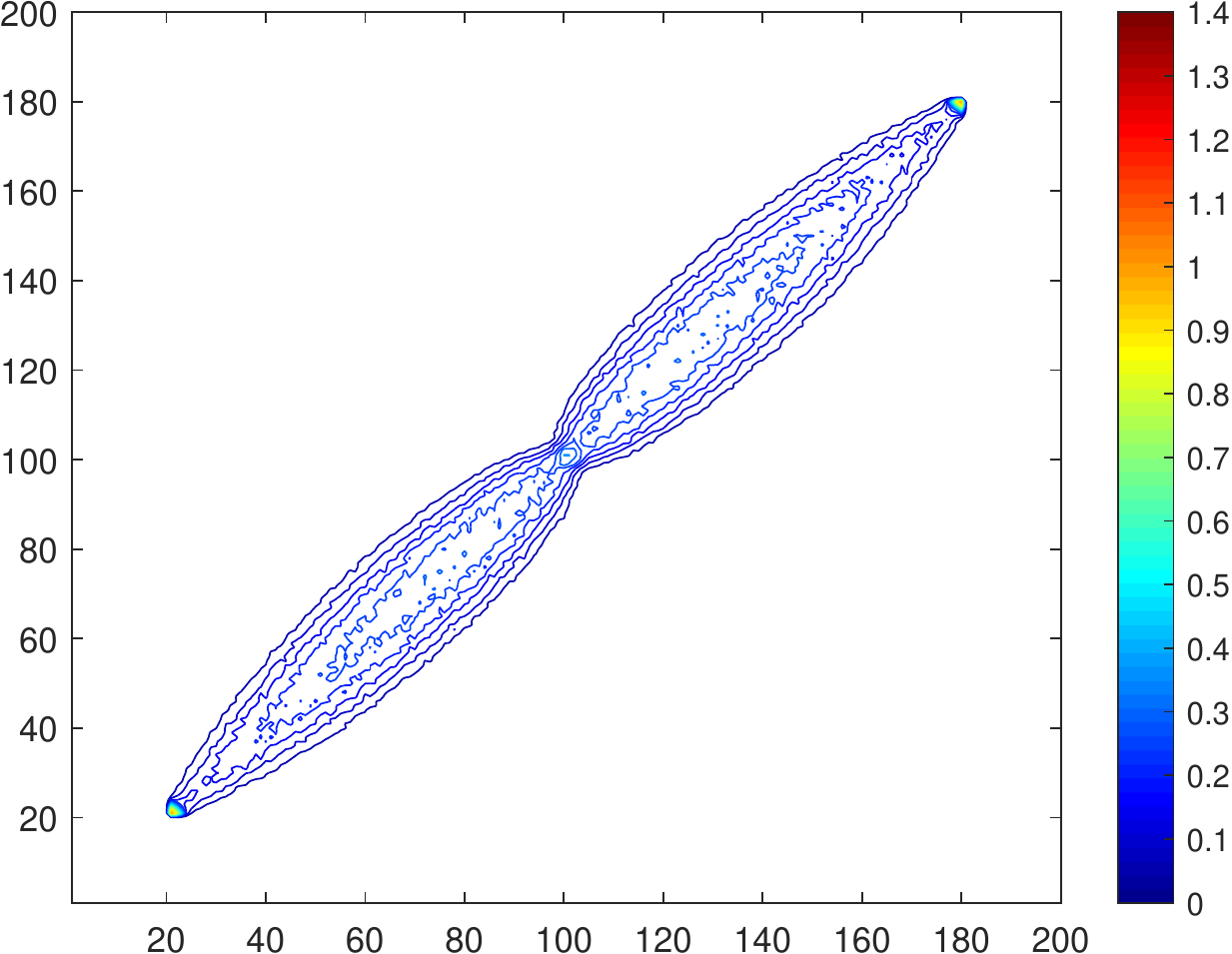}
\hspace{1em}
\includegraphics[width=68mm]{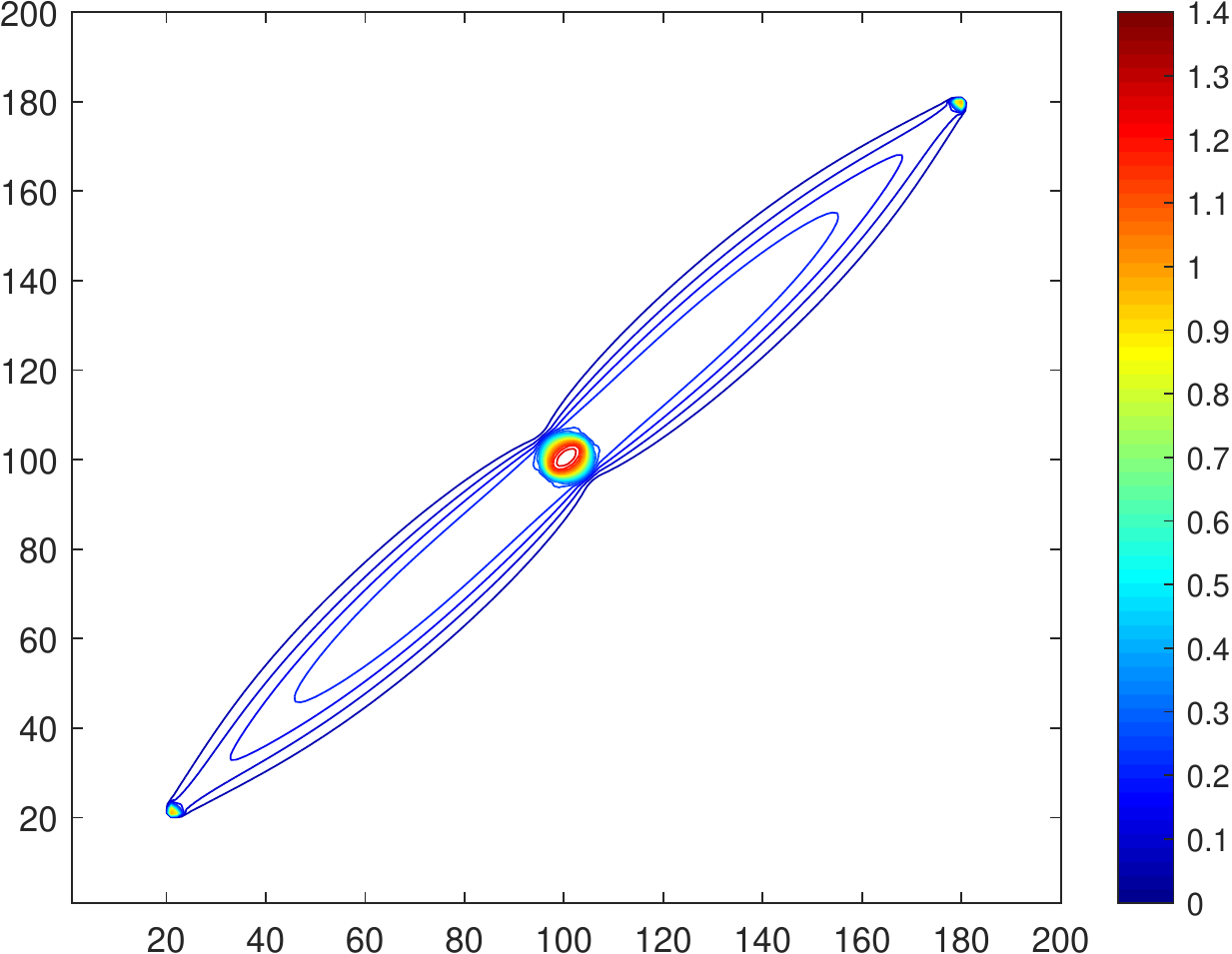}\\
\includegraphics[width=68mm]{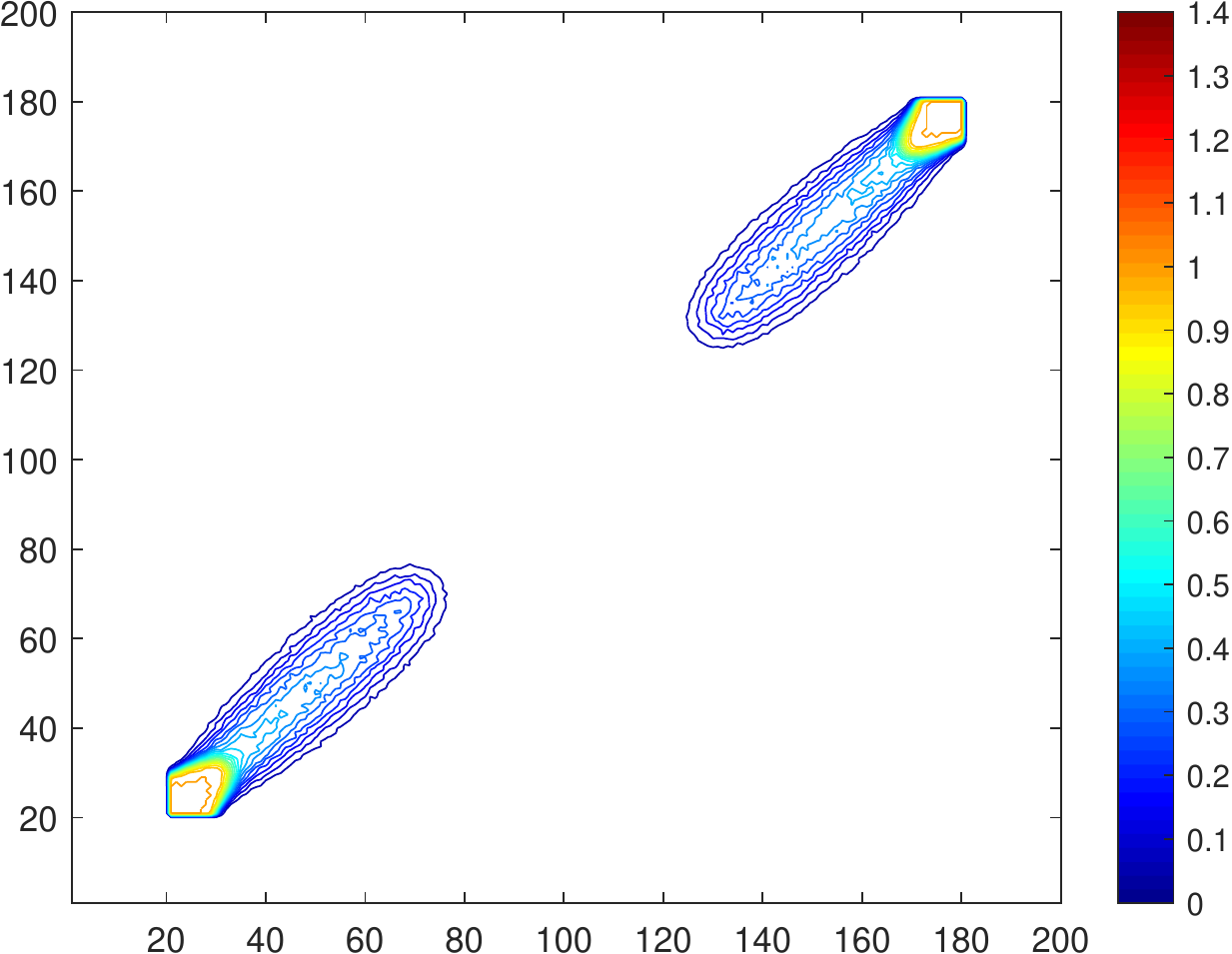}
\hspace{1em}
\includegraphics[width=68mm]{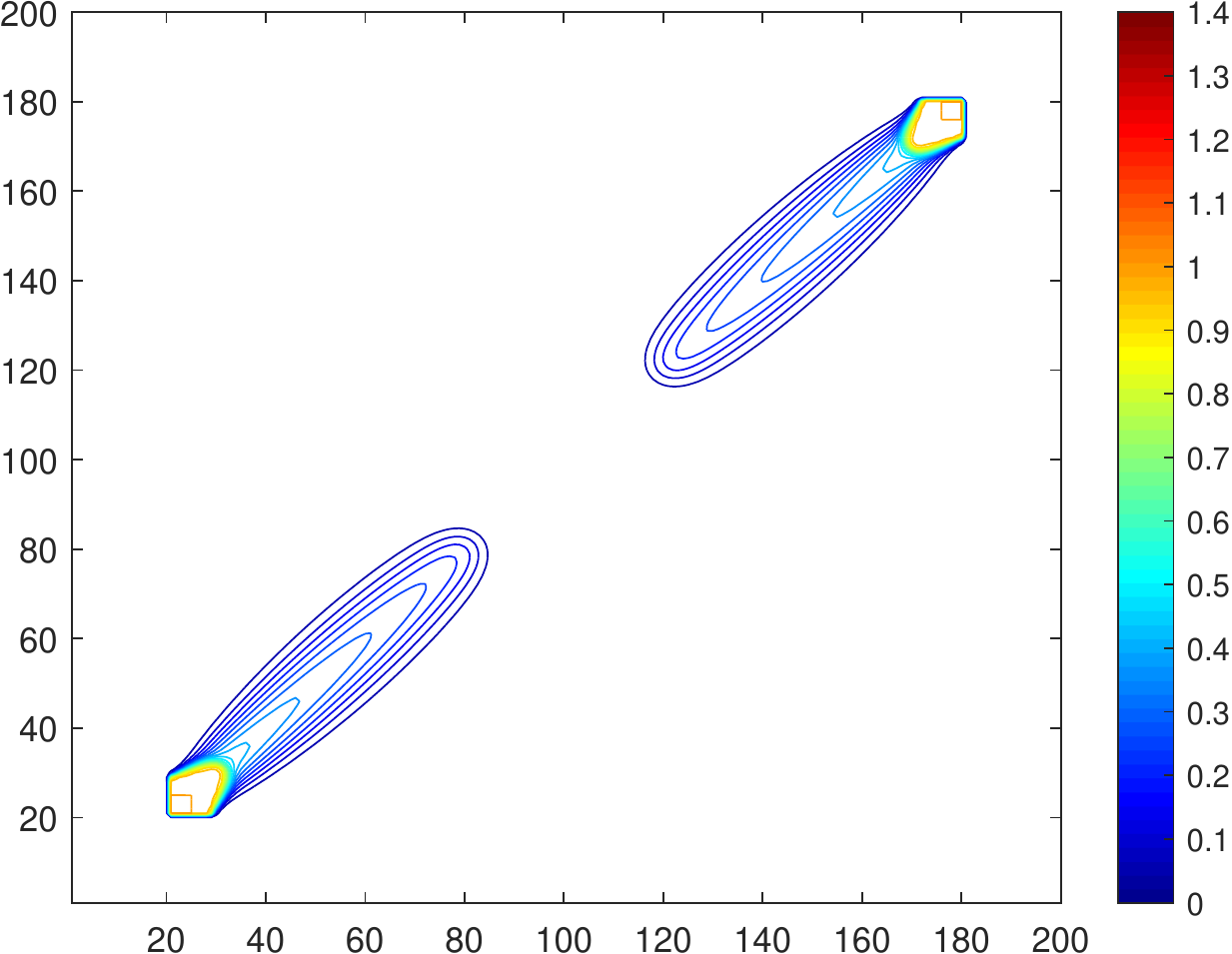}
\caption{Two group densities, initially distributed uniformly over disjoint squares, pass through one another under mild slowdown interaction strength.
A finite-difference simulation of the macroscopic PDEs~\eqref{PDEs} (right column) closely matches an average of 1000 realizations of the microscopic stochastic model (left column).
Slowdown strength: $\alpha=2$.
Times: 35, 105, 175, 245.}
\label{fig1alpha2}
\end{figure}

\begin{figure}[t]
\centering
\includegraphics[width=70mm]{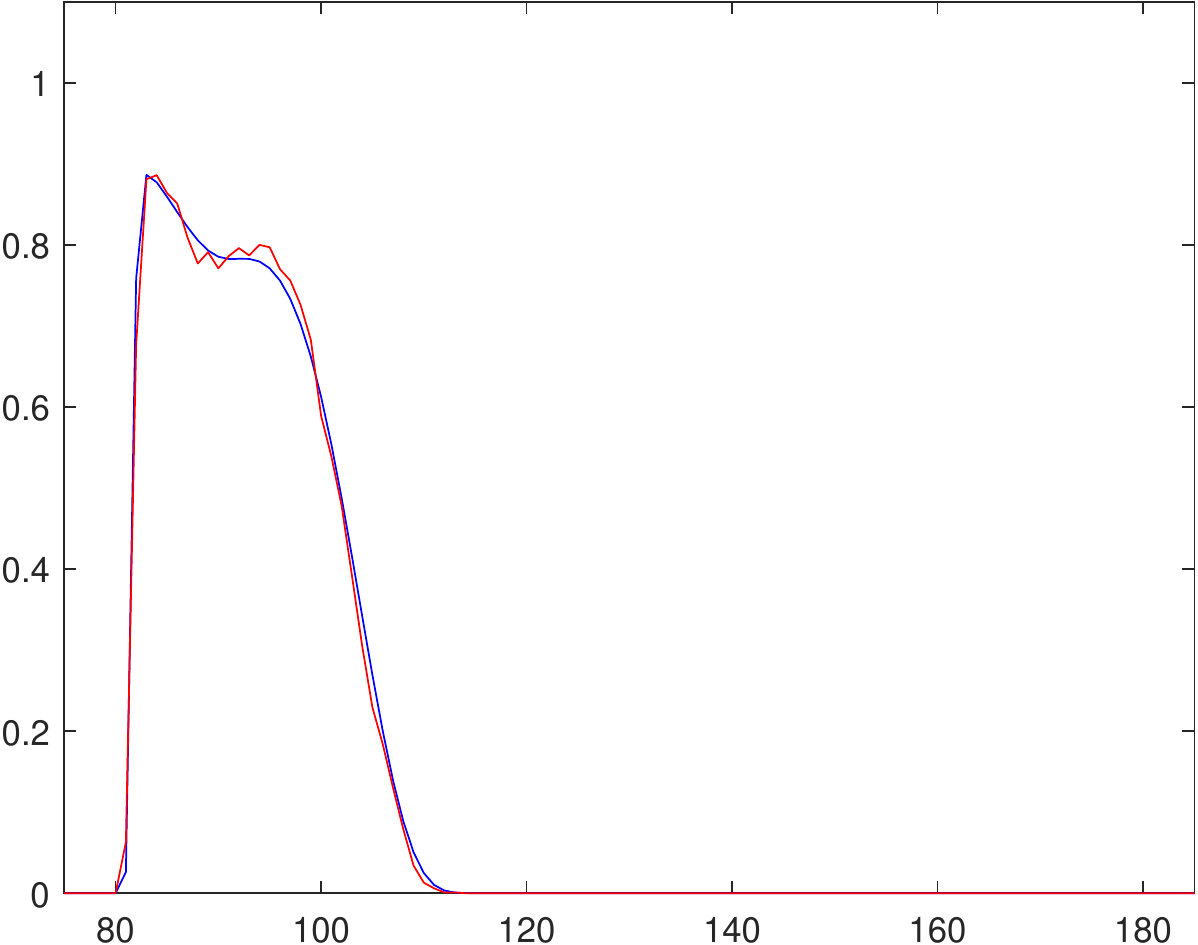}
\hspace{0.5em}
\includegraphics[width=70mm]{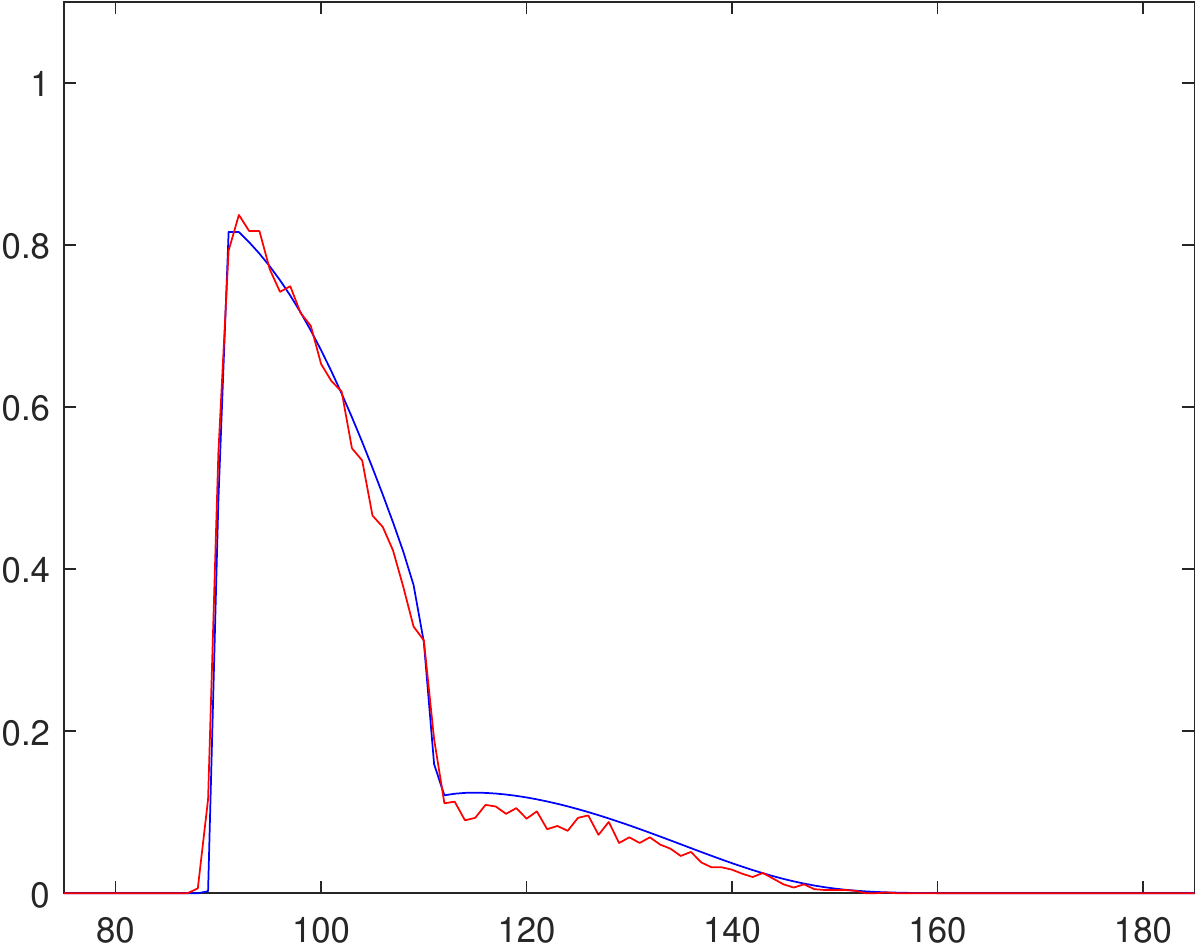}\\
\vspace{1ex}
\includegraphics[width=70mm]{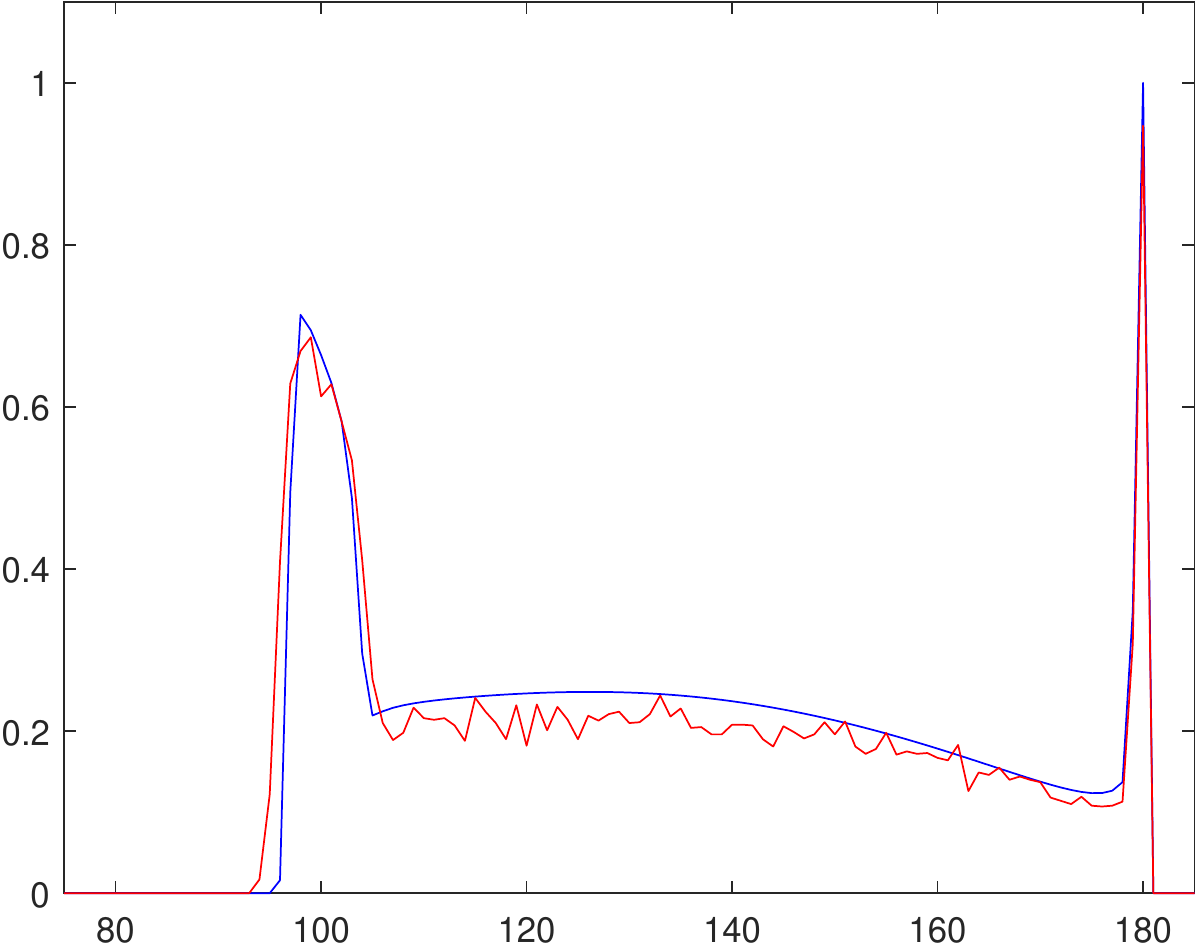}
\hspace{0.5em}
\includegraphics[width=70mm]{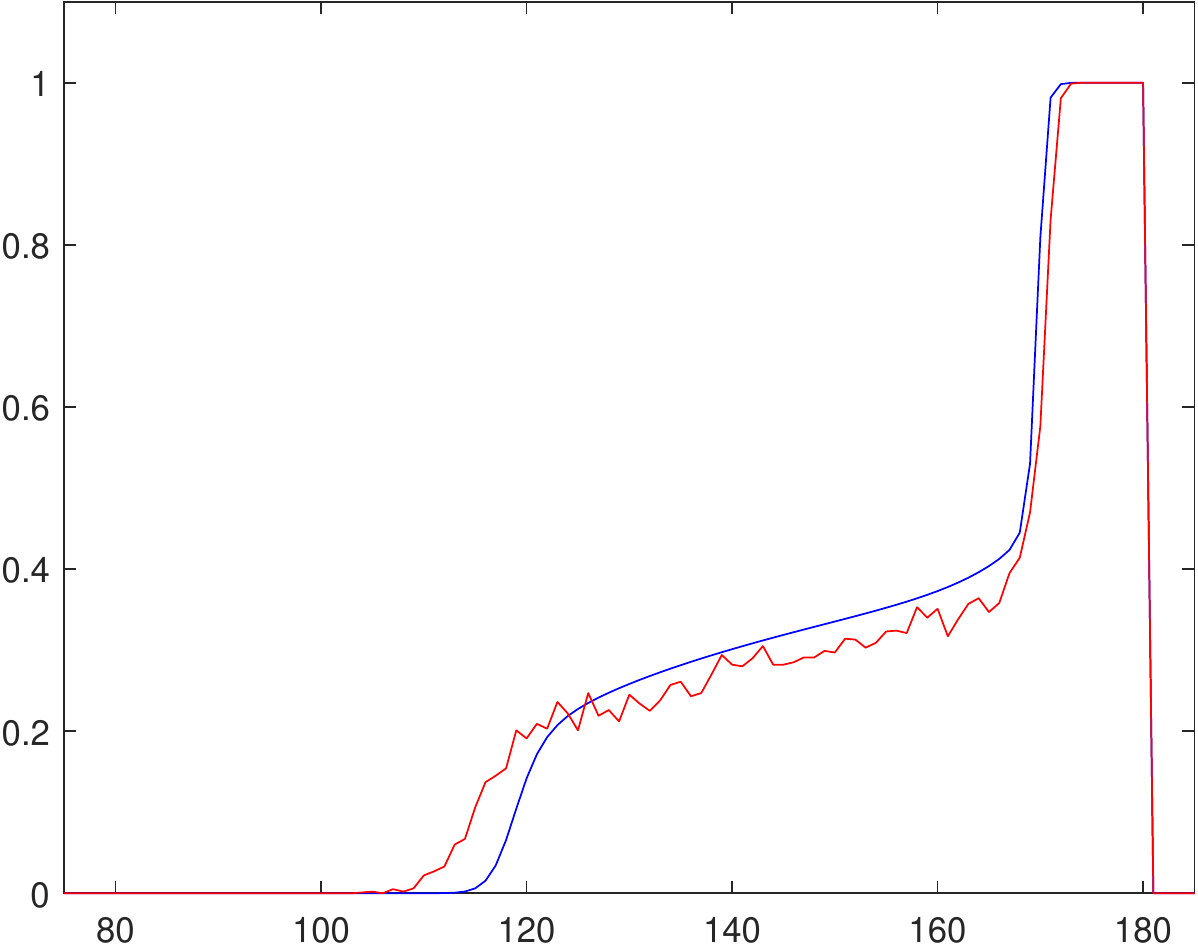} 
\caption{Evolution of group A density along the diagonal $j=k$ of the lattice.
Setup is as in Figure~\ref{fig1alpha2}.
Blue: Macroscopic model~\eqref{PDEs}.
Red: Microscopic stochastic model.
Slowdown strength: $\alpha=2$.
Times (left to right): 35, 105, 175, 245.}
\label{fig3alpha2}
\end{figure}

\begin{figure}[t]
\centering
\includegraphics[width=68mm]{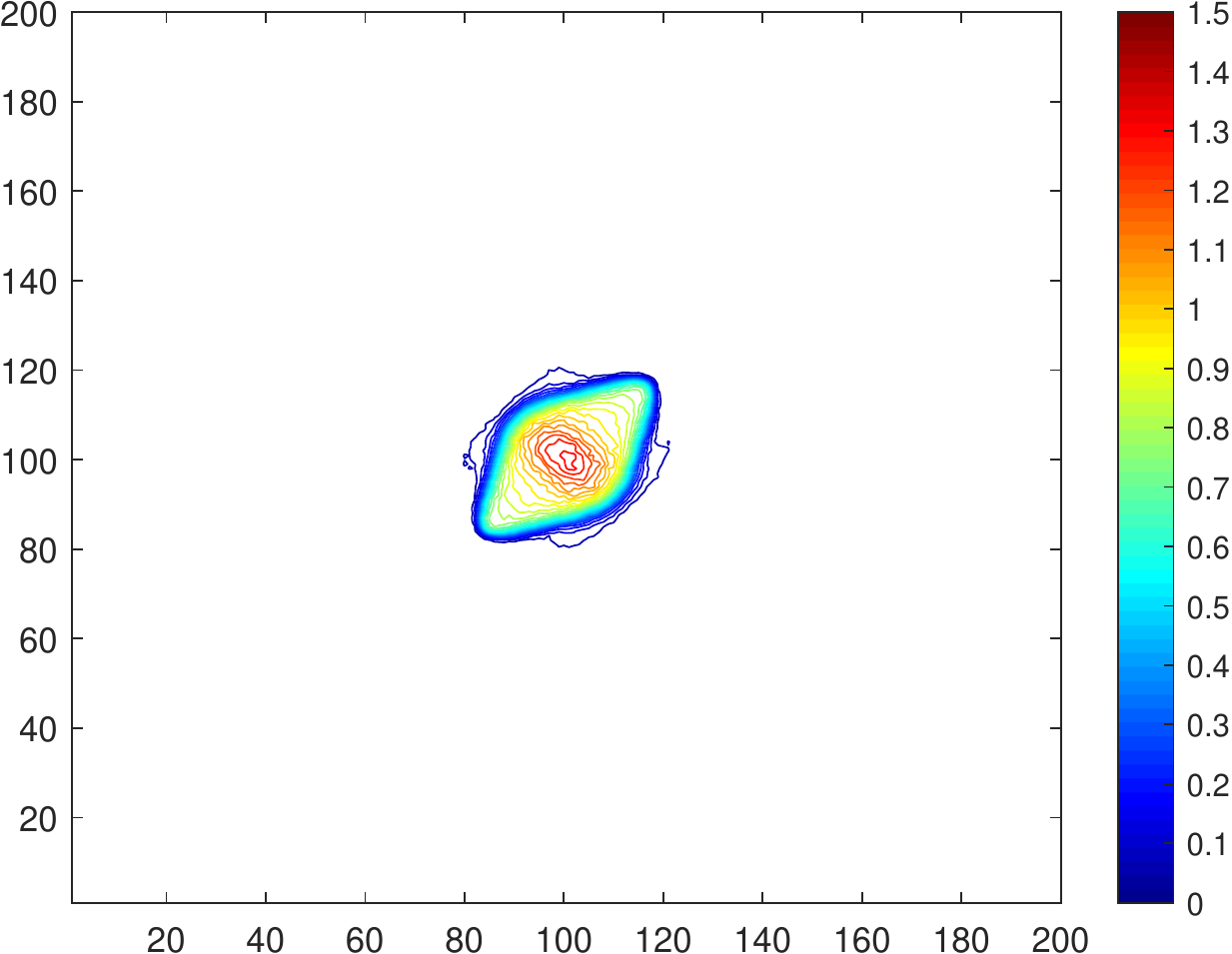}
\hspace{1em}
\includegraphics[width=68mm]{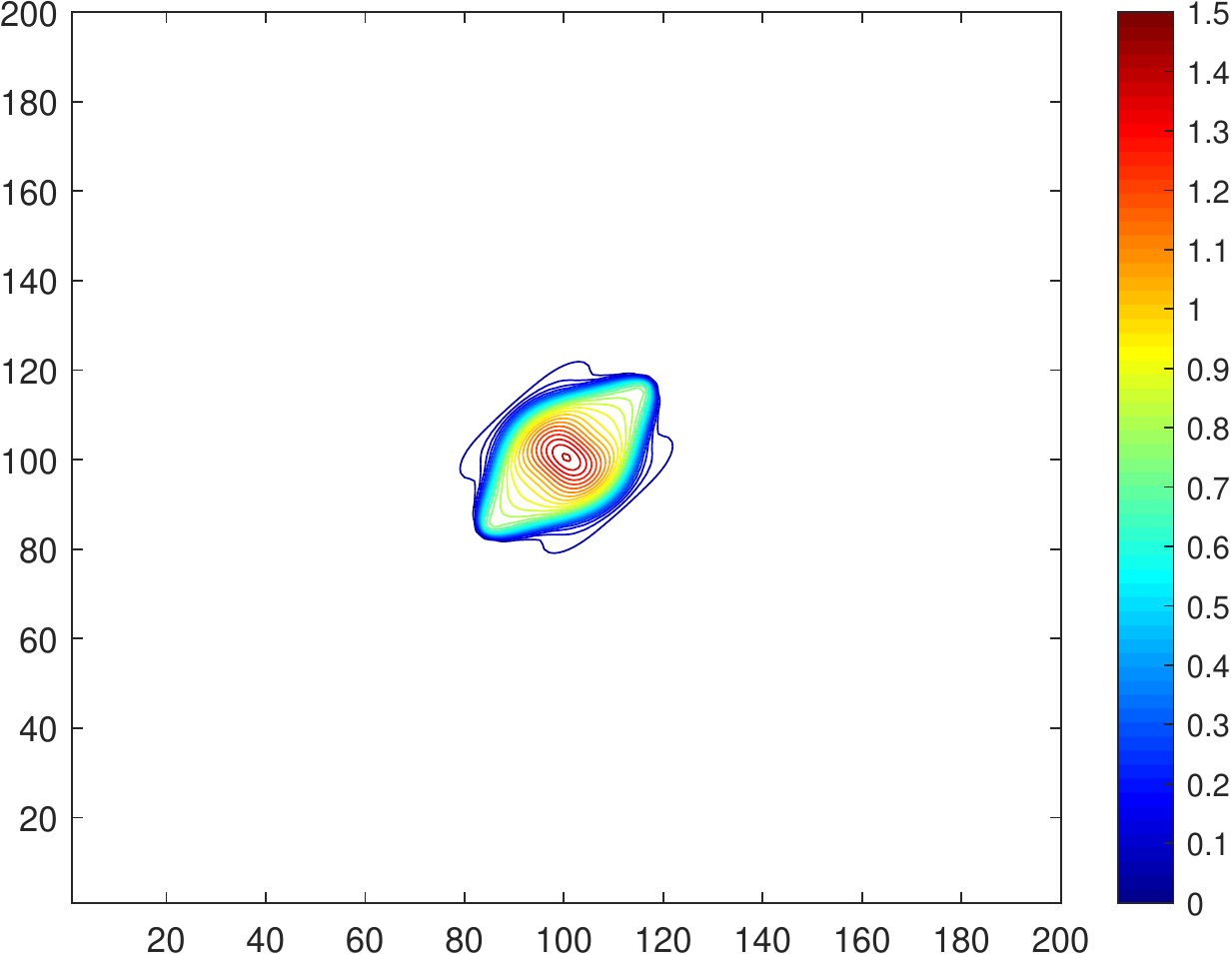}\\
\includegraphics[width=68mm]{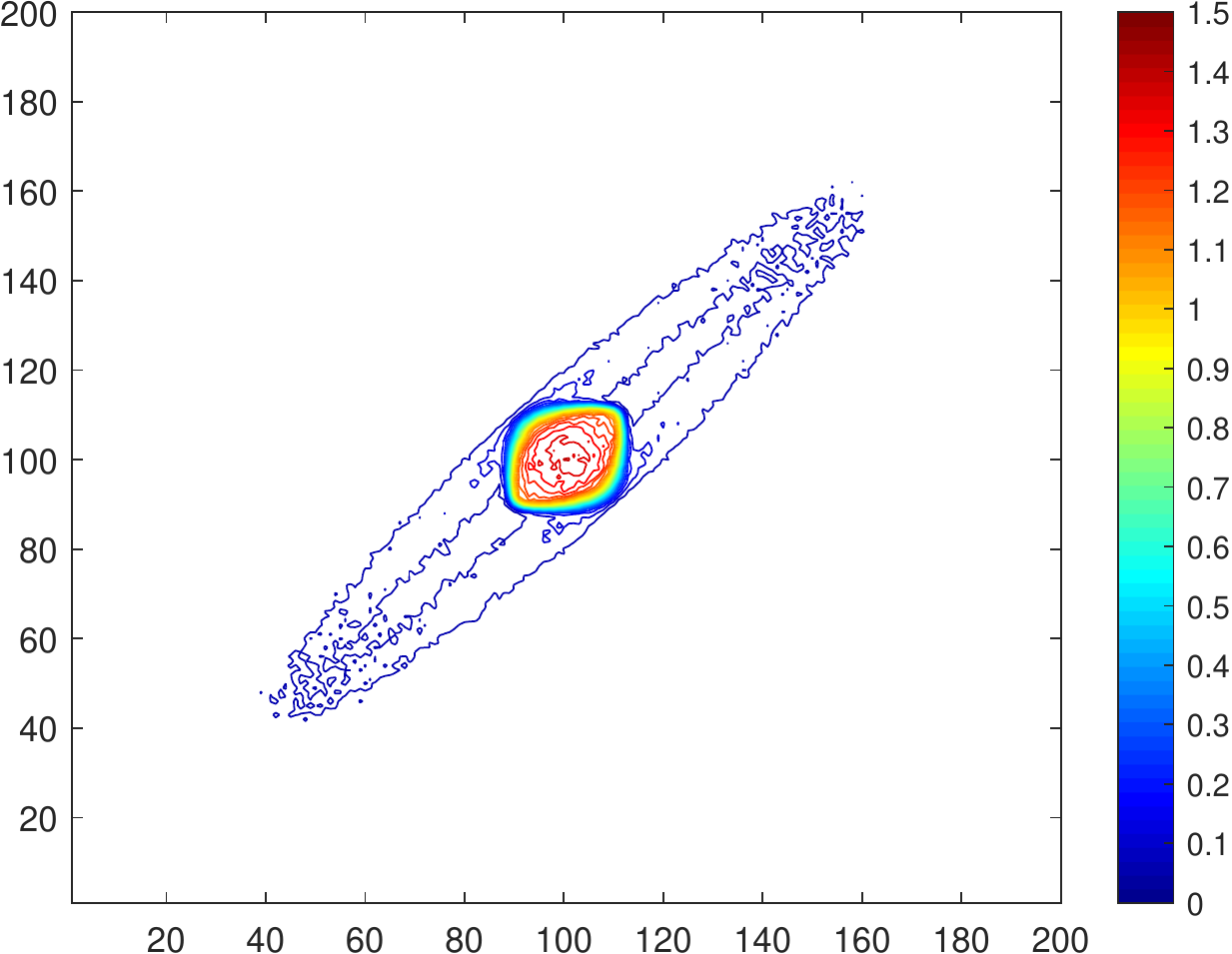}
\hspace{1em}
\includegraphics[width=68mm]{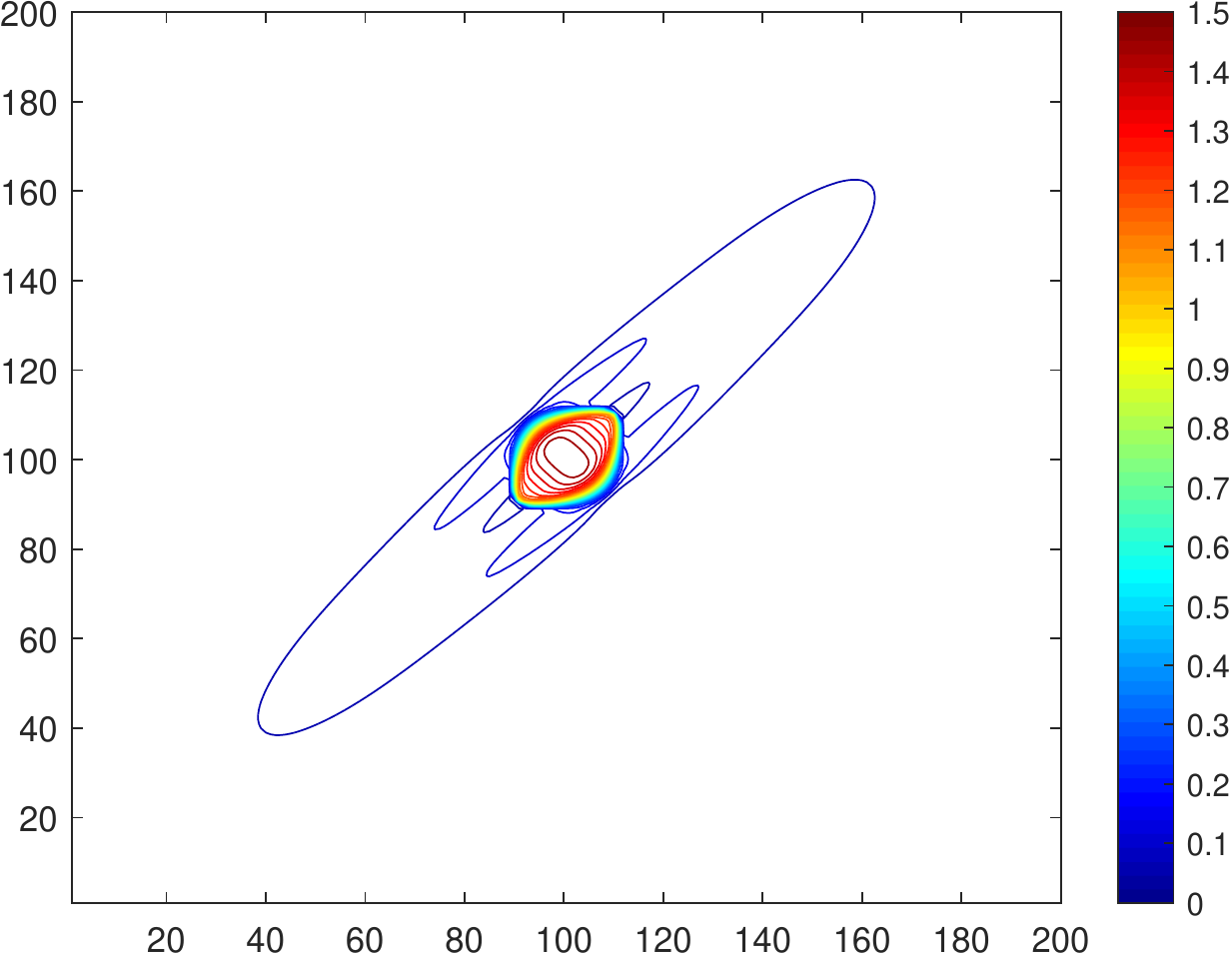}\\
\includegraphics[width=68mm]{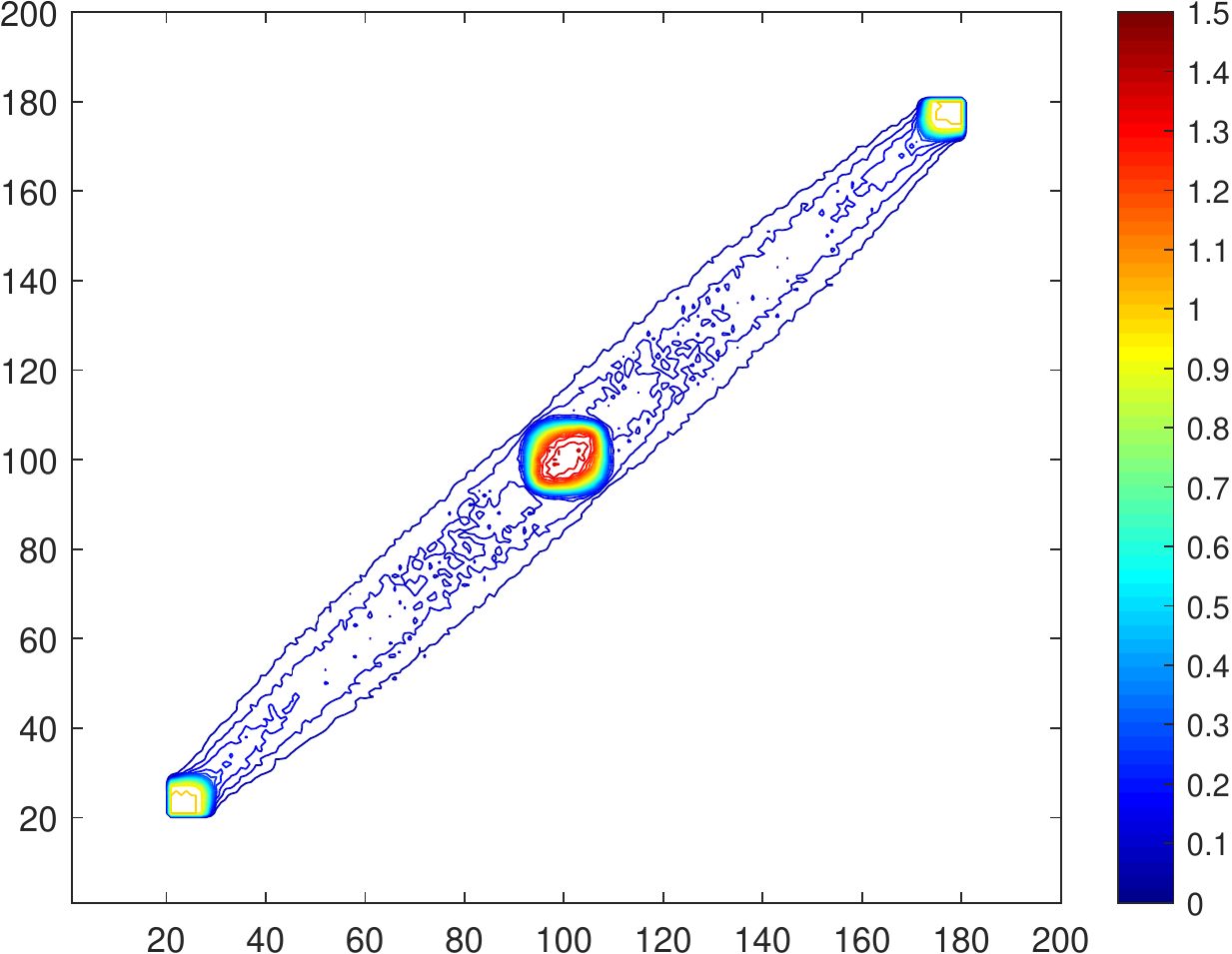}
\hspace{1em}
\includegraphics[width=68mm]{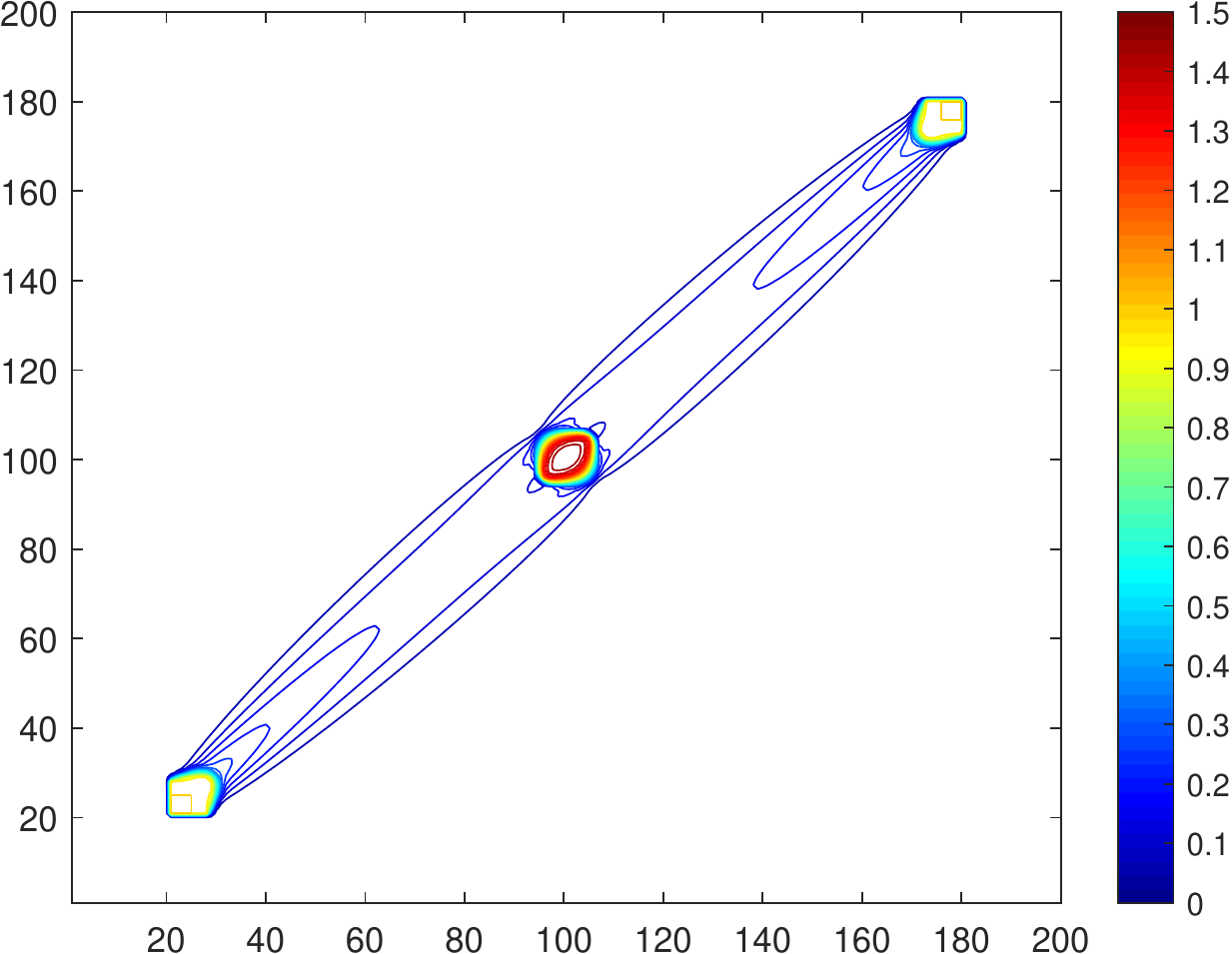}\\
\includegraphics[width=68mm]{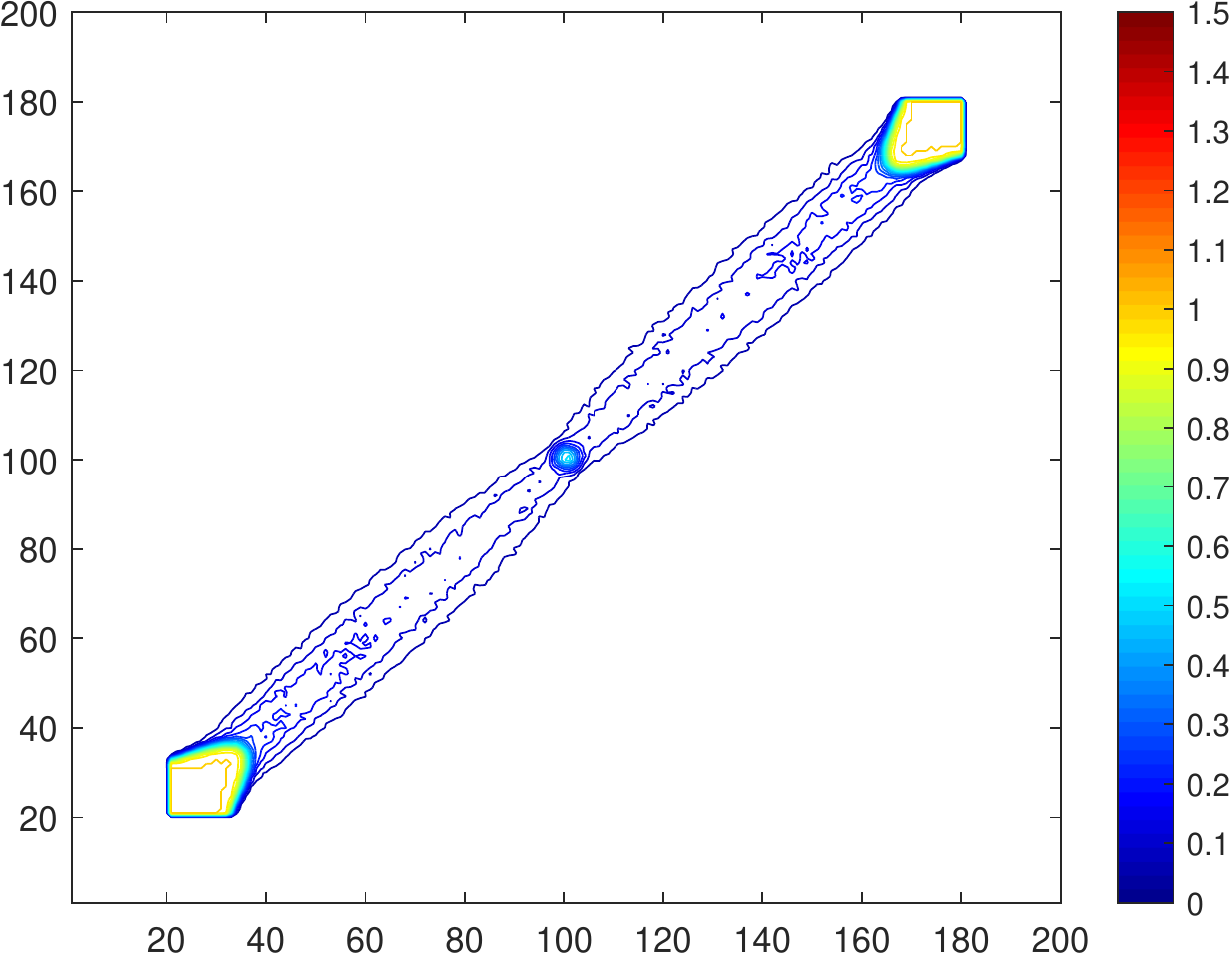}
\hspace{1em}
\includegraphics[width=68mm]{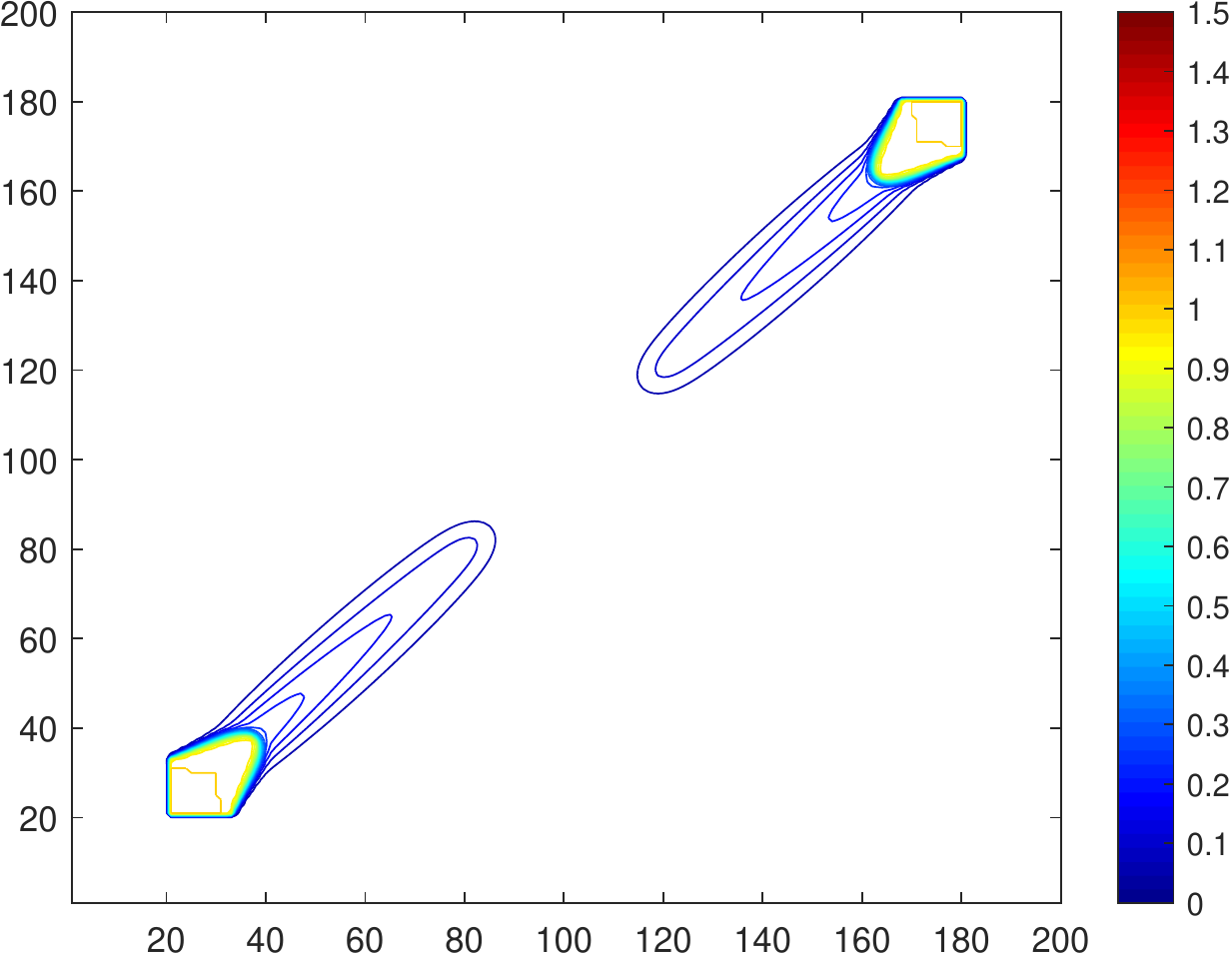}
\caption{Two group densities, initially distributed uniformly over disjoint squares, pass through one another under strong slowdown interaction strength.
The macroscopic model~\eqref{PDEs} (right column) quantitatively agrees with the microscopic stochastic model (left column), except for differences in the trailing fronts at $t=350$.
Slowdown strength: $\alpha=4$.
Times: 50, 150, 250, 350.}
\label{fig1alpha4}
\end{figure}

\begin{figure}[t]
\centering
\includegraphics[width=70mm]{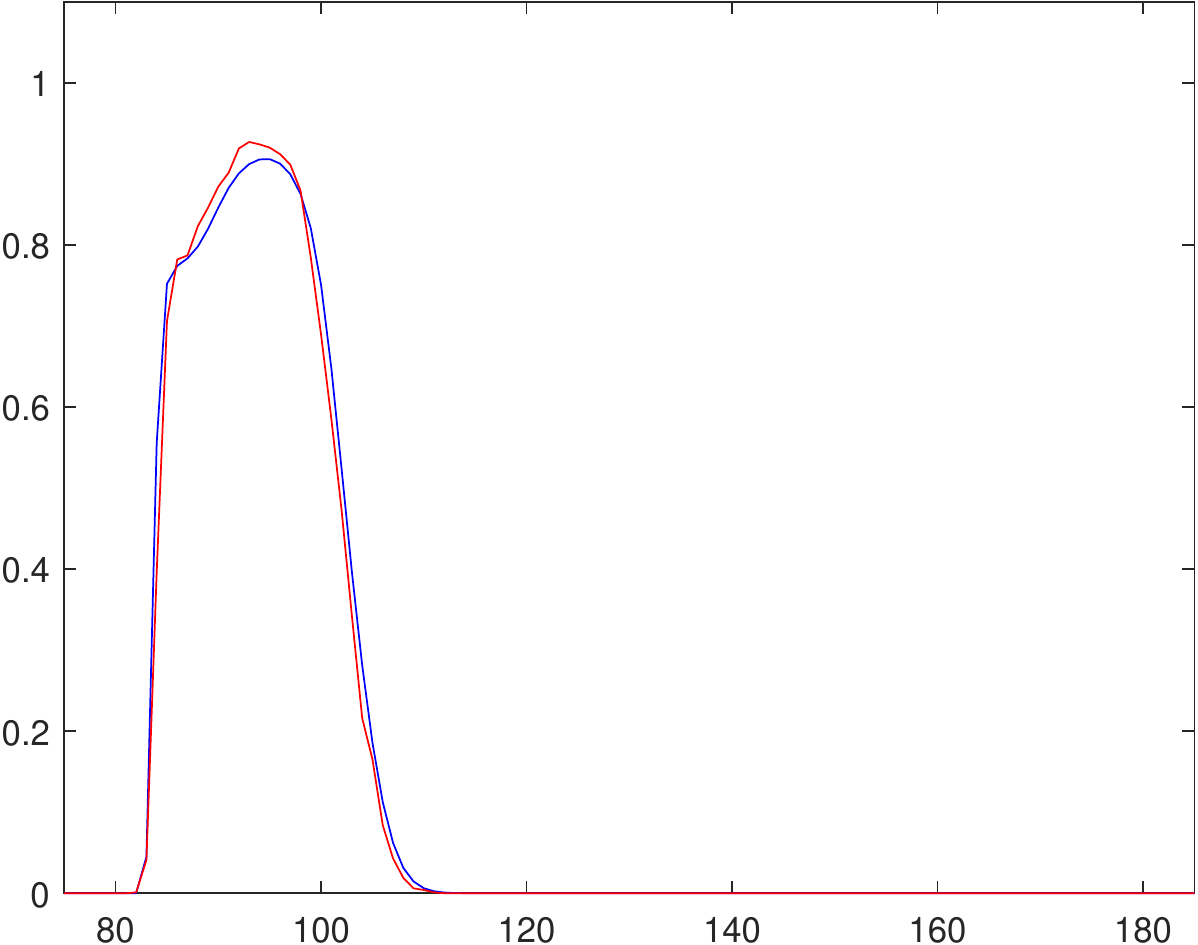}
\hspace{0.5em}
\includegraphics[width=70mm]{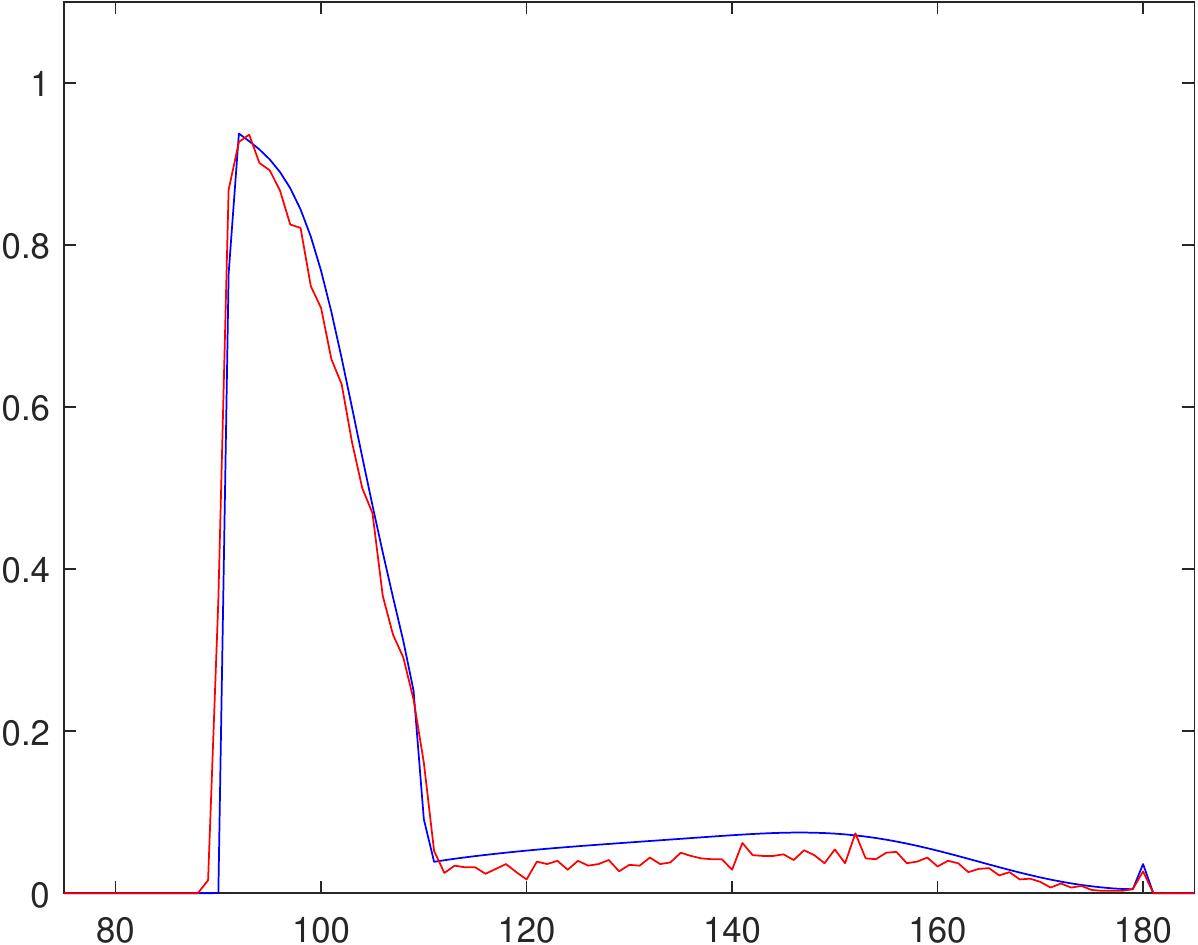}\\
\vspace{1ex}
\includegraphics[width=70mm]{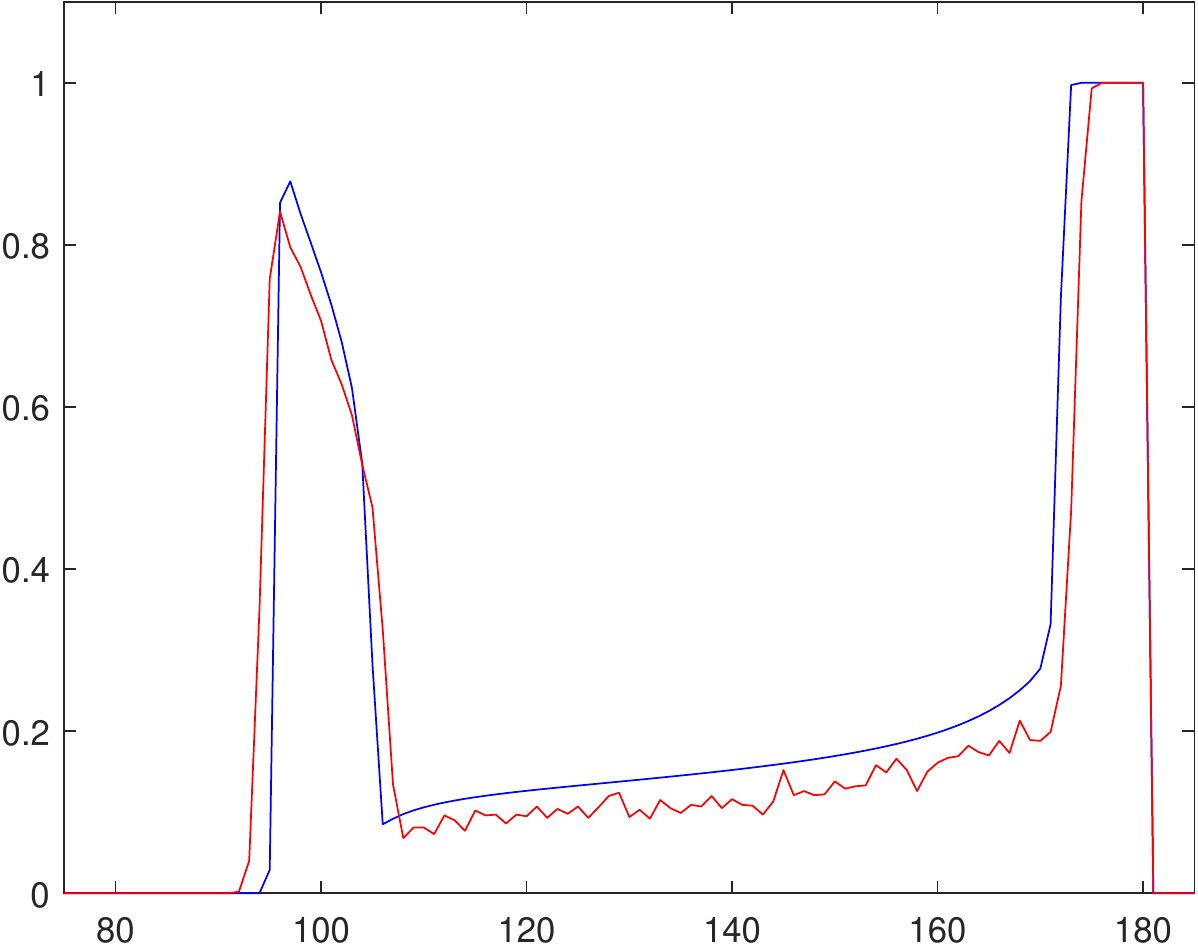}
\hspace{0.5em}
\includegraphics[width=70mm]{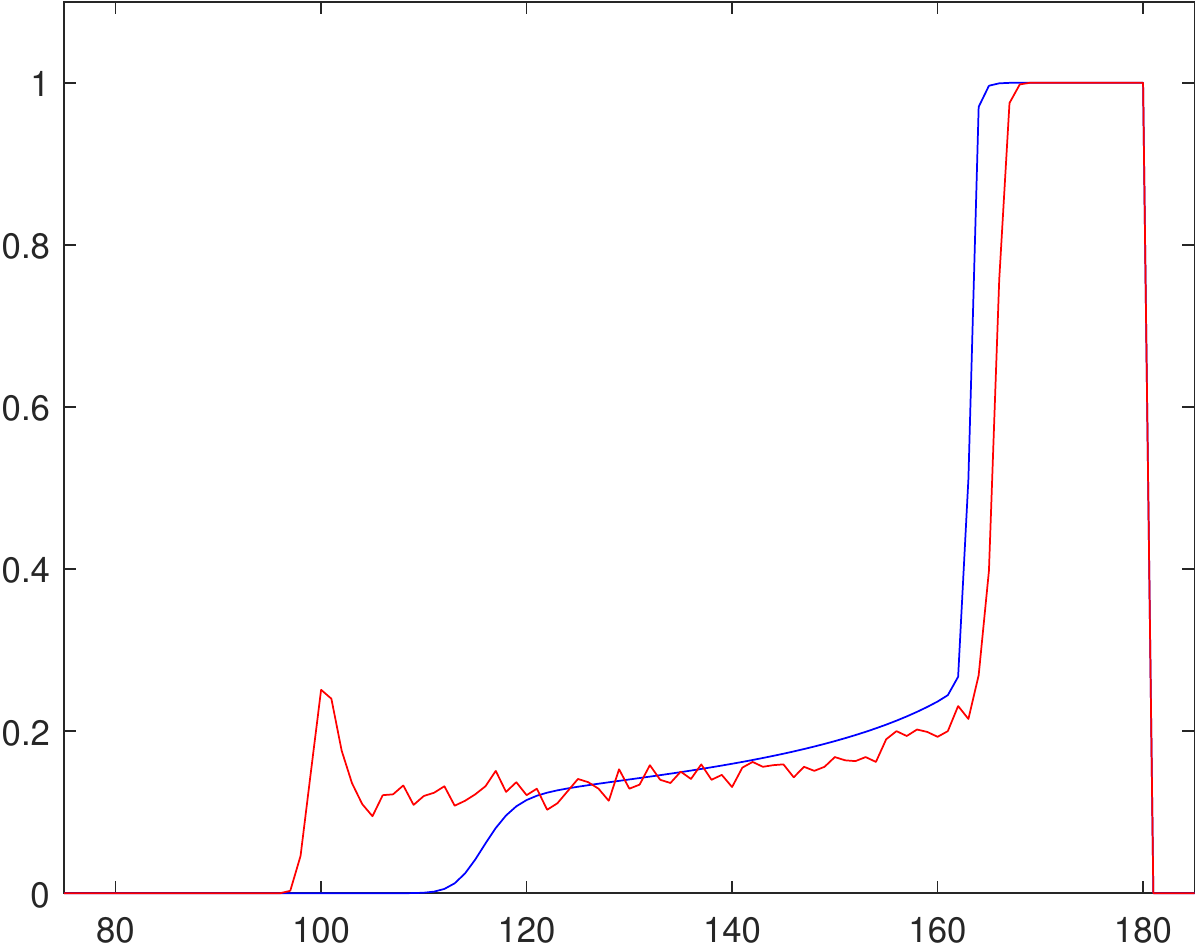} 
\caption{Evolution of group A density along the diagonal $j=k$ of the lattice.
Setup is as in Figure~\ref{fig1alpha4}.
Blue: Macroscopic model~\eqref{PDEs}.
Red: Microscopic stochastic model.
Slowdown strength: $\alpha=4$.
Times (left to right): 50, 150, 250, 350.}
\label{fig3alpha4}
\end{figure}

We simulate the microscopic model with timestep $\Delta t =.05$ and average over 1000 simulations to obtain a Monte Carlo approximation for the evolution of the group densities, $\rho^A$ and $\rho^B$. 
We then compare these results to those produced by numerically solving the mesoscopic model using a Runge-Kutta fourth-order method with variable timestep. 
A comparison of the evolutions of $\rho^A$ and $\rho^B$ for the microscopic and mesoscopic models is depicted in Figures~\ref{fig1alpha2},\ref{fig3alpha2} and Figures~\ref{fig1alpha4},\ref{fig3alpha4} for $\alpha=2$ and $\alpha=4$, respectively. 
Two-dimensional density plots are presented in Figures \ref{fig1alpha2} and \ref{fig1alpha4}. 
To examine the fine details of the group interactions, we illustrate the evolution of the density $\rho^A$ along the diagonal $j = k$ in
Figures~\ref{fig3alpha2} and~\ref{fig3alpha4}.

In Figure~\ref{fig1alpha2}, the two groups almost completely overlap in space at time $t=35$.
At $t=105$, the groups still overlap significantly, but approximately 15\% of the agents from each group have passed through the complementary group. 
The groups have nearly passed through one another at time $t=175$, and have completely done so by time $t=245$.
When assessing how well the deterministic model approximates the stochastic dynamics, times $t=175$ and $t=245$ provide the most demanding test.
Figures~\ref{fig1alpha2} and~\ref{fig3alpha2} demonstrate that the deterministic model is in excellent agreement with the results of the stochastic simulations. 
At time $t=175$, the deterministic model accurately captures the effect of the slowdown mechanism, a mechanism that results in complex profiles for the densities of the moving agents.  
Importantly, the deterministic model captures not only the overall shapes of the densities, but also the magnitudes of the peaks, as well as the leading and trailing fronts.

In the stronger slowdown interaction regime ($\alpha = 4$), the deterministic model once again accurately captures the behavior of the stochastic model, except for discrepancies in the trailing fronts (these fronts occur after the groups have passed through one another).
These results are depicted in Figures~\ref{fig1alpha4} and~\ref{fig3alpha4}.
Analyzing the simulations in detail, we note that due to the stronger slowdown interaction strength (i.e. $\alpha=4$ vs $\alpha=2$), it takes the two groups much longer to pass through one another.
This results in longer numerical simulations for both the deterministic and stochastic models.
Consequently, small discrepancies between the two have more time to grow when $\alpha=4$ than when $\alpha = 2$.
Nevertheless, we observe good agreement between the deterministic and stochastic models for times $t=50$, $150$, and $250$.
At time $t=150$, the positions of the peaks, as well as the leading and trailing fronts, are accurately captured by the deterministic model.

For $t=250$, the density of agents that have already passed through the other group is slightly higher in the deterministic case (see $110 \lessapprox k \lessapprox 170$ in the bottom left subplot of Figure~\ref{fig3alpha4}). 
This small overestimation of the density by the deterministic model
accumulates over time (from $t=250$ to $t=350$), and results in the two groups passing through one another earlier (at $t \approx 320$ for the deterministic model versus $t=360$ for the stochastic model).
Consequently, the trailing fronts have slightly separated by time $t=350$ (bottom right subplot of Figure~\ref{fig3alpha4}).

Summarizing our results when the initial group densities are uniform, we find excellent quantitative agreement between the macroscopic system of conservation laws~\eqref{PDEs} and the microscopic stochastic model.
Even when the slowdown interactions are strong ($\alpha = 4$), we see excellent agreement modulo small differences between the group pass-through times and trailing fronts.


\subsection{Simulations with Non-Uniform Initial Density}
\label{sec:numsumnonu}

In the second set of simulations, we analyze the performance of the macroscopic model when the initial agent densities vary in space. 
The stochastic model is more diffusive by nature than our derived deterministic models that approximate it. 
We believe that conservation law system~\eqref{PDEs} can produce and propagate sharp gradients in the solution, while such sharp gradients are typically `smoothed out' rather quickly in stochastic simulations. 
Further, as pointed out in~\cite{ckpt14,goatin}, conservation laws which are conditionally hyperbolic can potentially 
develop high-frequency, non-physical oscillations in the non-hyperbolic regime.
Therefore, this set of simulations is designed to test how well the macroscopic model performs in the non-hyperbolic regime with large initial gradients.

We consider a $100 \times 100$ lattice $\mathcal{L}$ and initial conditions for the deterministic model given by
\begin{equation}
	\rho^A_{j,k,0} = \begin{cases} \frac{1}{4}\left\lceil 2\left(1+\cos{\left(\frac{4\pi}{19}(j+k-62)\right)}\right)\right\rceil, & 31 \leq j \leq 50 \text{ and } 31 \leq k \leq 50,\\
			0, & \text{otherwise}, \end{cases}
\end{equation}
and
\begin{equation}
	\rho^B_{j,k,0} = \begin{cases} \frac{1}{4}\left\lceil 2\left(1+\cos{\left(\frac{4\pi}{19}(j+k-102)\right)}\right)\right\rceil, & 51 \leq j \leq 70 \text{ and } 51 \leq k \leq 70,\\
				0, & \text{otherwise}. \end{cases}
\end{equation}
To replicate this initial configuration in the stochastic model, we average over 20000 realizations, 
where we sample $1$ in each lattice cell with probabilities given by the expressions above.
Therefore, the expected value of the solution at time zero in stochastic simulations 
is given by the expressions above.

We specify floor velocity fields for the two groups using the potential functions
$\psi^A(j,k)$ and $\psi^B(j,k)$ on $\mathcal{L}$ defined by
\[
	\psi^A(j,k) = (80-j)^2 + (80-k)^{2}, \qquad
	\psi^B(j,k) = (21-j)^2 + (21-k)^{2},
\]
respectively. We obtain the floor velocity fields by computing the $\ell^{1}$-normalized gradients of the potentials:
\[
	\phi^A(j,k) = 
\left( \frac{80 - j}{|80 - j| + |80 - k|},  \frac{80 - k}{|80 - j| + |80 - k|}\right)
\]
and
\[
	\phi^B(j,k) =
\left(\frac{21 - j}{|21 - j| + |21 - k|},  \frac{21 - k}{|21 - j| + |21 - k|}\right).
\]
Under these floor fields, agents from group $A$ move toward the point $(80,80)$, while agents from group $B$ move toward $(21,21)$. 

Here we examine the mildly-strong slowdown interaction regime $\alpha = 2$ for the sake of brevity.
Since the two groups interact only for a short period of time in this regime, we expect the deterministic model to accurately capture the stochastic dynamics, and this does indeed happen.
That said, we will more closely examine how the nonuniformity in the initial data propagates forward.
We therefore concentrate on the early stages of the group-group interaction, as the system quickly loses memory of the initial nonuniformity.

\begin{figure}[t]
\centering
\includegraphics[width=2.5in]{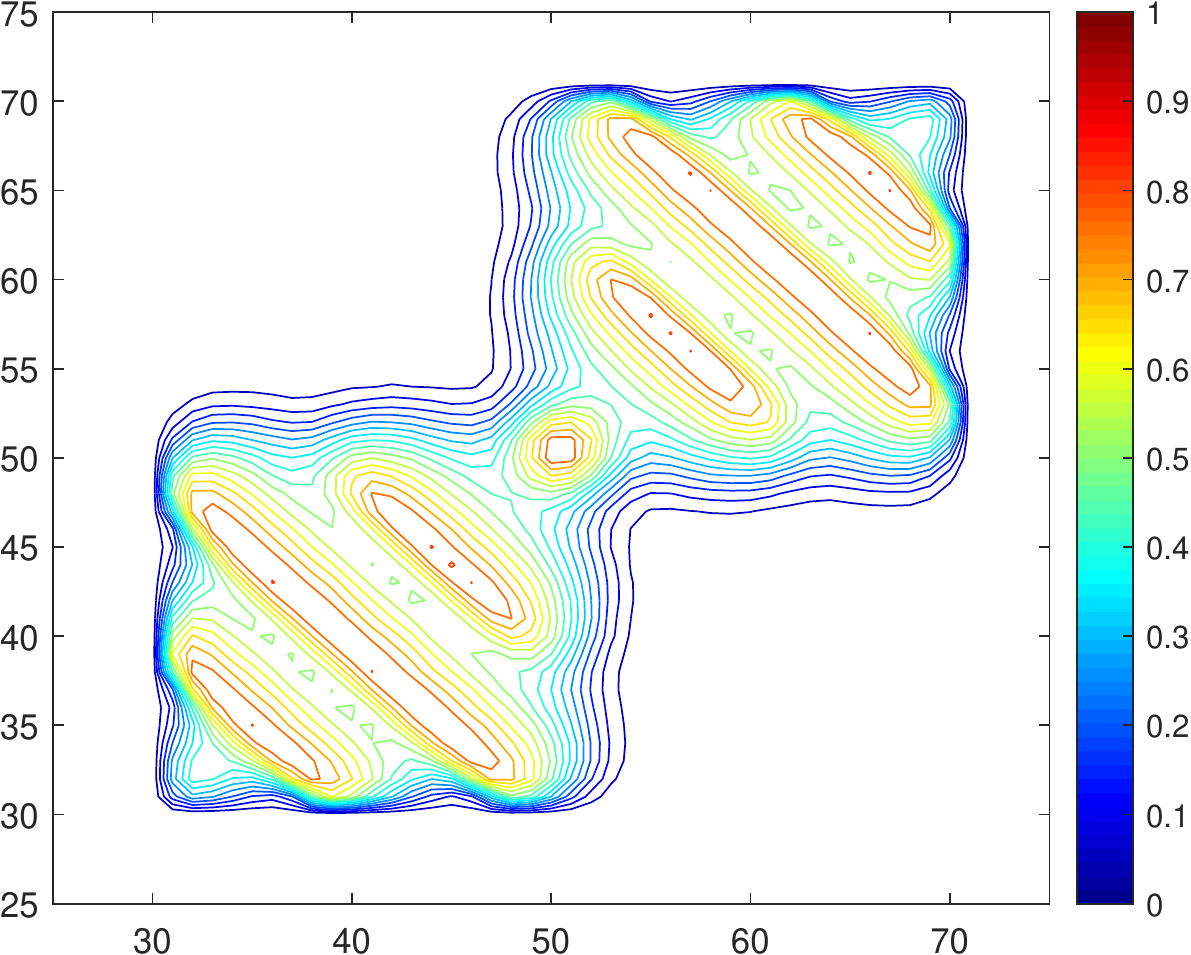}
\hspace{1em}
\includegraphics[width=2.5in]{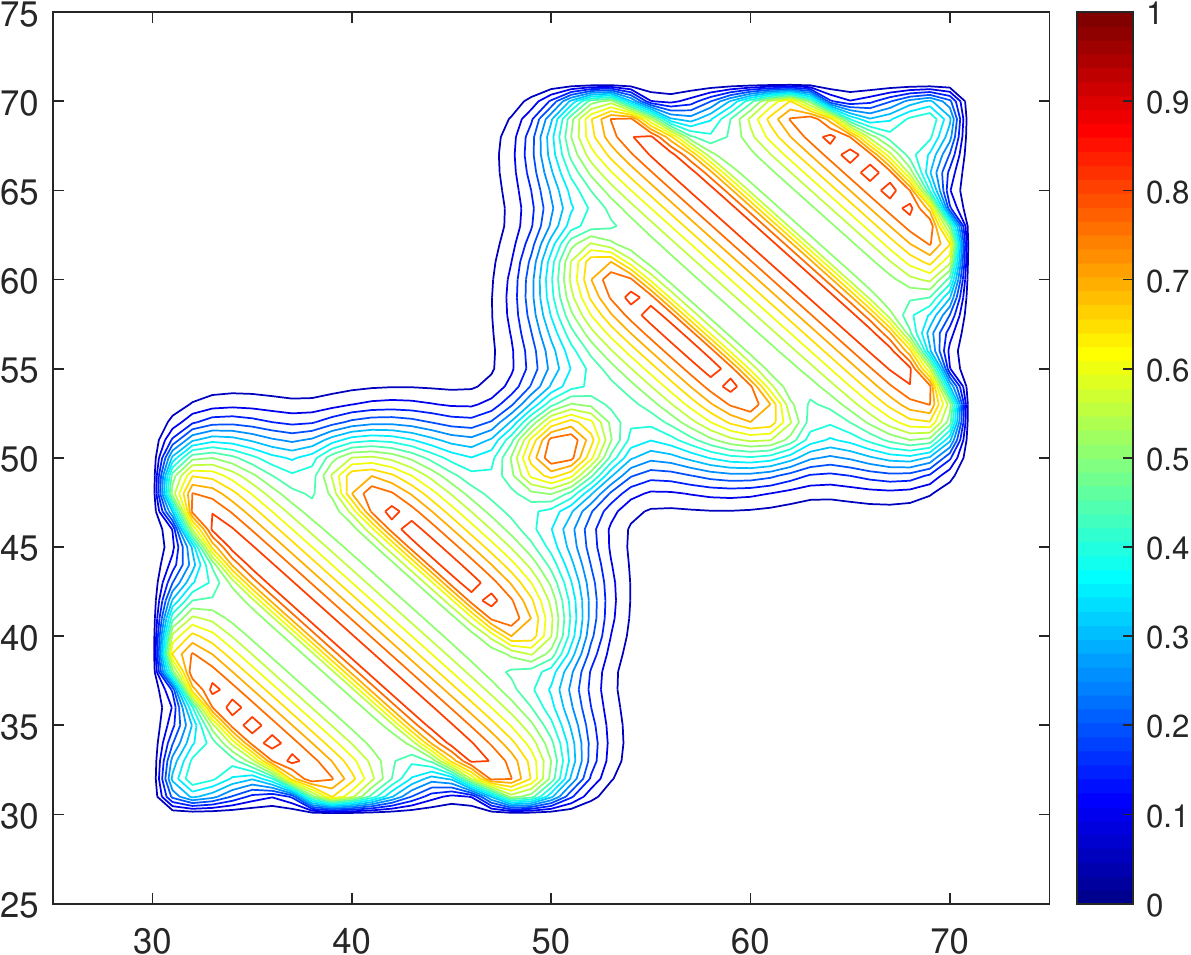}\\
\includegraphics[width=2.5in]{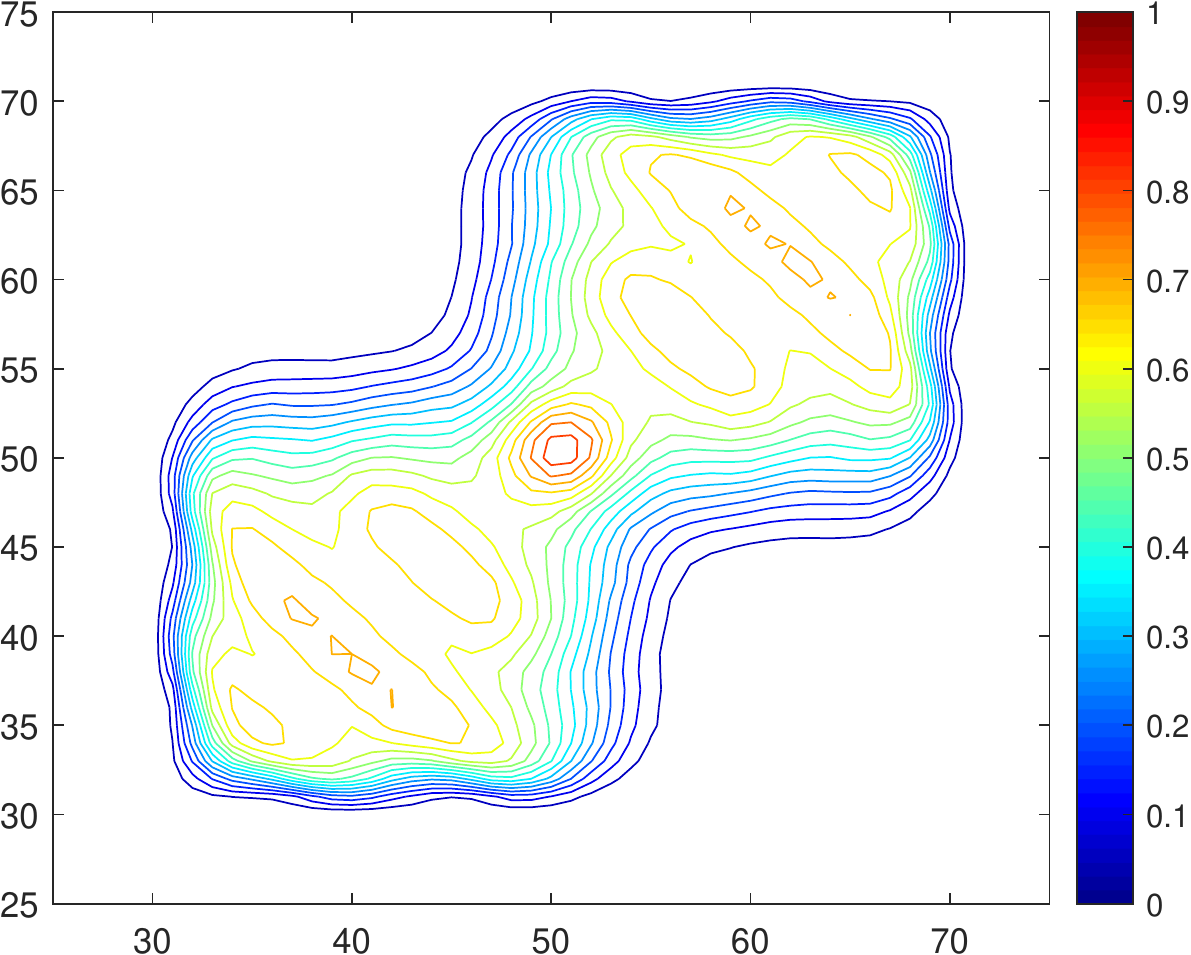}
\hspace{1em}
\includegraphics[width=2.5in]{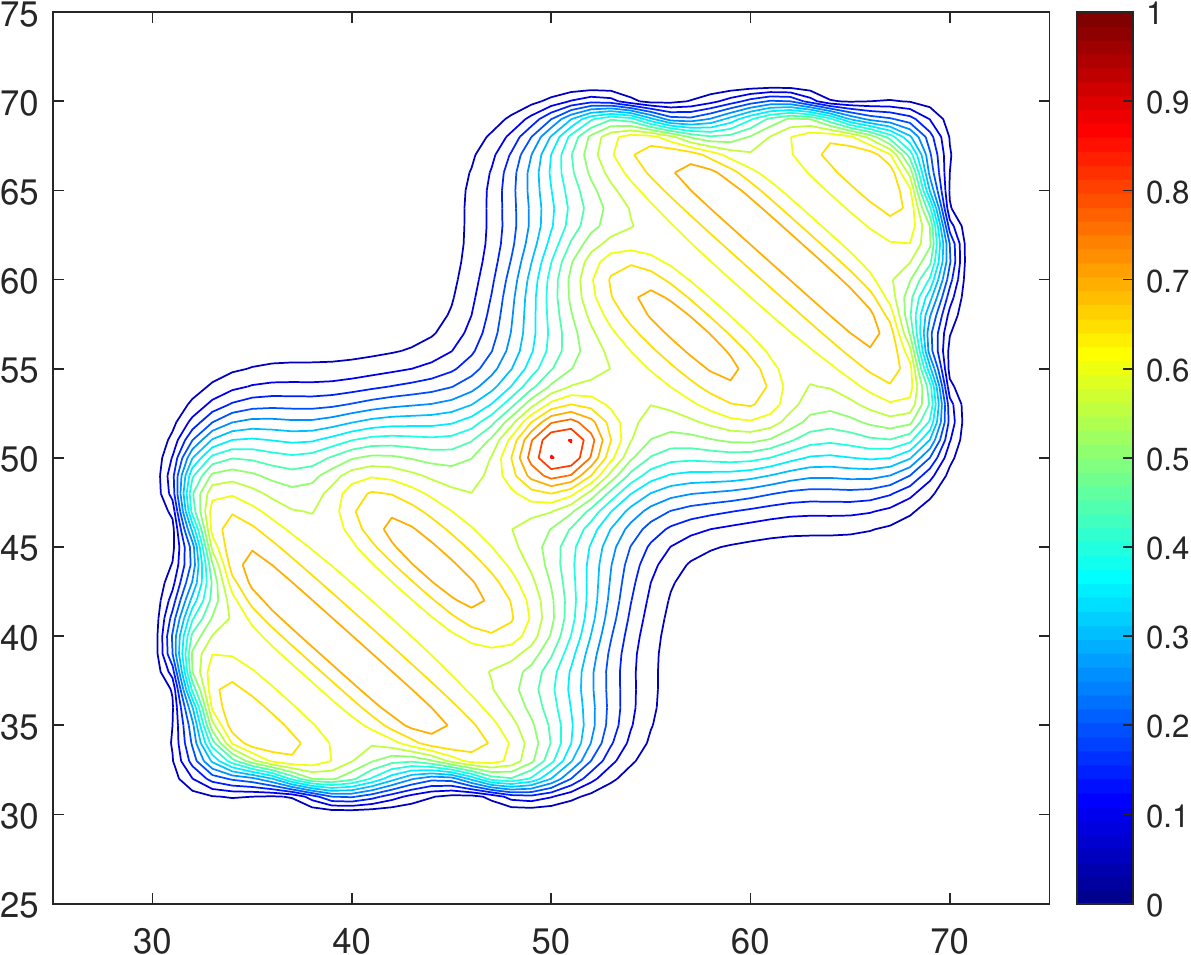}\\
\includegraphics[width=2.5in]{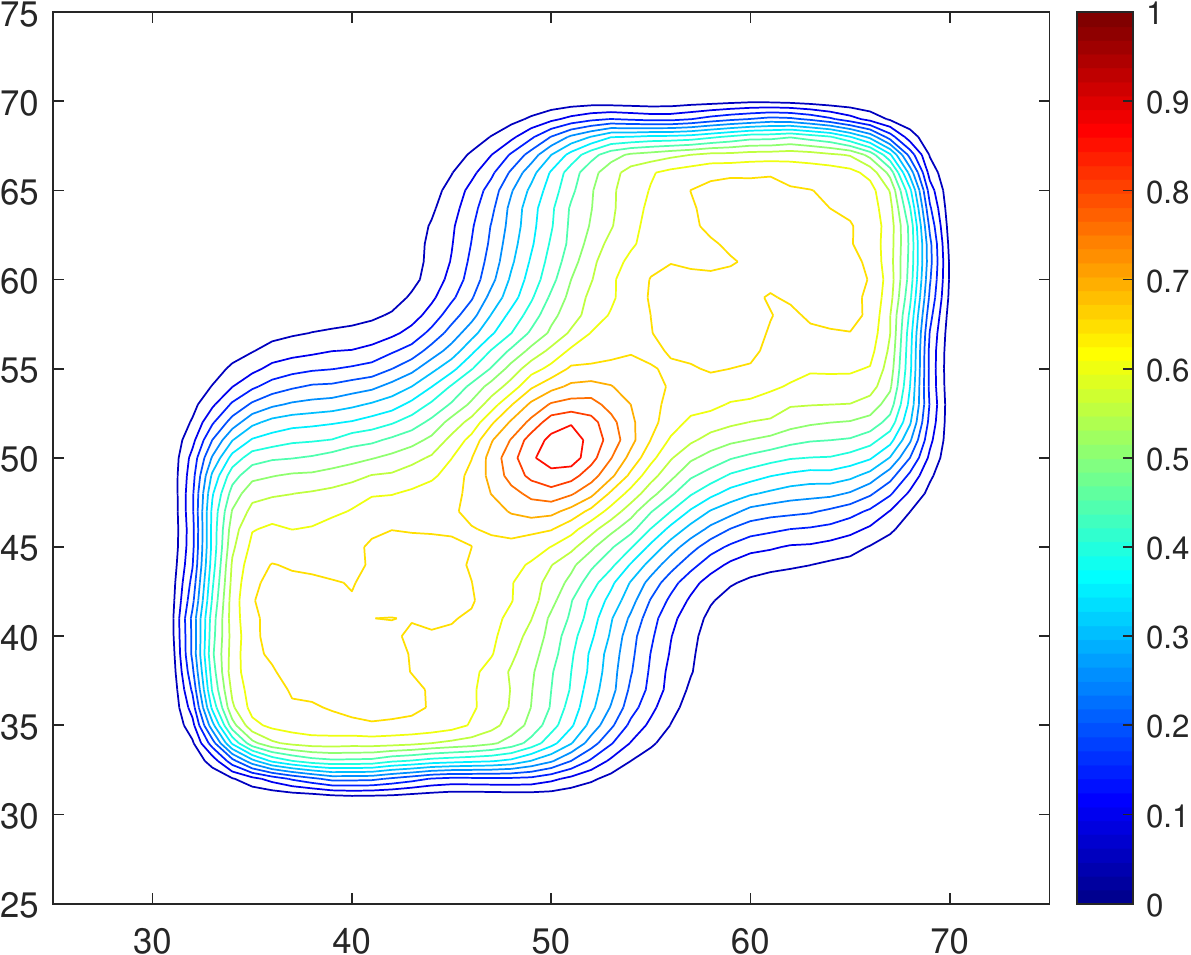}
\hspace{1em}
\includegraphics[width=2.5in]{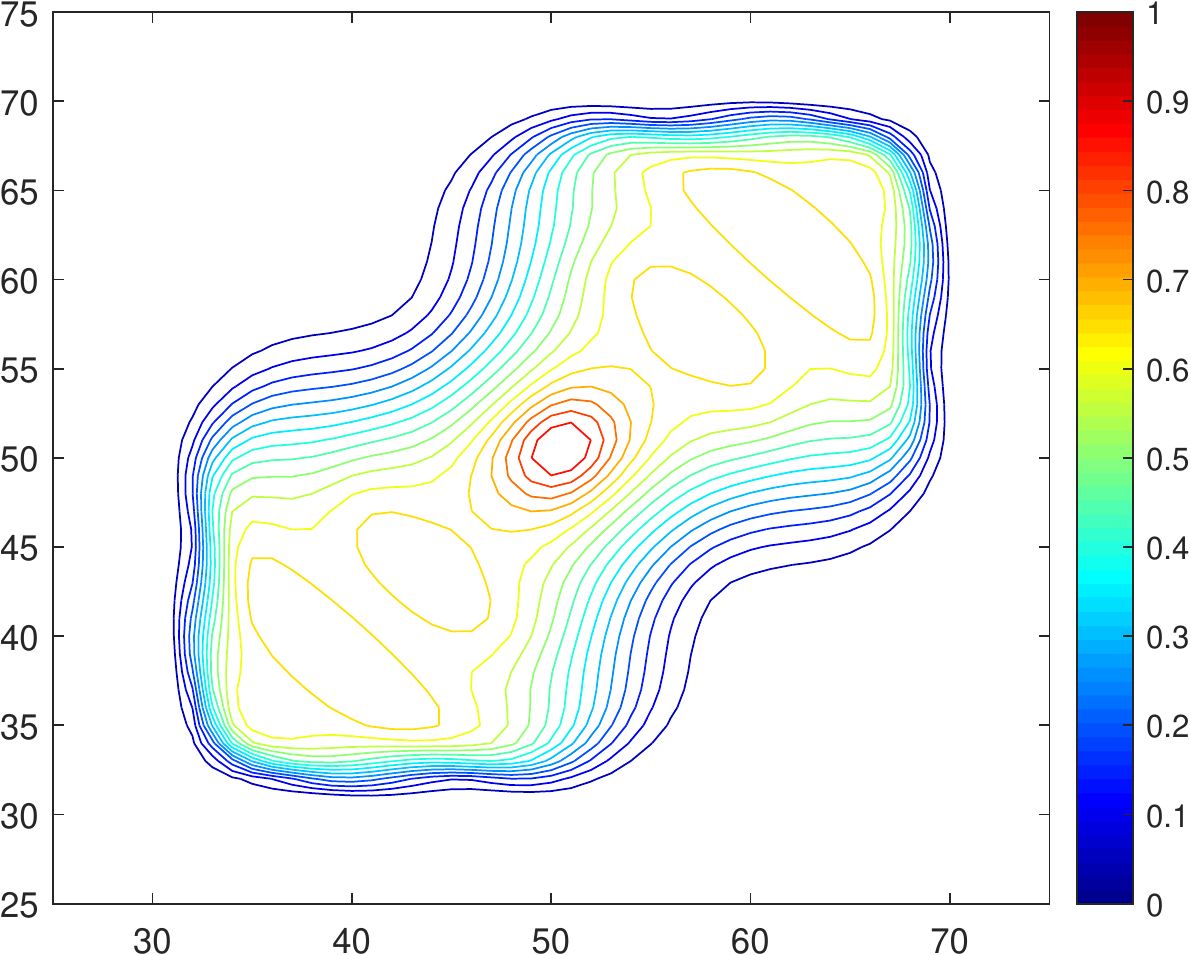}\\
\includegraphics[width=2.5in]{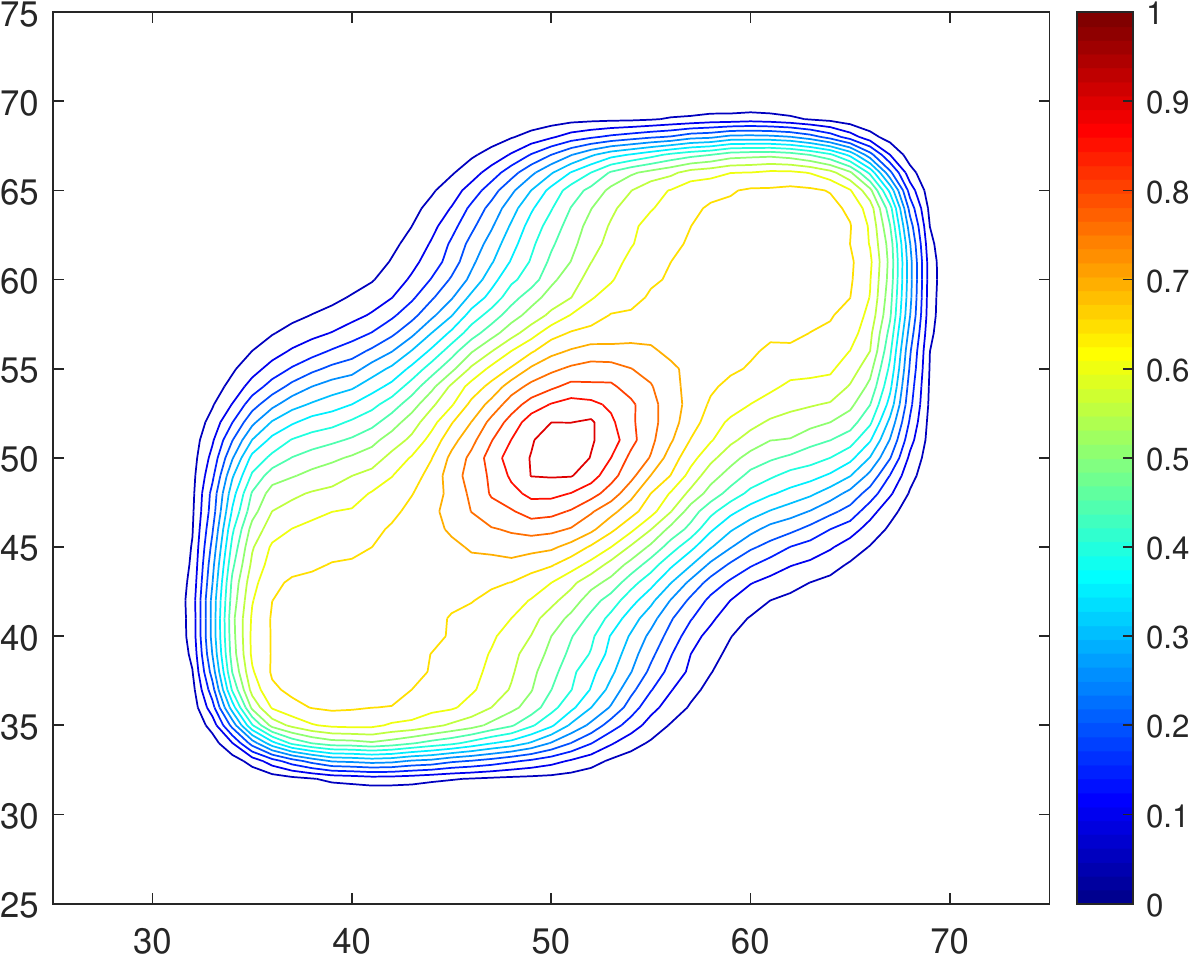}
\hspace{1em}
\includegraphics[width=2.5in]{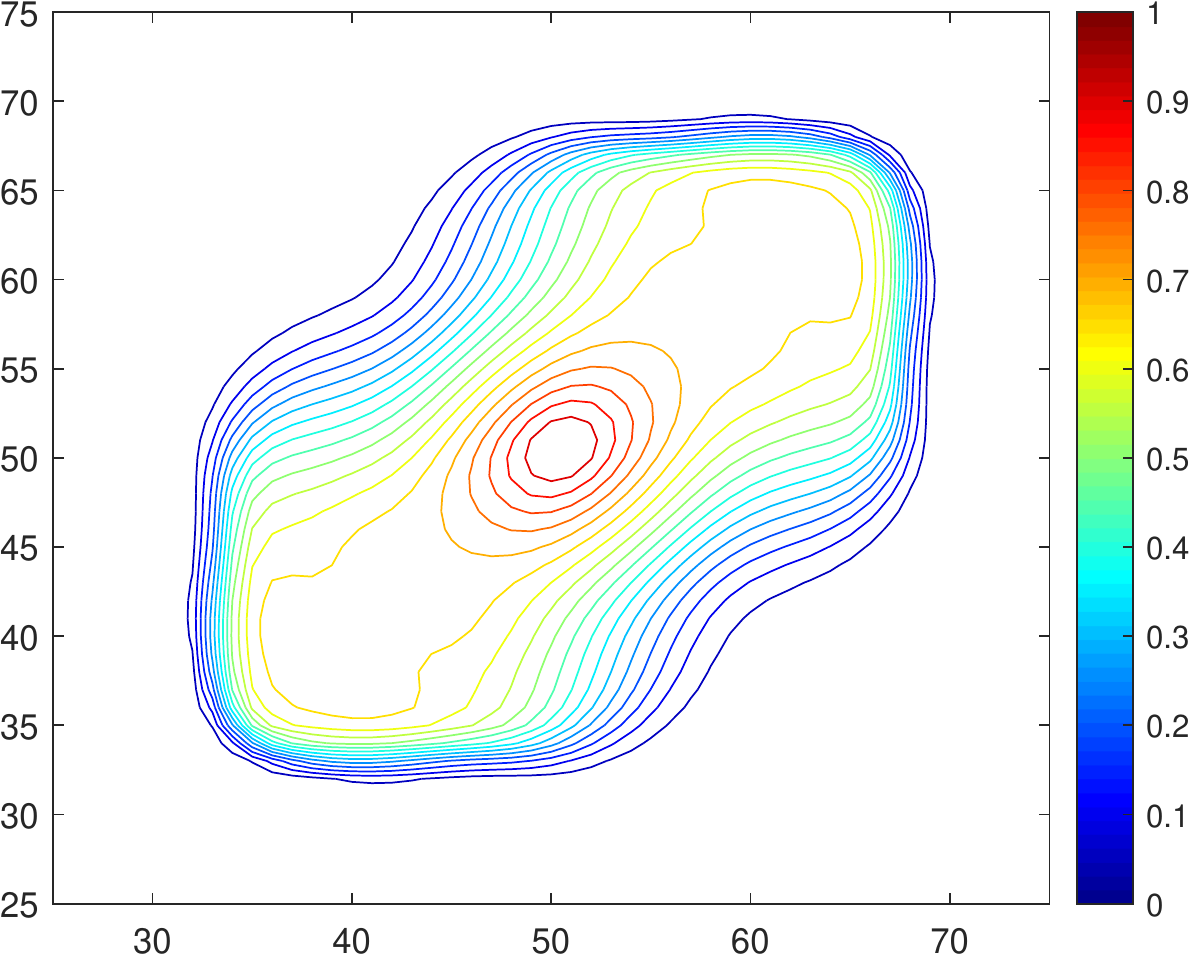}
\caption{Two group densities, initially distributed {\itshape nonuniformly} over disjoint squares, pass through one another under mild slowdown interaction strength.
We observe excellent agreement between the microscopic stochastic model (left column) and the approximating conservation laws~\eqref{PDEs} (right column).
Initial nonuniformity in the densities is quickly lost.
Slowdown strength: $\alpha = 2$.
Times: $4,8,12,16$ (top to bottom).}
\label{fig:levelf21}
\end{figure}

\begin{figure}[t]
\centering
\includegraphics[width=2.5in]{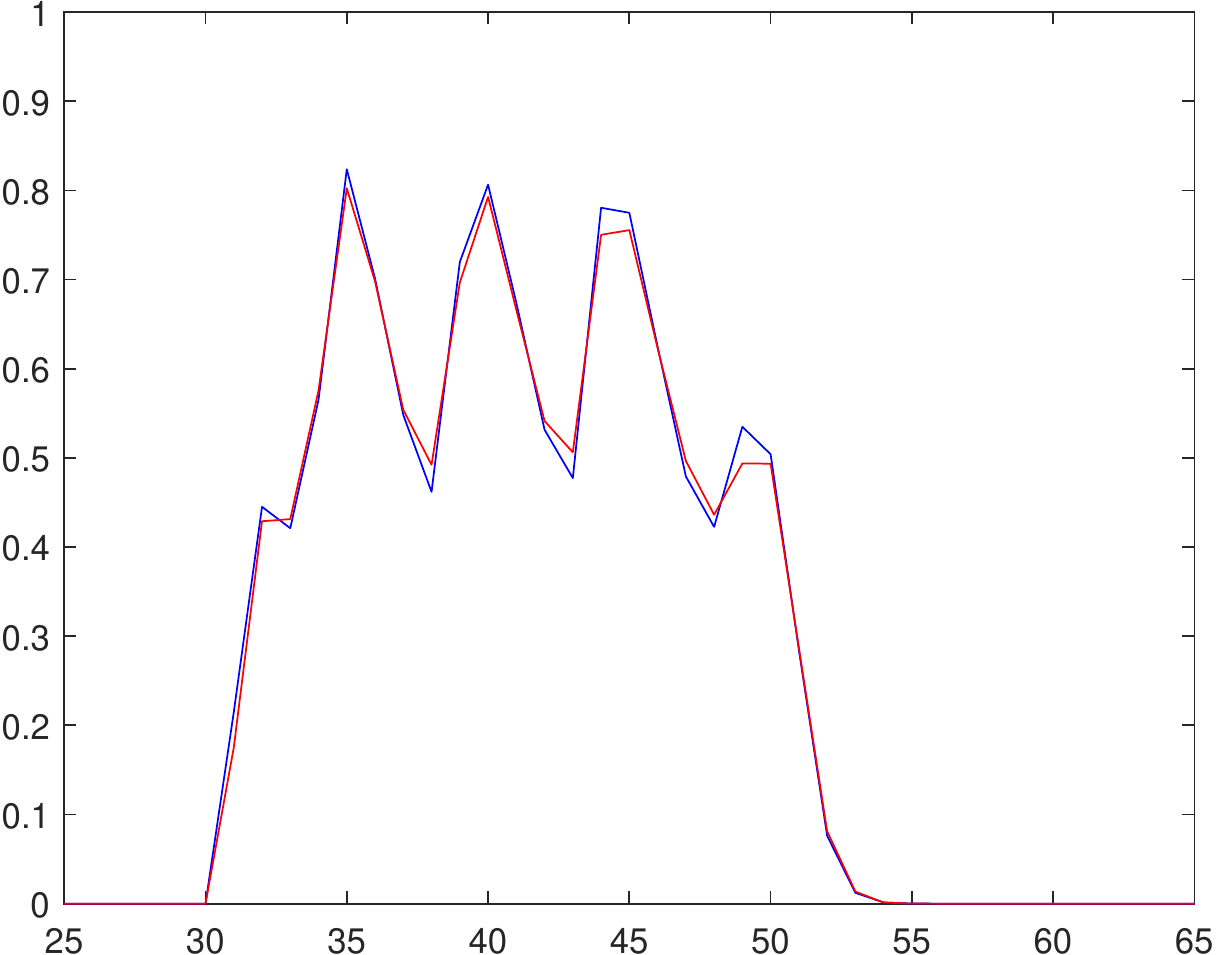}
\hspace{0.5em}
\includegraphics[width=2.5in]{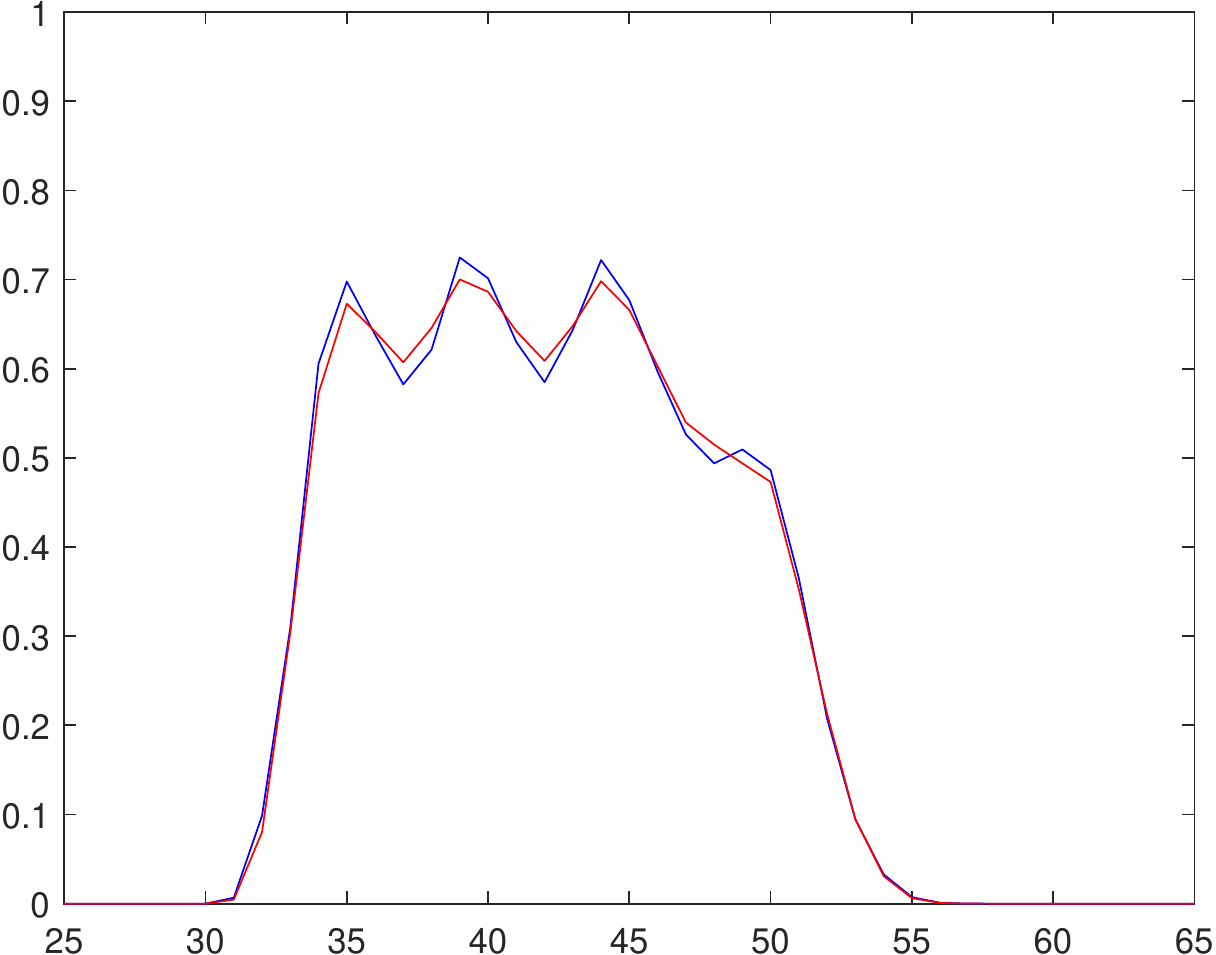}\\
\vspace{1ex}
\includegraphics[width=2.5in]{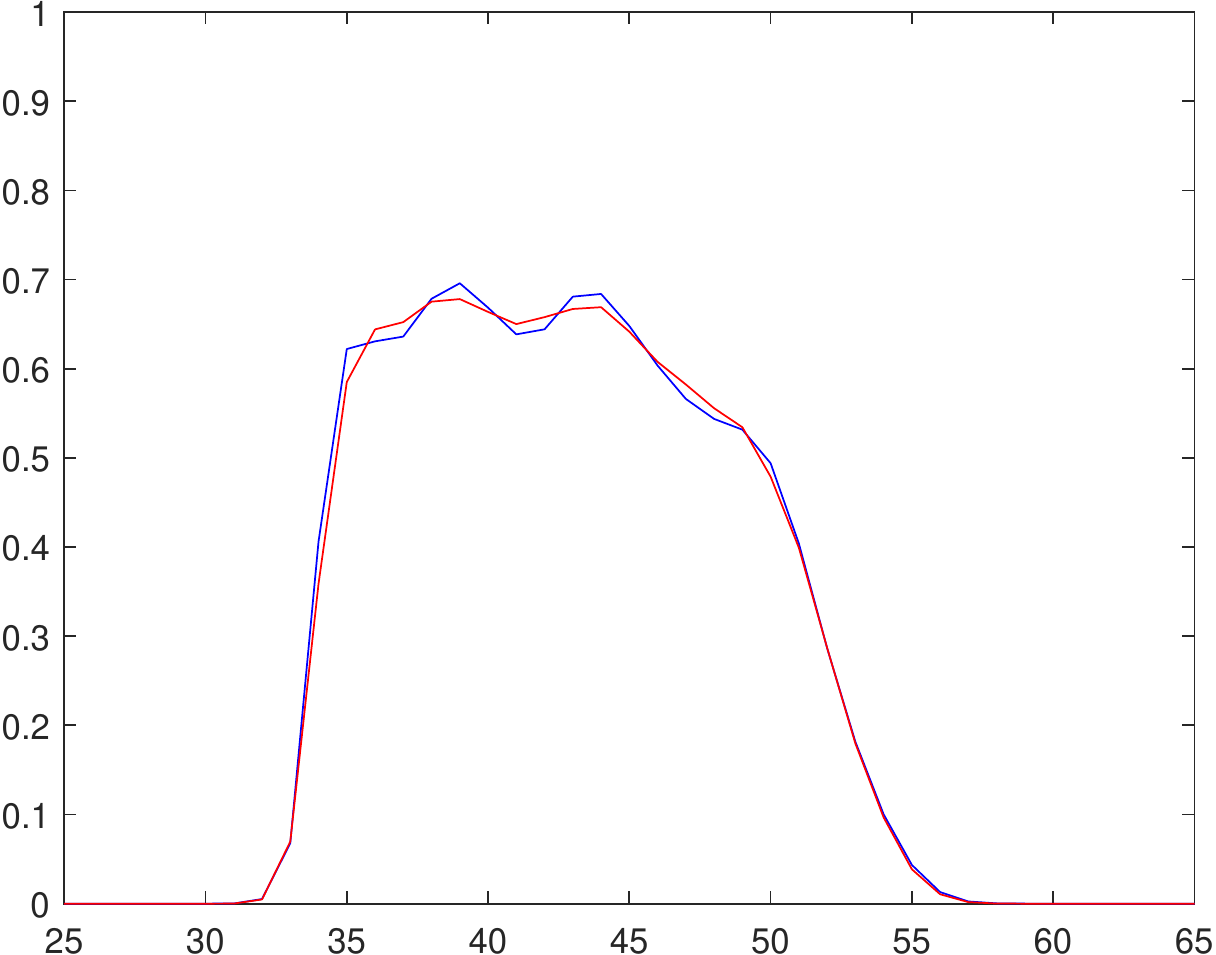}
\hspace{0.5em}
\includegraphics[width=2.5in]{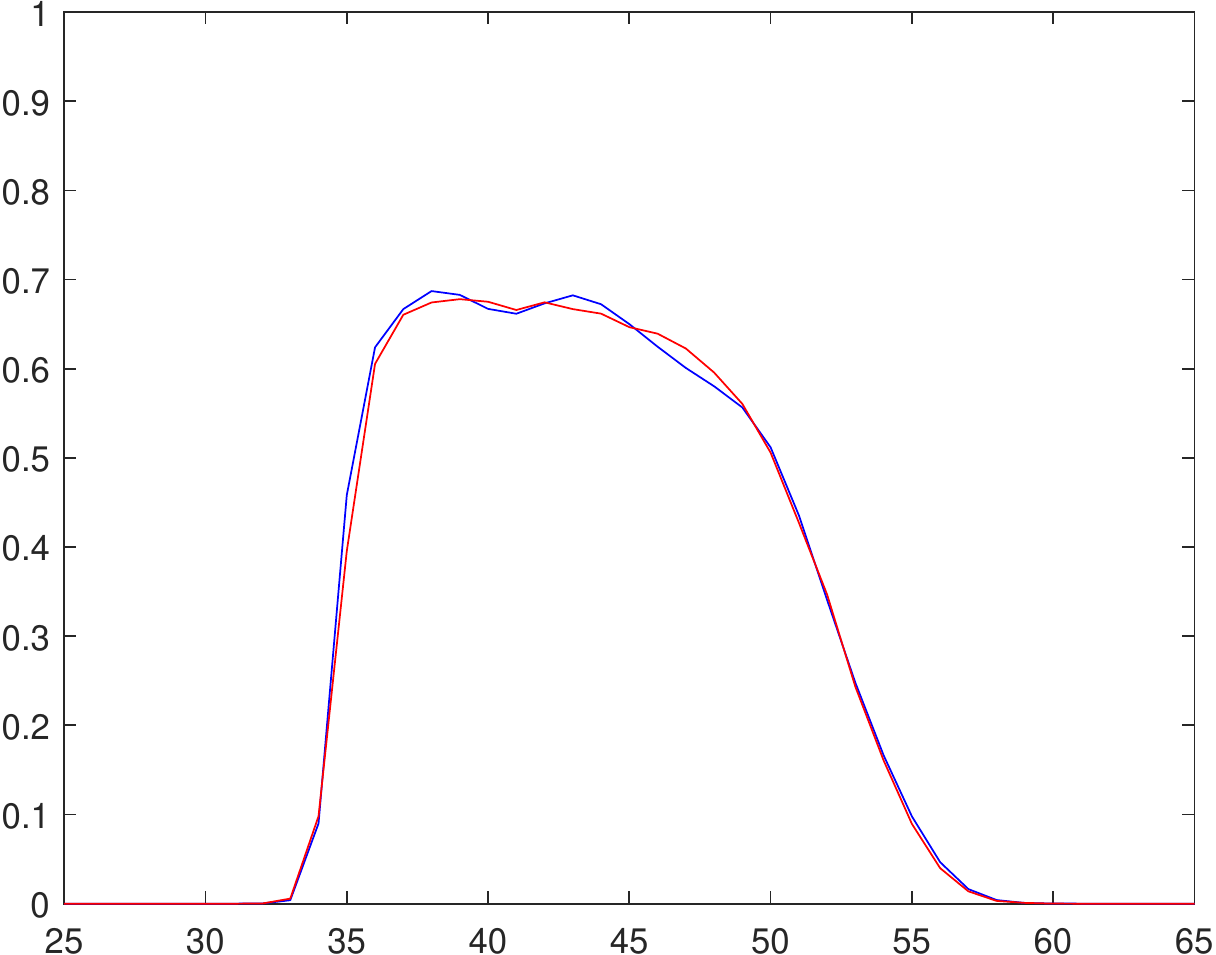}
\caption{Time evolution of the diagonal of the group A density, $\rho^A_{j,j,t}$, for the simulation illustrated in Figure~\ref{fig:levelf21}.
Density fluctuations are sharper and persist longer for the approximating conservation laws~\eqref{PDEs} (blue curves) than for the microscopic stochastic dynamics (red curves).
Times: $4,8,12,16$ (left to right).}
\label{fig:levelf22}
\end{figure}

\begin{figure}[t]
\centering
\includegraphics[width=2.5in]{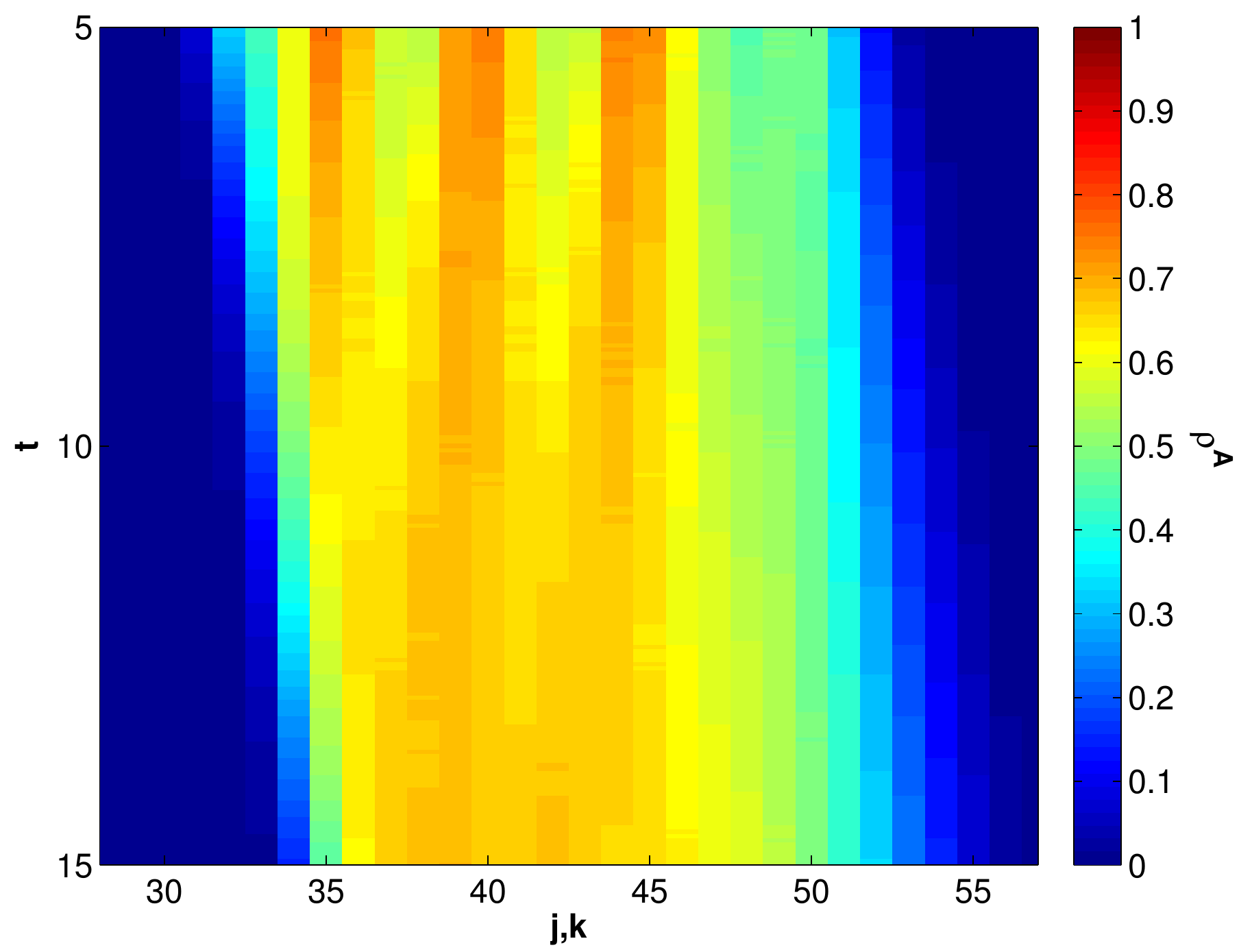}
\hspace{1em}
\includegraphics[width=2.5in]{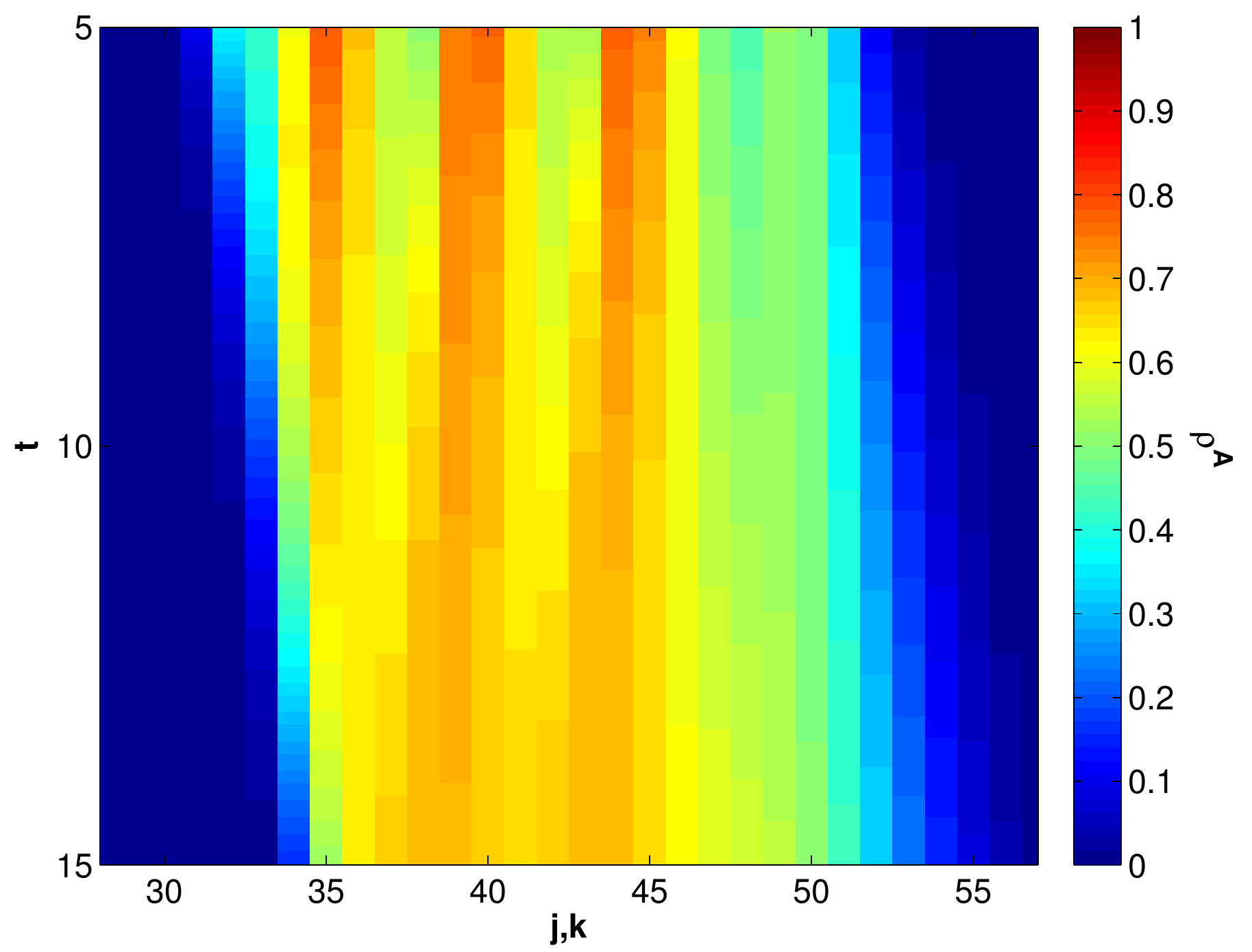}
\caption{Time evolution (top to bottom) of the diagonal of the group A density, $\rho^A_{j,j,t}$, for the simulation illustrated in Figure~\ref{fig:levelf21}.
Density fluctuations are sharper and persist longer for the approximating conservation laws~\eqref{PDEs} (right column) than for the microscopic stochastic dynamics (left column).}
\label{fig:levelf23}
\end{figure}

We compare the behavior of the deterministic and stochastic systems in Figures~\ref{fig:levelf21},~\ref{fig:levelf22}, and~\ref{fig:levelf23}.
As with the first set of simulations, the two systems quantitatively agree over short time intervals (Figure~\ref{fig:levelf21}).
Figure~\ref{fig:levelf21} also illustrates that memory of the nonuniformity in the initial densities is quickly lost, with regions of high density and low density quickly equilibrating.
However, before the nonuniformity is lost due to equilibration, density fluctuations are sharper and persist longer in the deterministic case (Figures~\ref{fig:levelf22} and~\ref{fig:levelf23}).
This indicates that the stochastic model is more diffusive than the macroscopic conservation law approximation, as was also the case in~\cite{ckpt14}.
We will address this problem in Section~\ref{sec:diff} by deriving a diffusive correction to~\eqref{PDEs}.

\section{Second-Order Corrections to the Macroscopic PDE Model}
\label{sec:diff}

As a matter of general philosophy, one expects that any stochastic model will be at least slightly more diffusive than its deterministic counterpart.
In particular, sharp fronts with large gradients are not natural for stochastic models. 
On the other hand, deterministic coarse-grained models tend to be hyperbolic (or conditionally hyperbolic in our case), and therefore should produce fronts resembling shocks. 
Our numerical simulations confirm this picture for our model.

As discussed in Section~\ref{sec:numsumnonu}, simulations performed with non-uniform initial pedestrian densities confirm that our stochastic model is more diffusive than the macroscopic system of conservation laws.
This is most evident in Figures~\ref{fig:levelf22} and~\ref{fig:levelf23}. 
In Figure~\ref{fig:levelf22}, we compare the diagonal cross-section of the group A density, $\rho^A_{j,j,t}$. 
We see that the stochastic model `smoothes out' density fluctuations faster than the macroscopic PDEs (most prominent at times $t=8, 12$).
Figure~\ref{fig:levelf23} depicts the time evolution of the diagonal cross-section in a continuous fashion. 
Once again, we see that density gradients are sharper and persist longer for the approximating conservation laws~\eqref{PDEs} than for the microscopic stochastic dynamics.
Here, we derive a second-order correction to the system of conservation laws~\eqref{PDEs} in order to mitigate this discrepancy in diffusiveness.

To derive the second-order model, we substitute the Taylor expansions
\begin{equation}
\label{eq:t}
\begin{aligned}
	\rho^A_{j\pm h,k} &= \rho^A_{j,k} \pm h\frac{\partial}{\partial x} \rho^A_{j,k} + \frac{h^2}{2}\frac{\partial^2}{\partial x^2} \rho^A_{j,k} + \mathcal{O}(h^3), \\
	\rho^A_{j,k\pm h} &= \rho^A_{j,k} \pm h\frac{\partial}{\partial y} \rho^A_{j,k} + \frac{h^2}{2}\frac{\partial^2}{\partial y^2} \rho^A_{j,k} + \mathcal{O}(h^3), \\
	\rho^B_{j\pm h,k} &= \rho^B_{j,k} \pm h\frac{\partial}{\partial x} \rho^B_{j,k} + \frac{h^2}{2}\frac{\partial^2}{\partial x^2} \rho^B_{j,k}+ \mathcal{O}(h^3), \\
	\rho^B_{j,k\pm h} &= \rho^B_{j,k} \pm h\frac{\partial}{\partial y} \rho^B_{j,k} + \frac{h^2}{2}\frac{\partial^2}{\partial y^2} \rho^B_{j,k} + \mathcal{O}(h^3),
\end{aligned}
\end{equation}
into the flux equation \eqref{e:flux-form}.
Keeping $h$ fixed and neglecting terms of order $h^3$ or higher, we arrive at the second-order PDE system 
\begin{equation}
\label{PDEs2}
\begin{aligned}
	\rho^A_t &+ [\phi^A_1f(\rho^A)g(\rho^B)]_x + [\phi^A_2f(\rho^A)g(\rho^B)]_y = \frac{\eps}{2} \left( H^{11}_x + H^{12}_y \right), \\
	\rho^B_t &+ [\phi^B_1f(\rho^B)g(\rho^A)]_x + [\phi^B_2f(\rho^B)g(\rho^A)]_y = \frac{\eps}{2} \left( H^{21}_x + H^{22}_y \right). 
\end{aligned}
\end{equation}
Here, $\eps$ denotes a small parameter which corresponds to the $\mathcal{O} (h^2)$ terms retained in the expansions \eqref{eq:t}.
The right side of~\eqref{PDEs2} is given by $H^{11}_x  =S^{11} + D^{11}$, $H^{21}_x  =S^{21} + D^{21}$, $H^{12}_y  =S^{12} + D^{12}$, and $H^{22}_y  =S^{22} + D^{22}$, with
\begin{equation}
\label{eq:S}
\begin{aligned}
S^{11} &= \left( \phi^A_{1,x} f(\rho^A)g(\rho^B) \right)_x + \phi^A_{1,x} \left( \rho^A_x g(\rho^B) + (c_1-c_2)f(\rho^A)\rho^B_x \right), \\
S^{12} &= \left( \phi^A_{2,y} f(\rho^A)g(\rho^B) \right)_y + \phi^A_{2,y} \left( \rho^A_y g(\rho^B) + (c_1-c_2)f(\rho^A)\rho^B_y \right), \\
S^{21} &= \left( \phi^B_{1,x} f(\rho^B)g(\rho^A) \right)_x + \phi^B_{2,x} \left( \rho^B_x g(\rho^A) + (c_1-c_2)f(\rho^B)\rho^A_x \right), \\
S^{22} &= \left( \phi^B_{2,y} f(\rho^B)g(\rho^A) \right)_y + \phi^A_{2,y} \left( \rho^B_y g(\rho^A) + (c_1-c_2)f(\rho^B)\rho^A_y \right),
\end{aligned}
\end{equation}
and
\begin{equation}
\label{eq:D}
\begin{aligned}
D^{11} &= \phi^A_{1} \left( \rho^A_x g(\rho^B) + (c_1-c_2)f(\rho^A)\rho^B_x \right)_x, \quad
D^{12}   = \phi^A_{2} \left( \rho^A_y g(\rho^B) + (c_1-c_2)f(\rho^A)\rho^B_y \right)_y, \\
D^{21} &= \phi^B_{2} \left( \rho^B_x g(\rho^A) + (c_1-c_2)f(\rho^B)\rho^A_x \right)_x, \quad
D^{22}   = \phi^A_{2} \left( \rho^B_y g(\rho^A) + (c_1-c_2)f(\rho^B)\rho^A_y \right)_y.
\end{aligned}
\end{equation}
Here the $D^{ij}$ terms correspond to the diffusion operator and the $S^{ij}$ terms represent second-order nonlinear corrections.
The diffusion terms $D^{ij}$ are analogous to the diffusive corrections derived in the one-dimensional case~\cite{ckpt14}.

We now examine the behavior of the second-order system~\eqref{PDEs2} in the context of the settings we have simulated in Section~\ref{sec:num}.
Recall that for each group, the floor velocity field is given by the $\ell^{1}$-normalized gradient of a potential that grows quadratically outward from the target point.
Consequently, the floor fields exhibit very little spatial variation far from the corresponding target points.
It follows that when far from both target points, the $S^{ij}$ terms may be neglected, reducing~\eqref{PDEs2} to a second-order diffusive correction to conservation law system~\eqref{PDEs}.
But the interesting dynamics in our simulations occur precisely when this reduction is valid:
The groups pass through one another far from both target points.

The nonlinear corrections $S^{ij}$ become important near the target point for each group.
This can be intuitively understood because in the $t\to\infty$ limit, agents from each group accumulate tightly around the corresponding target point.
Near the target points, the nonlinear corrections $S^{ij}$ must therefore counter-balance the diffusive terms $D^{ij}$ (which cause agents to spread out away form the target point).

\section{Discussion}
\label{sec:conc}
%

In this paper, we have introduced a novel stochastic microscopic model for the evolution of interacting groups of particles in two-dimensional geometries.
This model features two interaction mechanisms - the global floor-field mechanism and a local slow-down interaction mechanism.
We have derived a system of (conditionally hyperbolic) conservation laws that describe the effective dynamics on a macroscopic level, as well as a second-order correction to this system. Simulations show excellent agreement between the microscopic and macroscopic descriptions of the dynamics.
Importantly, the combination of the two interaction mechanisms survives when we pass to the limiting PDEs.
Numerical simulations of the effective PDEs are more than an order of magnitude faster than simulations of the stochastic counterpart. 
Consequently, the effective PDEs can be used to quickly assess averaged behavior of agent groups in complex geometries.

The stochastic model can be extended in several nontrivial ways. 
In particular, to make this model even more relevant to pedestrian dynamics, one should include avoidance mechanisms. 
It has been recognized that humans tends to avoid collisions, and more generally interactions, with high-density groups (see \textit{e.g.}~\cite{degond13,avoidmodel1,avoidmodel2} and references therein).
To model avoidance of high-density groups and obstacles, it is possible to introduce a mathematical `change of direction' mechanism based on look-ahead potentials~\cite{soka06,hst14,sunti14}.
For instance, the floor-field velocity can depend on the look-ahead potential.
However, if this dependence is strong, then correlations in the system may become considerable~\cite{hst14}, thereby making the derivation of macroscopic PDEs more challenging. 
Nevertheless, weak dependence of the floor-field velocity on the look-ahead potential can be incorporated into the model with relative ease.
In addition to avoidance mechanisms, time-dependent floor velocity fields can be included in both stochastic and effective PDE models.

Memory effects can be quite important in biological contexts, for instance when modeling bacterial motion (see \textit{e.g.}~\cite{lewe14}). 
A mathematical mechanism similar to that introduced in~\cite{lewe14} can be included in our stochastic model and should survive when passing to the mean-field PDEs.
This mechanism relies on introducing additional groups of agents and including probabilistic rules for group switching. 
We expect that for the derivation of the effective PDEs, the group-switching mechanism can depend strongly on nearest neighbors, but will depend only weakly on the look-ahead potential.

We will study these additional interaction mechanisms in subsequent papers.
Motivation will arise from concrete applications, such as the study of evacuation scenarios or complex crossings (\textit{e.g.} the Shibuya crossing) in pedestrian dynamics.
When considering interaction mechanisms of ever greater complexity, the crucial challenge will be balancing the complexity of microscopic stochastic models against the degree to which macroscopic PDE descriptions faithfully capture the underlying microscopic dynamics.

Macroscopic PDEs for the evolution of bulk quantities provide an effective tool for assessing the behavior of interacting agents in complex environments.
Such PDEs can be used to efficiently tune systems with many parameters, since numerical simulation of them outpaces Monte-Carlo simulation of their stochastic microscopic counterparts.
Such tuning can uncover optimal agent behavior (\textit{e.g.} optimal evacuation dynamics).

\section*{Funding}
This research has been partially supported by NSF grants DMS 1413437 (William Ott) and DMS 1620278, {ONR N00014-17-1-2845}, and
{DOE DE-SC0019130} (Ilya Timofeyev).


\end{document}